\documentclass[fleqn,10pt]{wlscirep}
\usepackage[utf8]{inputenc}
\usepackage[T1]{fontenc}
\usepackage{lineno}

\usepackage{mathrsfs}
\usepackage{graphicx}
\usepackage{multicol}
\usepackage{csquotes}
\usepackage{comment}
\usepackage{soul}
\usepackage{color}
\usepackage{algorithmicx}
\usepackage{algorithm}
\usepackage{algpseudocode}
\usepackage{placeins}
\usepackage{amssymb}
\usepackage{caption}
\usepackage{subcaption}
\usepackage{xcolor}
\usepackage{xpatch}
\usepackage[mathscr]{euscript}

\algrenewcommand\alglinenumber[1]{{\sffamily\footnotesize#1}}
\makeatletter
\xpatchcmd{\algorithmic}{\itemsep\z@}{\itemsep=1ex plus0pt}{}{}
\makeatother

\definecolor{MMColor}{HTML}{00F9DE}
\definecolor{Jade}{HTML}{00A36C}

\newcommand\T{\rule{0pt}{7ex}}       
\newcommand\B{\rule[-4ex]{0pt}{0pt}} 

\title{High resolution synthetic residential energy use profiles for the United States}

\author[1,3]{Swapna Thorve}
\author[1]{Young Yun Baek}
\author[1]{Samarth Swarup}
\author[1,2]{Henning Mortveit}
\author[1]{Achla Marathe}
\author[1,3]{Anil Vullikanti}
\author[1,3]{Madhav Marathe}
\affil[1]{Network Systems Science and Advanced Computing, Biocomplexity Institute and Initiative, University of Virginia}
\affil[2]{Engineering Systems and Environment, University of Virginia}
\affil[3]{Department of Computer Science, University of Virginia}

\affil[*]{corresponding author(s): Madhav Marathe (marathe@virginia.edu), Swapna Thorve (st6ua@virginia.edu)}

\begin{abstract}
Efficient energy consumption is crucial for achieving sustainable energy goals in the era of climate change and grid modernization. Thus, it is vital to understand how energy is consumed at finer resolutions such as household in order to plan demand-response events or analyze impacts of weather, electricity prices, electric vehicles, solar, and occupancy schedules on energy consumption. However, availability and access to detailed energy-use data, which would enable detailed studies, has been rare. In this paper, we release a unique, large-scale, digital-twin of residential energy-use dataset for the residential sector across the contiguous United States covering millions of households. 
The data comprise of hourly energy use profiles for synthetic households, disaggregated into Thermostatically
Controlled Loads (TCL) and appliance use. 
The underlying framework is constructed using a bottom-up approach. 
Diverse open-source surveys and first principles models are used for end-use modeling.
Extensive validation of the synthetic dataset has been conducted through comparisons with reported energy-use data. We present a detailed, open, high resolution, residential energy-use dataset for the United States. 

\end{abstract}

\begin{document}
\flushbottom
\maketitle
\thispagestyle{empty}

\section*{Background \& Summary}

Modernization of the U.S. electric grid is occurring at a noteworthy rate due to the installation of new technologies within the grid such as smart meters. They enable two-way communication between the customer and utilities, providing information and granular control of power usage for individual households~\cite{Hart2008,Mohassel2014}.
The grid is also witnessing rapid transformations due to increasing penetration of electric vehicles (EV) and distributed energy resources (DER) such as rooftop photovoltaics (PV), community solar, and wind energy.
While this wave of modernization is beneficial, the electric grid is simultaneously facing a sharp increase in crisis situations as a result of climate change phenomena~\cite{hailegiorgis2018,Auffhammer2017} such as extreme weather events and global warming.
One example of extreme weather is the February 2021 North American cold wave that caused a tremendous strain on the power grid especially in Texas where millions lost power for days~\cite{Rai2021}. 
Another example is where global warming impacts household HVAC energy use. Although the rise of~1$^\circ$ to~2$^\circ$C in winter temperatures is expected to decrease heating requirements, a similar rise in summer temperatures is expected to increase cooling needs significantly~\cite{Petri2015}. 

In the face of these challenges, achieving sustainable energy goals has become paramount for maintaining a healthy grid.
To this end, the research community is faced with important questions regarding reduction of carbon footprints~\cite{Goldstein19122,NAP2021,Gillingham2021,Berrill2021_iop,Berrill2021_acs}, incentivizing DER adoption~\cite{MIT2011}, studying benefits of building energy retrofit~\cite{DEB2021116990,Nutkiewicz2021,Gillingham2021}, integration of electric vehicles~\cite{Muratori2018} and consumer behavior~\cite{Mahdavi2021} in the grid, and mechanisms for designing electricity pricing~\cite{Tanaka2021,Tsaousoglou2019} to create efficient residential consumption patterns. 
Answering many of these questions requires comprehensive knowledge of energy-use patterns, building stock, the structure of distribution networks, consumer behaviors, and so on.
However, such exhaustive datasets are rarely freely available (or available at all) for research use, making it hard for the research community to pursue these endeavours~\cite{nas2016}.
Reasons for unavailability of such data range from privacy concerns to the lack of a system for making data available to researchers.

\begin{figure*}[!h]
    \centering
    \includegraphics[width=17cm]{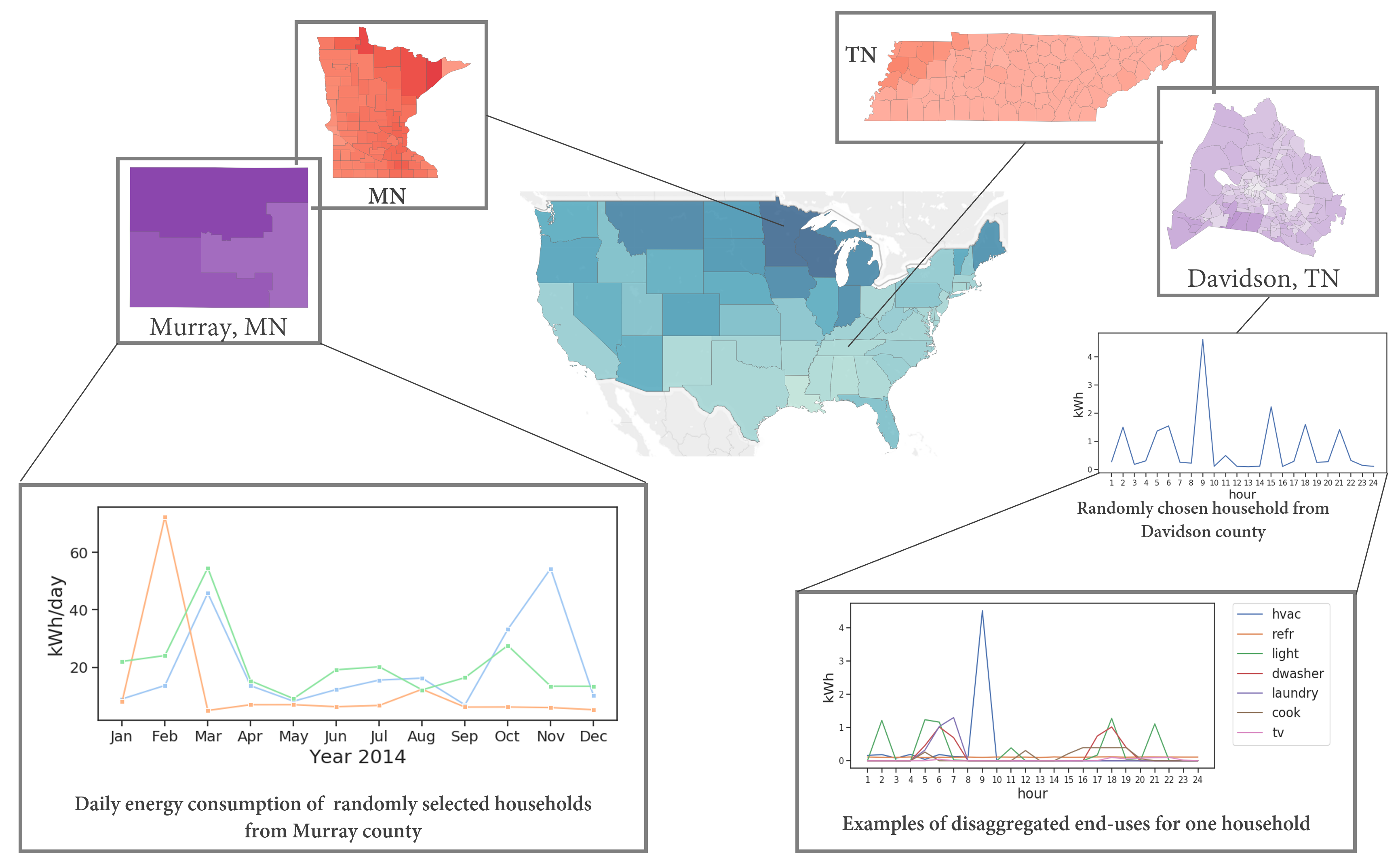}
    \caption{\textbf{Data overview.}~This figure shows examples of the spatio-temporal resolutions of multiple facets of the dis-aggregated synthetic energy demand data. The figure shows sample data at state, county, and household level at different temporal granularities. The data is generated for all households in the U.S.}
    \label{fig:pitch_picture}
\end{figure*}

\begin{table}[!ht]\caption{Energy-use datasets published in the residential sector.}
    \centering
   \begin{tabular}{l l}
   \hline
    \textbf{Authors/Dataset} &  \textbf{Description} 
    \rule{0pt}{3ex} \rule[-1.5ex]{0pt}{0pt} \\
    \hline
    
    Klemanjak et al.~\cite{Klemenjak2020,SynD_DOI} & \parbox{13cm}{
    A synthetic energy demand dataset was released for~21 appliances in Austria in~2020. Data collected from two households was used to train models and then appropriate noise was added for appliance start times and durations to mimic variations in actual consumption patterns.
    }
    \rule{0pt}{5ex} \rule[-1.5ex]{0pt}{0pt} \\

    Kolter et al.~\cite{Kolter11redd,REDD_DOI} & \parbox{13cm}{The Reference Energy Disaggregation Data Set (REDD) is published by MIT. The dataset contains high-frequency current/voltage waveform data of the power mains in households along with labeled circuits in the house.}
    \rule{0pt}{6ex} \rule[-1.5ex]{0pt}{0pt} \\

    Makonin et al.~\cite{RAE-Makonin-17} & \parbox{13cm}{The Rainforest Automation Energy (RAE) dataset was published by Harvard in~2017. The dataset contains~1Hz data (mains and sub-meters) from two residential houses.}
    \rule{0pt}{5.5ex} \rule[-1.5ex]{0pt}{0pt} \\

    Murray et al.~\cite{Murray2017,Murray2017_doi}& \parbox{13cm}{Load measurements from~20 households of UK from a two year longitudinal study. 
    } 
    \rule{0pt}{4ex} \rule[-1.5ex]{0pt}{0pt} \\

    Pecan Street\cite{pecanstreet,pecan_doi} & \parbox{13cm}{Labeled circuit data for households across major cities in the U.S. This is said to be the most comprehensive dis-aggregate energy data available for the U.S. }
    \rule{0pt}{4ex} \rule[-1.5ex]{0pt}{0pt} \\

    Rashid et al.~\cite{Rashid2019,iblend_doi} & \parbox{13cm}{The I-blend dataset has recorded  minute-level consumption of all the buildings at an academic institute in India over a period of~52 months}
    \rule{0pt}{5ex} \rule[-1.5ex]{0pt}{0pt} \\

    Paige et al.~\cite{Paige2019,flEECe_doi} & \parbox{13cm}{The flEECe dataset provides energy data at a~1Hz sampling rate for four circuits for six net-zero energy senior housing units in Virginia, USA for nine months}
    \rule{0pt}{5ex} \rule[-1.5ex]{0pt}{0pt} \\

    Shin et al.~\cite{Shin2019,enertalk_doi} & \parbox{13cm}{The first Korean dataset measuring appliance-level energy data was released in~2019 for~22 houses in Korea.}
    \rule{0pt}{5ex} \rule[-1.5ex]{0pt}{0pt} \\

    Kelly et al.~\cite{UK-DALE,ukdale_doi} & \parbox{13cm}{Power demand is recorded from five houses UK houses at two levels -- whole house and individual appliances. This dataset is referred to as the UK-Dale dataset. Two versions of this dataset have been released.}
    \rule{0pt}{6ex} \rule[-1.5ex]{0pt}{0pt} \\

    Anderson et al.~\cite{Anderson2012BLUED,blued_doi} & \parbox{13cm}{Building-Level fUlly-labeled dataset for Electricity Disaggregation (BLUED) for one household in Pittsburg U.S. for one week.  State transition of appliances are labeled and time-stamped, providing the necessary ground truth for the evaluation of NILM algorithms. }
    \rule{0pt}{6ex} \rule[-1.5ex]{0pt}{0pt} \\

    Barker et al.~\cite{Barker12anopen,smartstar_doi} & \parbox{13cm}{Electricity usage data is monitored every minute from nearly every plug load from~400 anonymous homes.
    }
    \rule{0pt}{4.5ex} \rule[-1.5ex]{0pt}{0pt} \\

    Beckel et al.~\cite{Beckel2014_ECO} & \parbox{13cm}{Electricity consumption is monitored via smart plugs for six households in Switzerland over a period of~8 months.}
    \rule{0pt}{4.5ex} \rule[-1.5ex]{0pt}{0pt} \\

    Pereira et al.~\cite{Pereira2014SustDataAP,SustData_doi,Pereira2022} & \parbox{13cm}{Power usage for~44 apartments and~6 homes in Portugal is collected for~264 days at~30 minute intervals. The advanced version of this dataset `SustDataED2' dataset contains 96 days of aggregated and individual appliance consumption from one household in Portugal.}
    \rule{0pt}{6ex} \rule[-1.5ex]{0pt}{0pt} \\
    
    Monacchi et al.~\cite{GRREND,greend_doi} & \parbox{13cm}{Common household devices are monitored for power consumption in Austria and Italy (GREEND dataset).}
    \rule{0pt}{4.5ex} \rule[-1.5ex]{0pt}{0pt} \\

    Pullinger et al.~\cite{Pullinger2021,ideal_doi} & \parbox{13cm}{1-second electricity data is gathered over a period of~23 months from 255 UK homes (IDEAL household energy dataset).}
    \rule{0pt}{4.5ex} \rule[-1.5ex]{0pt}{0pt} \\

     Ruhnau et al.~\cite{Ruhnau2019,Ruhnau_doi} & \parbox{13cm}{Synthetic national time series of heat demand that covers over~16 countries in the EU from~2008 to~2018.}
    \rule{0pt}{4.5ex} \rule[-3ex]{0pt}{0pt} \\
    \hline
    
   \end{tabular}
   \label{tab:lit-syn-data}
\end{table}

Most of the published energy use data are metered data, a result of longitudinal studies conducted by researchers (Table~\ref{tab:lit-syn-data}) with relatively small samples of households that may not be representative of the wider geographical region and demographics. 
Some of these studies monitor households over a longer period of time (e.g. two years), however, the downside of such experiments is that it takes a considerable amount of time (e.g. participant consent, equipment setup, monitoring) and manual effort (e.g., data cleaning, imputing missing values) before such data is usable.
Although these studies release energy data for free use, many of them limit publishing participant details (e.g. building characteristics and location, household level demographics). 
Participant details are usually withheld due to privacy reasons/participant consent, lack of information, or unavailability of these attributes in the free version of the data.
Literature has attempted to address some of these issues by creating appropriate data structures for releasing appliance metadata information for households alongwith their energy use data~\cite{Kelly_2014_metadata,UK-DALE}.
However, we observe that many of the issues still persist in the U.S. context. One such example is the Pecan Street Dataport~\cite{pecan_doi}. 
Pecan Street Inc.~\cite{pecanstreet} is the largest publisher of energy-use data in the U.S. through their portal -- \emph{Dataport}. They collect energy-use data in California (CA), Texas (TX), New York (NY), and Colorado (CO). This is a potentially very useful data set.
However, only a small sample ($\sim$25 households in CA and TX) of energy-use data is freely available for public use and do not contain sufficient (or any) demographic or building information.

A dataset synthesized over a larger spatial scope offers the opportunity to study regional and temporal differences in energy use while a smaller region dataset offers studying energy use patterns that may be particular to the region.
Irrespective of spatial scope, small sample size makes it difficult to get a good representation of the population variation in the region (e.g. explaining/exploiting role of household demographics, behavior, and building characteristics in energy use).
In addition to the spatial scope and number of samples,  many of the datasets do not release sufficient (or any) participant details.
Such limited data restricts the usage of these energy-use data for detailed practical analyses or studying scenario interventions and equity questions in the grid (e.g., which type of demographic and building stock is best suited for EV adoption, or how much carbon footprint can be reduced by retrofitting buildings). 
Thus, we observe that there is a general sparsity of large scale high resolution energy use datasets along with detailed metadata information at household level such as appliance ownership, building data, important demographic features.

We summarize key drawbacks of energy datasets for the U.S. as follows -- limited spatial scope, small sample size, lack of sufficient household, appliance, \& building metadata.
Given these wide array of problems with the state-of-art energy-use data availability, we introduce synthetic energy use datasets that are able to address many of these issues. 
Synthetic data is defined as data generated by models that provide accurate statistical representations of the real world. 
Examples of such data for the smart grid are synthetic power distribution networks~\cite{Meyur2020}, energy consumption profiles for offices and commercial buildings~\cite{LiLBNL2021} and for residential buildings~\cite{Thorve2018,Klemenjak2020,ROTH2020,Tong2021}.
Our work specifically addresses the data scarcity gap in energy use research for the U.S. residential sector.
We propose a synthetic framework for modeling large-scale high resolution energy use data by integrating diverse datasets and end-use models for bottom-up dis-aggregate energy modeling.
This results in a novel synthetic energy use dataset (i.e., a digital twin of household level energy demand) comprising hourly electrical energy demand profiles for 
U.S. households.
The total electrical energy use is published as a composition of eight primary end-uses in a household -- heating/air-conditioning (HVAC), lighting, dishwashing, cooking, laundry (clothes washer and  clothes dryer), refrigeration, hot water, and miscellaneous plug load (vacuuming, computer use, TV).
A detailed data-intensive bottom-up framework is developed to generate synthetic energy-use profiles by integrating multiple open-source surveys and a synthetic population for the U.S~\cite{SPEW}.
A mixture of methods (stochastic, machine learning, physics-based engineering methods) is used to model different end-uses in all households that consume electricity as a primary fuel across the~48 contiguous states and Washington, D.C. in North America.
To the best of our knowledge, this synthetic energy-use dataset is the first detailed, large-scale, freely available household-level electricity consumption behaviors dataset for the U.S.
Our synthetic energy-use infrastructure is well-suited to solve the newer smart grid problems mentioned earlier.
We publish the dis-aggregated energy use timeseries for all the synthetic households.
The published data is representative of the U.S. households, provide household level metadata, and are a good representation of the real world energy use.

\begin{table*}[!h]
\centering
  \caption{List of primary datasets used for constructing the residential demand models.} \label{tab:datasets}
  
\begin{tabular}{l l}
    \hline
    
    \textbf{Dataset}  & \textbf{Description} 
    \rule{0pt}{3ex} \rule[-1.5ex]{0pt}{0pt} \\
    
    \hline
    
    \begin{tabular}{@{}l@{}} 
    American Time
    Use \\Survey
    (ATUS 2015)
    \end{tabular} 
    & \parbox{13cm}{
    ATUS provides nationally representative estimates of how, where, and with whom
    people in the U.S. spend their time, and is the only federal survey providing
    data on the full range of activities, from childcare to volunteering. This survey
    provides demographic information as well as information on energy-related activities~\cite{ATUS2015}. 
    24-hour data is recorded for~5115 participants.
    }
    
    \rule{0pt}{8ex} \rule[-7.25ex]{0pt}{0pt} \\
    
    \begin{tabular}{@{}l@{}} 
    Synthetic Populations \\and Ecosystems of \\the World (SPEW)
      \end{tabular} 
    & \parbox{13cm}{ 
    SPEW~\cite{Gallagher2018,SPEW} is a framework that produces synthetic populations for various countries. We used the open-sourced version of the synthetic population available for the U.S. constructed for the year~2013.
    The sampled base population is the byproduct of American Community Survey (ACS) Public Use Microdata Sample (PUMS) data.
    Statistical methods such as Simple Random Sampling (SRS) and Iterative Proportional Fitting~(IPF)~\cite{deming40ipf,Fienberg} are used to estimate joint distributions of population characteristics given their marginal distributions at a small geographic level (e.g. PUMA-level for the U.S.).
    Data records are available at household level for all of U.S. Descriptors are available for mapping records from PUMS data onto the base synthetic population.   
    } 
	
    \T\rule[-12.8ex]{0pt}{0pt} \\
    
    \begin{tabular}{@{}l@{}} 
    Public Use 
    Microdata\\ Sample 
    (PUMS 2013)
    \end{tabular}  
    & 
    \parbox{13cm}{
    PUMS is a~5\% representative sample for a larger region than block group referred to as a Public Use Microdata Area (PUMA)~\cite{Pums2013}. PUMAs are described by the Census 
    as ``a collection of counties or tracts within counties with more than~100,000 
    people''. These statistical areas are defined for the circulation of PUMS data.
    PUMS contains individual records of the characteristics for a~5\% sample of 
    people and their households. One PUMS record is a complete Census record. 
    }

    \T\rule[-7ex]{0pt}{0pt} \\
    
    \begin{tabular}{@{}l@{}} 
    North American Land\\
    Data Assimilation \\System
    (NLDAS)
    \end{tabular}  
    & \parbox{13cm}{ 
    Hourly temperature data for North America.
    Data resolution is at~1/8th-degree grid over North America~\cite{NLDAS}.
    }   
    
    \T\rule[-5.5ex]{0pt}{0pt} \\
    
    \begin{tabular}{@{}l@{}} 
    Residential Energy\\
    Consumption Survey\\
    (RECS 2015)
    \end{tabular} 
    & \parbox{13cm}{ 
    U.S. Energy Information Administration (EIA) Residential Energy Consumption
    Survey (RECS)~\cite{RECS2015} data is a national sample survey that collects energy-related
    data for housing units. For~2015, data was collected from~5,686 households
    to represent~118.2 million U.S. households. We use this dataset to obtain housing unit-specific information  
    such as floor area, main heating fuel, fuel equipment, indoor temperature setting,
    presence of air conditioner, dishwasher, washer, dryer, refrigerator,
    water heater fuel, water heater size, water heater age, number of lighting units, etc,.
    }

    \T\rule[-10ex]{0pt}{0pt} \\
    
    \begin{tabular}{@{}l@{}} 
    National Solar Radiation\\ Database (NSRDB)
    \end{tabular}
    &
    \parbox{13cm}{ 
    NREL provides solar radiation data for the U.S. We use hourly data that comes from the physics-based approach called the Physical Solar Model (PSM). Data is available for the U.S. for~1998--2014~\cite{NSRDB2020}.
    The GHI variable is used as an indicator of irradiance level in the lighting model. GHI is modeled solar radiation on a horizontal surface received from the sky.
    This is measured in $\frac{\textrm{watt}}{\textrm{meter}^2}$. 
    }
    \T\B \\
    
    Miscellaneous
    &
    \parbox{13cm}{
    Appliance power and efficiencies, gallons of hot water required for activities, and any other input data required for models is drawn from surveys and data collected from ground and/or testing~\cite{NIST_cook,NIST_dw,NRELhotwater129,NRELhotwater61}.
    }
    \T\B \\
    \hline
    
\end{tabular}
\end{table*}

\begin{table}[!h]\caption{Notations}
    \centering
   \begin{tabular}{c l}
   \hline
    \textbf{Notation} &  \textbf{Description} 
    \rule{0pt}{3ex} \rule[-1.5ex]{0pt}{0pt} \\
    \hline
    
    $H_i$ & Household~$i$ drawn from the synthetic population 
    \rule{0pt}{3ex} \rule[-1.5ex]{0pt}{0pt} \\
    
    $P_{i,j}$ & Synthetic household member~$j$ of household~$H_i$
    \rule{0pt}{0ex} \rule[-1.5ex]{0pt}{0pt} \\
    
    $A_k$ & Respondent~$k$ from ATUS survey  
    \rule{0pt}{0ex} \rule[-1.5ex]{0pt}{0pt} \\
    
    $S_l$ & Household~$l$ from RECS survey
    \rule{0pt}{0ex} \rule[-1.5ex]{0pt}{0pt} \\
    
    ${\mathsf{Irr}}^{\mathsf{i}}$ & Irradiance threshold for~$H_i$. Drawn from a Normal distribution $\mathsf{Normal}(60,10)$
    \rule{0pt}{0ex} \rule[-1.5ex]{0pt}{0pt} \\
    
    $\langle O_{i,0},\ldots,O_{i,t},\ldots,O_{i,23}\rangle$ & Occupancy time series of synthetic household $i$ over~24 hours, $t \in \{0,1,\ldots,23\}$
    \rule{0pt}{0ex} \rule[-1.5ex]{0pt}{0pt} \\
    
    $\langle \mathrm{Irr}_{0},\ldots, \mathrm{Irr}_{t},\ldots, \mathrm{Irr}_{23}\rangle$ & Hourly irradiance time series of a census tract for a given day in the year~2014 
    
    \rule{0pt}{0ex} \rule[-1.5ex]{0pt}{0pt} \\
    
    $\langle T^{\mathsf{out}}_0,\ldots, T^{\mathsf{out}}_t,\ldots, T^{\mathsf{out}}_{23}\rangle$ & Hourly temperature series of the outside environment for a given day~($^{\circ}F$)
    
    \rule{0pt}{0ex} \rule[-1.5ex]{0pt}{0pt} \\
    
    $\langle T^{\mathsf{in}}_0,\ldots, T^{\mathsf{in}}_t,\ldots, T^{\mathsf{in}}_{23}\rangle$ & Thermostat setpoint~($^{\circ}F$) 
    
    \rule{0pt}{0ex} \rule[-1.5ex]{0pt}{0pt} \\
    
    $\eta$ & Efficiency of the HVAC equipment and water heaters 
    
    \rule{0pt}{0ex} \rule[-1.5ex]{0pt}{0pt} \\
    
    $R^{\mathsf{roof}}$ , $R^{\mathsf{wall}}$ & Thermal resistance coefficient for roof and wall for different climate zones
    
    \rule{0pt}{0ex} \rule[-1.5ex]{0pt}{0pt} \\
    
    $T^{\mathsf{hot}}_{v}$ & \parbox{12cm}{Temperature ($^{\circ}F$) of hot water end-point category $v$,\\ where $v\;\in\;\{\mathsf{shower,bath,cwasher,dishwasher}\}$} 
    
    \rule{0pt}{1ex} \rule[-1.5ex]{0pt}{0pt} \\
    
    $T^{\mathsf{cold}}_{m,z}$ & Mains water temperature ($^{\circ}F$) for month $m$ and climate zone $z$   
    
    \rule{0pt}{1ex} \rule[-1.7ex]{0pt}{0pt} \\
    
    
    
    $d\in\mathsf{D}$ & \parbox{12cm}{End-use $d\in\mathsf{D}$ where $\mathsf{D} = \{\mathsf{hvac}, \mathsf{h2o}, \mathsf{light},  \mathsf{refr},  \mathsf{dwasher}, \mathsf{cook}, \mathsf{cwasher}, \mathsf{cdryer},  \mathsf{TV},\\ \mathsf{computer}, \mathsf{cleaning} \}$}
    
    \rule{0pt}{1ex} \rule[-3.5ex]{0pt}{0pt} \\
    
     $\langle E^d_{i,0},E^d_{i,t},\ldots,E^d_{i,23} \rangle$ & 
     \parbox{12cm}{
     Hourly energy use profile of~$H_i$ for a end-use ${d}$ and $t\in\{0,\ldots,23\}$ }

    \rule{0pt}{0ex} \rule[-2.9ex]{0pt}{0pt} \\
    
    $E^{d}_{i}$ & \parbox{12cm}{Daily energy consumed over 24 hours by end-use $d$ in household~$H_i$. $E^{d}_{i} = \sum^{23}_{t=0} E^{d}_{i,t}$ and $d\in\mathsf{D}$ and $t\in\{0,1,\ldots,23\}$}
   
   \rule{0pt}{0ex} \rule[-3.7ex]{0pt}{0pt} \\
   
   $\langle G^{\mathsf{h2o}}_{i,0},G^{\mathsf{h2o}}_{i,t},\ldots,G^{\mathsf{h2o}}_{i,23} \rangle$ &
   \parbox{12cm}{
   Hourly profile of hot-water use (gallons per hour) of~$H_i$ for a end-use $\mathsf{h2o}$ and $t\in\{0,\ldots,23\}$. $G^{\mathsf{h2o}}_{i,t} = \sum^{}_{v\in\mathsf{V}}G^{\mathsf{h2o}}_{i,t,v}$ where $\mathsf{V} = \{\text{shower}, \text{bath}, \text{dishwasher}, \text{clothes washer}\}$ 
   }
    
    \rule{0pt}{0ex} \rule[-2.7ex]{0pt}{0pt} \\   
    
   ${G}^{\mathsf{h2o}}_{i}$ & 
    \parbox{12cm}{
   Daily amount of hot water consumed (in gallons) by a household~$H_i$ in a day.\\ ${G}^{\mathsf{h2o}}_i = \sum^{23}_{t=0} G^{\mathsf{h2o}}_{i,t}$}
   
   \rule{0pt}{1ex} \rule[-2.8ex]{0pt}{0pt} \\
   
   ${G}^{\mathsf{h2o}}_{i,v}$ & 
    \parbox{12cm}{
   Daily amount of water consumed (in gallons) by a household~$H_i$ in a day by an event $v$.\\ ${G}^{\mathsf{h2o}}_{i,v} = \sum^{23}_{t=0} G^{\mathsf{h2o}}_{i,t,v}$}
   
    \rule{0pt}{3ex} \rule[-3ex]{0pt}{0pt} \\
    \hline
    
   \end{tabular}
   \label{tab:notations}
\end{table}

\section*{Methods}

\begin{figure*}[!h]
    \centering
    \includegraphics[width=17cm, height=18cm]{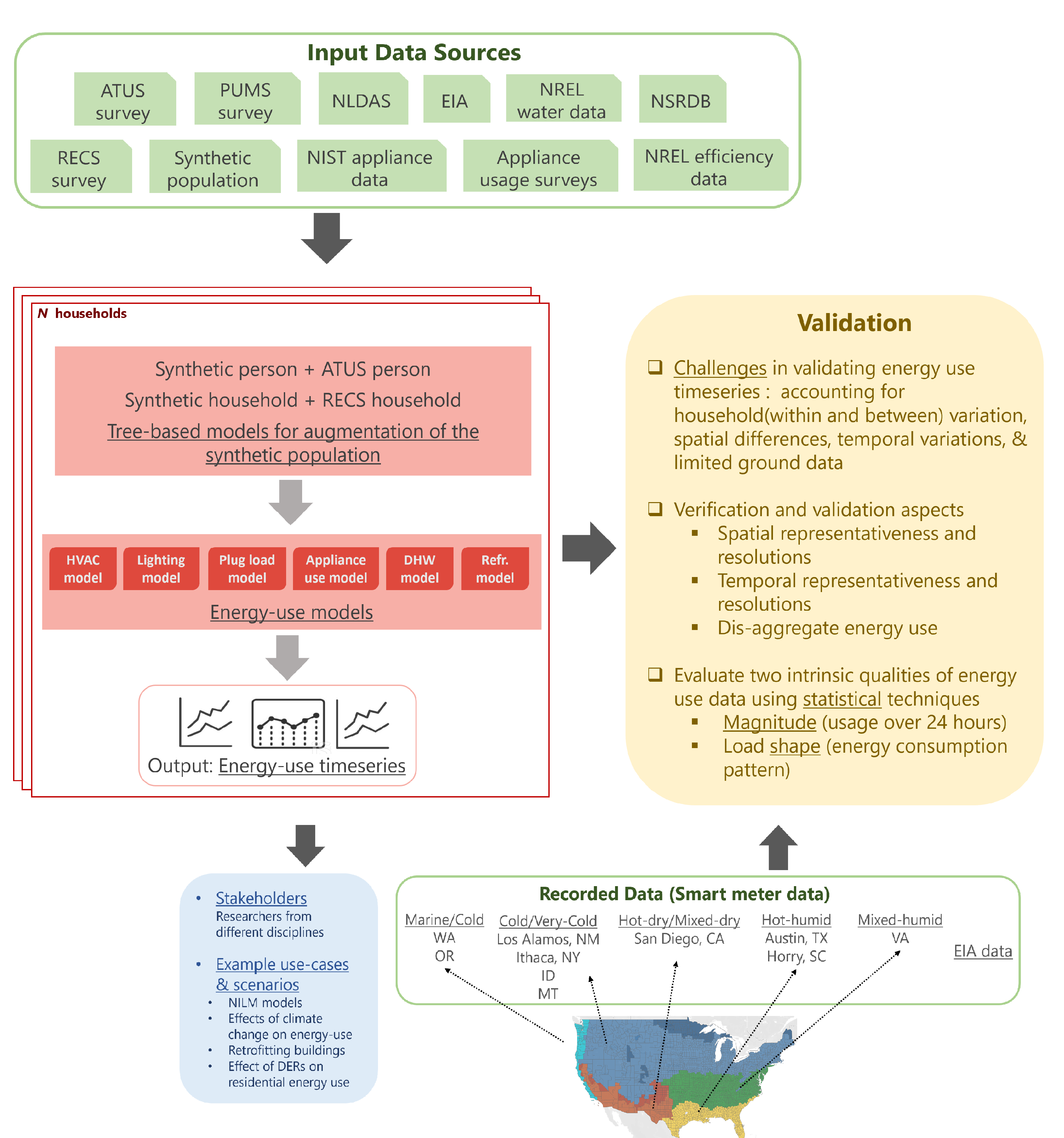}
    \caption{\textbf{Overview of the energy modeling infrastructure.} Many different types of input data are used in the proposed modeling framework. These are shown at the top. For complete description of input datasets refer to Table~\ref{tab:datasets}. These datasets are input to different modeling components of the framework. Some datasets support augmentation of the synthetic population while others are input to the energy-use models. All the models are described in the \emph{Methodology} section.  The bottom rectangle describes the recorded data/smart meter data from different climate zones of the U.S. These datasets are used for validation of the synthetic energy-use timeseries. The validation block (yellow backdrop) describes three components of V\&V -- regional, magnitude, and structural/shape comparisons. This line of validation covers (a) different temporal aspects (hourly and daily), (b) spatial aspects in terms of regions and seasons, (c) diversity aspect of the large-scale synthetic data. The blue text refers to the V's of big data. Each colored block possesses the given V characteristic.}
    \label{fig:big_picture_energy}
\end{figure*}

This section describes the datasets and models employed to generate synthetic energy use time series at the household level, see Table~\ref{tab:datasets}. All notations used in the paper are described in Table~\ref{tab:notations}.

\noindent
The presented framework is composed of a synthetic representation of the U.S. population,  regression models for surveys, and bottom-up energy use models.
A synthetic population is composed of households and people in households. 
The synthetic households are generated using census surveys and statistical methods such that the synthetic population is \emph{statistically similar} to the original population.
An open-source version of the U.S. synthetic population -- Synthetic Populations and Ecosystems of the World (SPEW)~\cite{Gallagher2018,SPEW} is used in our framework.
The SPEW synthetic population is comprised of demographic characteristics of synthetic households and synthetic individuals. The synthetic population is created using U.S. census data such as PUMS (Table~\ref{tab:datasets}) and statistical methods such as sampling and the Iterative Proportional Fitting (IPF) method~\cite{BECKMAN1996415}. 

The SPEW households are made of basic demographic (e.g., income, age) and locality information.
Although the SPEW population is representative of the U.S. population on a finer spatial resolution, it is not equipped with energy and activity related information (e.g., building characteristics, time spent at home, number of cooking activities) necessary for estimating energy use at household level or person level.
Building stock, energy and activity related information is collected by national surveys in the U.S. -- Residential Energy Consumption Survey RECS~\cite{RECS2015} and American Time Use Survey ATUS~\cite{ATUS2015} respectively.
The basic synthetic population is augmented with energy and activity related attributes by building machine learning models. This augmentation is called as the \emph{enrichment step}. 
The enriched synthetic population along with other freely available data sources can be used together as inputs to the energy use modeling framework. The energy use modeling framework has six models for representing nine energy uses -- HVAC, lighting, domestic hot-water, refrigerator, dishwasher, cooking, clothes washer, clothes dryer, and  miscellaneous plug load such as TV, computer use, cleaning activities (e.g., vacuuming).
The first subsection describes the modeling details of the \emph{enrichment step} and the following subsection describes energy demand models.

\subsection*{Enrichment models}
The enrichment models support creating comprehensive synthetic structures for calculating residential energy usage. This step is called as the \emph{enrichment step}. Refer to Figure~\ref{fig:big_picture_energy} for a pictorial representation of the overview of the framework. Datasets used in this workflow are described in Table~\ref{tab:datasets}.
Since the demographic features available in the synthetic population are not sufficient for computing energy usage, it is made richer by adding layers of information related to building stock and energy consumption from the RECS survey such as building characteristics, appliance ownership, and thermostat set-point behaviors. 
This mapping of features is made by building inference tree models.
Activity schedules for a normative day of an ATUS survey respondent  are attached to synthetic individual by building a multivariate random forest regression model.
These models are described below.

\hfill \break
\noindent
\textbf{The ATUS model.}~The ATUS data provides nationally representative surveys of people's activities in different location types such as childcare in or outside the house, time spent at work, laundry time at home, waiting times in hospital, and so on, see Table~\ref{tab:datasets} for a description.
The time-use diaries of the survey individuals can be attached to synthetic individuals by matching an appropriate survey individual to a synthetic individual.
In our work, we consider \emph{appropriate matching} based on amount of time a person spends in different location types such as home, work, school, shopping, and other miscellaneous locations.
This seems a reasonable approach because we are interested in learning how an individual spends 24 hours of the day by categorizing the amount of time spent  at important location types -- for e.g., the time spent in different location types for a person works full-time is quite different than a house bound senior citizen or a college student. 
This rationale of assigning survey respondents to synthetic individuals is also presented in prior work by Lum et al~\cite{Lum2016}.

\begin{figure}[!h]
    \centering
    \includegraphics[width=15cm]{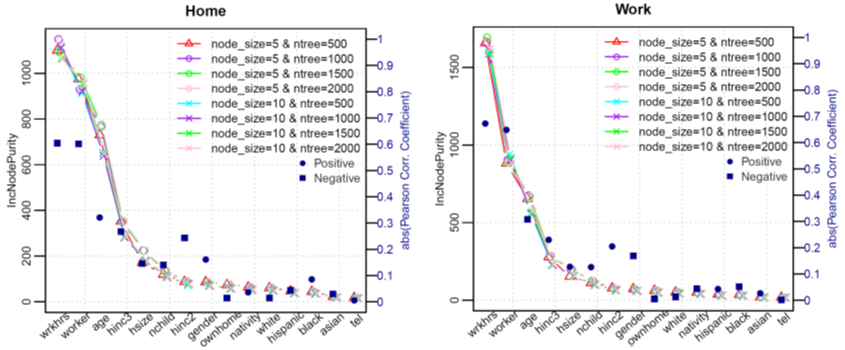}
    \caption{\textbf{Impurity-based feature importance and correlation.}~Each plot shows Gini importance of features for two dependent variables -- home and work. The x-axis shows independent variables in order of importance based on \emph{IncNodePurity}. The selection of the parameters for `ntree' (number of decision trees) and `node size' (minimum size of terminal nodes). Eight conditions are tested for the combination of the two parameters: ntree=500, 1000, 1500, and 2000; node size=5, and 10. The plots show robust results across the different conditions. According to the plots, the following five independent variables -- $\mathsf{wrkhrs, worker, age, hinc3, hsize}$ mostly affect all the dependent variables. The right-hand y-axis shows the absolute Pearson Correlation Coefficient. The positive and negative coefficients are distinguished by blue dots and squares, respectively. Except $\mathsf{wrkhrs, worker}$, all other independent variables weakly correlated with the dependent variables.}
    \label{fig:atus_fi_maintext}
\end{figure}

Random forest regression method is used to build a model that predicts the amount of time a person spends in locations types such as home, work, shopping, other, school, and trip counts during the day.
Thus, six dependent variables are modeled -- trip count during the day and time spent at each location type - home, work, shopping, other, school.
Independent variables used to build the  model are as follows -- number of members in the household ($\mathsf{hsize}$), number of children ($\mathsf{nchild}$), age ($\mathsf{age}$), working hours ($\mathsf{wrkhrs}$), gender ($\mathsf{gender}$), income modeled as a categorical variable ($\mathsf{hinc2, hinc3}$), and binary variables such as an American citizen or not ($\mathsf{nativity}$), worker or not ($\mathsf{worker}$), owns home or not ($\mathsf{ownhome}$), has a phone or not ($\mathsf{tel}$), and race related variables such as if person is white, Hispanic, black, or Asian ($\mathsf{white}$, $\mathsf{hispanic}$, $\mathsf{black}$, $\mathsf{asian}$). Figure~\ref{fig:atus_fi_maintext} shows example of feature importance for two dependent variables. 

Once the model is trained on ATUS respondents, a synthetic person~$P_{i,j}$ is randomly assigned a survey individual from the leaf nodes in the trained ensemble model.
Thus, the result gives every synthetic individual a time-use diary.
The energy-use models will extract home activities from a time-diary and also build a household-level occupancy schedule over the 24-hour duration, denoted as~$\langle O_{i,0},O_{i,1},\ldots,O_{i,23}\rangle$.
These are used as an input to the energy use models.
Synthetic household member activity scheduling conflicts are handled in the activity model.

\hfill \break
\noindent
\textbf{The RECS mapping model.}
The baseline synthetic population does not have any building structural characteristics and appliance ownership information.
These salient features are important for modeling different categories of energy use and are available in the RECS survey.
We overlay RECS household attributes onto a synthetic household by building multivariate conditional inference trees~\cite{ctreeR,Hothorn2006ctree}. Conditional inference tree is a non-parametric class of regression trees that uses recursive partitioning of dependent variables based on the value of correlations.
Four dependent variables are modeled -- square footage of the dwelling, presence of laundry appliances, presence of air conditioner, presence of dishwasher.
The independent variables are year in which the house was built, occupancy time of the current tenants, own or rent the residence, total number of rooms, income, number of refrigerators, number of members in the household, dwelling type, dwelling is located in urban or rural area, primary heating fuel type.
The independent variables are common attributes between RECS survey records and synthetic household records.  
Conditional inference trees are trained on different census regions in the U.S. to tease out regional differences. 
A RECS household~$S_l$ is randomly selected from the appropriate leaf nodes of the conditional inference tree and assigned to the synthetic household~$H_i$ every time a new simulation is run. 
This dynamic assignment introduces stochasticity when the simulation is executed for same and/or different days.

\subsection*{Energy use modeling}

The enriched synthetic population (i.e., the output of the \emph{enrichment step}) enables encoding of behaviors (time spent in different energy related activities at home), normative attributes (e.g., square footage, age, income, gender), declarative attributes (e.g., individual activities as a sequence) and procedural attributes (e.g., behaviors capturing dependencies, interactions, frequency of performing activities) into the knowledge required for building  energy use profiles~\cite{Barrett:2018:HPS:3195636.3158342}. 
The synthetic infrastructure is leveraged to build six energy use models (Figure~\ref{fig:big_picture_energy}). 
Nine end-uses are synthesized for each household. These end-uses are divided into two parts -- Thermostatically Controlled Loads (TCL) and appliance use.
For a household $i$, nine end-uses published in the data are --
\begin{enumerate}
    \item \textbf{HVAC} ($E^\mathsf{hvac}$). This category includes heating and cooling electric load from central air conditioning during hot days and electric furnace/heater used during cold days. This is a TCL load.
    \item \textbf{Domestic hot water use} ($E^\mathsf{h2o}$). Energy consumed for heating water that is needed for personal grooming activities such as shower/bath, laundry activities such as using clothes washer, and dishwasher. This is a TCL load.
    \item \textbf{Dishwasher} ($E^\mathsf{dwasher}$). Energy used by dishwashers.
    \item \textbf{Clothes Washer} ($E^\mathsf{cwasher}$). Energy used by electric clothes washers.
    \item \textbf{Clothes Dryer} ($E^\mathsf{cdyer}$). Energy consumed by dryer.
    \item \textbf{Cooking} ($E^\mathsf{cook}$). Energy consumed by electric cooking range, oven, and other kitchen appliances such as coffee maker, microwave, toaster, etc.
    \item \textbf{Miscellaneous plug load}  ($E^\mathsf{misc}$). This type of energy indicates plug load attributed to cleaning activities and electronic devices such as TV, computers, other smaller electronic gadgets.
    \item \textbf{Refrigeration}  ($E^\mathsf{refr}$). Energy consumed by refrigerators.
    \item \textbf{Lighting}  ($E^\mathsf{light}$). Energy consumed by lighting units.
\end{enumerate}

Table~\ref{tab:notations} describe the notations used in the methodology sections.
The total energy summed over 24 hours (${E}^{\mathsf{total}}_{i}$) of a household $i$ is given by the equations below --

\begin{subequations}\label{eq:energy_eq}
\begin{align}
        {E}^{\mathsf{total}}_{i} &= {E}^{\mathsf{TCL}}_{i} + {E}^{\mathsf{appliances}}_{i}   \\
        {E}^{\mathsf{TCL}}_{i} &= {E}^{\mathsf{hvac}}_{i} + {E}^{\mathsf{h2o}}_{i} \\
        {E}^{\mathsf{appliances}}_{i} &= {E}^{\mathsf{dwahser}}_{i} + {E}^{\mathsf{cook}}_{i} + {E}^{\mathsf{cwasher}}_{i} + {E}^{\mathsf{cdryer}}_{i} + {E}^{\mathsf{light}}_{i} + {E}^{\mathsf{refr}}_{i} + {E}^{\mathsf{misc}}_{i}\\
        {E}^{\mathsf{misc}}_{i} &= {E}^{\mathsf{tv}}_{i} + {E}^{\mathsf{computer}}_{i} + {E}^{\mathsf{cleaning}}_{i}
\end{align}
\end{subequations}

\subsection*{HVAC model $E^{\mathsf{hvac}}$} 

According to the U.S. Energy Information Administration (EIA)~\cite{eia}, 
HVAC is responsible for the highest proportion of energy consumption in households.
The HVAC model calculates how much energy is required to maintain ambient/comfort temperature indoors.
This is dependent on factors ranging from the area of the house, outdoor temperature, efficiency of HVAC equipment, and so on.
Occupant behaviour of thermostat settings in different seasons and household occupancy during the day play an important role in understanding thermal comfort levels and how its effect on electricity consumption.
Engineering and statistical approaches~\cite{SWAN20091819} are presented in the literature to simulate energy consumption of heaters/furnace and air conditioners~\cite{MURATORI2013465,SHIMODA20071617,Tsuji2004,Subbiah2017}. 
We adopt the engineering based approach from Subbiah et al.~\cite{Subbiah2017} where the function of heating/cooling a household~$H_i$ at hourly intervals is defined as:
\begin{equation} \label{eq:hvac}
   E^{\mathsf{hvac}}_{i,t} = \frac{\Delta T}{\eta} \times \left(\frac{\textrm{FloorArea}_{i}}{R^{\mathsf{roof}}} \; +\; \frac{\textrm{WallArea}_{i}}{R^{\mathsf{wall}}} \right)
\end{equation}
Here~$E^{\mathsf{hvac}}_{i,t}$ is the energy consumed by household $H_i$ at the end of hour $t$ in kWh by heating/cooling equipment to maintain thermal comfort.
$\textrm{FloorArea}_{i}$ is the floor area and $\textrm{WallArea}_{i}$ is the wall area (extrapolated from floor area~\cite{Subbiah2017}) of $H_i$.
The quantities~$R^{\mathsf{roof}}$ and~$R^{\mathsf{wall}}$ are~R-values (insulation level) for households in different climate zones, while~$\eta$ is defined in Table~\ref{tab:notations}. Next,
$\Delta T$ is the absolute difference between $T^{in}_t$ and $T^{\mathsf{out}}_t$, and~$T^{\mathsf{in}}_t$ is indoor thermostat temperature at hour~$t$.
The hourly outside temperature ($T^{\mathsf{out}}_t$) is obtained from NOAA NLDAS data mentioned in Table~\ref{tab:datasets}. 
Efficiency and insulation data is obtained from guidelines published by EIA. 
All other household attributes are obtained from the enriched synthetic population.
Depending upon occupancy patterns throughout the day, changes in thermostat behaviors are assigned to each household.
Heating and cooling threshold temperatures for appliance on/off times are taken from the thermostat study published by NREL in~2017~\cite{NREL2017-thermostat-study}.

\subsection*{Domestic Water Heating Model $E^{\mathsf{h2o}}$ }

The EIA~\cite{eia}
\href{https://www.eia.gov/todayinenergy/detail.php?id=37433}{\textcolor{black}{EIA}} 
shows that 17\%-32\% of the household energy use is attributed to domestic hot water use (DHW).
Literature shows models used for estimating hot water demand at multiple temporal resolutions -- annual, daily, hourly, and minute intervals.
One of the initial models for estimating load profiles of hot water demand was developed in~2001 by Jordan et al.~\cite{Vajen2001H2o} for a period of one year for temporal resolutions of~1 min,~6 min, and~1 hour. However, this work does not consider historical nor factual flow rates to determine how much hot water ($\mathrm{gallons/day}$) is used by a household. A follow-up paper was developed for synthesizing water demand profiles for Switzerland~\cite{DESANTIAGO2017SwissH2O} by calibrating this model using field data.
A model to simulate yearly DHW event schedule for a single-family household was developed by Hendron et al.~\cite{Hendron2010H2o} from the National Renewable Energy Laboratory (NREL) in~2010.
The simulator used two surveys that collected information about water demand in U.S. households for five categories: sink, bath, shower, clothes washer, and dishwasher.
This model has been widely accepted in the literature. One recent example of the adaptation of Hendron's model is for simulating hot water demand in Canadian households~\cite{ROULEAU2019H2oCanada}. The model is calibrated for survey data collected for Canada and appropriate adjustments are made with respect to Canadian lifestyles.

For our model, we use the distributions of duration and flow rates of activities involving hot water usage such as bath/shower, clothes washer, and dishwasher from Hendron et al. 
Note that duration and flow rates can take negative values (Table~\ref{tab:dhw-specifics}). 
The flow rate is capped to~0.05gpm and the duration is capped to~1 minute for any negative value~\cite{Hendron2010H2o}.
Table~\ref{tab:dhw-specifics} characterizes the average count of daily events, duration, and flow rates.
The values of hot water temperature for different uses and the cold water inlet temperature are obtained from studies conducted by NREL in different regions of U.S.~\cite{NRELhotwater129,NRELhotwater61,Hendron2008Building}
An engineering based approach is used to estimate hot water usage~\cite{NRELhotwater129,Subbiah2017} in household $i$ for event~$v$ at time $t$
\begin{equation}\label{eq:dhw}
    \begin{array}{l}
     E^{\mathsf{hot}}_v = \frac{G^{\mathsf{hot}}_{v,i,t}\;\times\;\Delta T}{\eta} \times 0.00189\;, \quad \text{where} \\
     G^{\mathsf{hot}}_{v,i,t}\;=\;\mathsf{duration}_v\;\times\;\mathsf{flow\_rate}_v\;, \quad \text{and} \quad 
     \Delta T\;=\;T^{\mathsf{cold}}_{m,z}\;-\;T^{\mathsf{hot}}_v\;.
    \end{array}
\end{equation}
The gallons of hot water $G^{\mathsf{hot}}_{v,i,t}$ consumed by event $v$ is computed as a product of $\mathsf{flow\_rate}$ (gpm) and $\mathsf{duration}$ (minutes). 
Both these characteristics are drawn from distributions in Table~\ref{tab:dhw-specifics}. 
$E^{\mathsf{hot}}_v$ is the energy consumed by the event $v$ to heat $G^{\mathsf{hot}}_v$ gallons of water.
Last four entries in the Table~\ref{tab:notations} shows summation of multiple events occurring across the time horizon.
Here~$\eta$ is the efficiency of the electric water heaters. 
Surveys conducted by NREL have shown that $\eta$ is a complex function of storage capacity of water heater, type of water heater, age of water heater. 
No distributions are available for $\eta$ in the current studies. 
Field data collected from NREL surveys~\cite{NRELhotwater61,NRELhotwater129,Hendron2008Building} show that the efficiency varies anywhere between 80\%-99\%. 
Here~0.00189 ($\frac{\text{kWh}}{\text{gal}\;^{\circ}\text{F}}$) is a conversion constant obtained from Subbiah et al.~\cite{Subbiah2017}, and~$\Delta T$ is the temperature difference ($^{\circ}\text{F}$) between mains (inlet) water temperature $T^{\mathsf{cold}}_{m,z}$ for a given month $m$ in a climate zone $z$ and the water temperature required for a particular end-point. The values for $T^{\mathsf{cold}}_{m,z}$ and $T^{\mathsf{hot}}_v$ are obtained from NREL surveys~\cite{NRELhotwater129,NRELhotwater61}. 
Whenever the activity model detects the presence of an event $v$, we calculate the energy used by hot-water for the event using Equation~\ref{eq:dhw}.
Note that we compute hot water energy usage only for synthetic households having electric water heaters.

\begin{table}[!h]
	\centering
	\caption{Hot water model characteristics}\label{tab:dhw-specifics}
	\begin{tabular}{c c c c}
		\hline
		\textbf{Event $v$} & \textbf{Range of $T^{\mathsf{hot}}_v$ (F)} & \begin{tabular}{@{}l@{}} \textbf{Flow rate (gpm)} \\ $\mu$, $\sigma$, distribution \end{tabular}  &  \begin{tabular}{@{}l@{}} \textbf{Duration (minutes)} \\ $\mu$, $\sigma$, distribution \end{tabular}  
		\rule{0pt}{3ex} \rule[.5ex]{0pt}{0pt} \\
		\hline
        
        Shower & [105,116] & 2.25, 0.68, $\mathsf{Normal}$  & 7.81, 3.52, $\mathsf{Normal}$ 
        \rule{0pt}{2ex} \rule[-1.5ex]{0pt}{0pt} \\

        Bath & [105,116] & 4.40, 1.17, $\mathsf{Normal}$ & 5.65, 2.09, $\mathsf{Normal}$ 
        \rule{0pt}{0ex} \rule[-1.5ex]{0pt}{0pt} \\
		
		Dishwasher & [120,140] & 1.39, 0.20, $\mathsf{Normal}$ & 1.53, 0.41, $\mathsf{LogNormal}$ 
		\rule{0pt}{0ex} \rule[-1.5ex]{0pt}{0pt} \\
		
		Clothes washer & [60,130] & 2.20, 0.62, $\mathsf{Normal}$ & 3.05, 1.62, $\mathsf{Normal}$ 
		\rule{0pt}{0ex} \rule[-1.5ex]{0pt}{0pt} \\
		\hline
	\end{tabular}
\end{table}

\subsection*{Lighting $E^{\mathsf{light}}$}

Lighting accounts for 5--10\%~\cite{eia} of the consumption
 with
lighting usage in residential setting mainly characterized by outdoor lighting conditions and occupancy schedules in households~\cite{Capasso1994}. 
A Markov-chain approach is adopted by Widen et al.~\cite{WIDEN2009Light} for modeling lighting demand in Swedish households using time use data in Sweden.
A stochastic model is developed for residential lighting estimation for the city of Cordova in Spain by Palacios-Garcia~\cite{PALACIOSGARCIA2015Light} based on a model developed by Stokes et al.~\cite{STOKES2004103} using measured lighting data for 100 UK homes.
Another stochastic model is developed by Richardson et al.~\cite{RICHARDSON2009Light} for UK households using time-use data and lighting data from the Energy Information Administration(EIA).

We build a stochastic model for lighting demand in U.S. dwellings by building on design concepts from work done by Richardson et al.~\cite{RICHARDSON2009Light}, Stokes et al~\cite{STOKES2004103}, and Paatero \& Lund et al.~\cite{PaateroLund2006}.
Richardson's model is particularly interesting since it supports important characteristics of light usage such as `co-use' and `relative weights'.
The model uses the concept of `co-use' of lighting, i.e., lighting in a dwelling is often shared by household members in the same space of the dwelling at the same time.
The model also considers that all lighting units are not used at the same frequency (e.g. frequently occupied rooms such as kitchen space and living area will use more lighting than other rooms) and employs a weighting scheme to indicate relative usage.

Outdoor lighting conditions are modeled using irradiance time series. It is obtained from NSRDB described in Table~\ref{tab:datasets}. Hourly irradiance data is collected using the NSRDB API
 for the~365 days of the year~2014 at census tract resolution for the U.S. Thus, all synthetic households in a census tract use the same irradiance time series for a given day. 
The household level hourly occupancy profile $\langle O_{i,0}, O_{i,1}, \ldots, O_{i,23} \rangle$ is developed by examining activities of awake synthetic household members of~$H_i$ at home. 
Presence of awake occupants in the dwelling support the decision making of light switch-on event.
The distribution of lighting units in households are derived from the RECS survey. In general, distribution of lighting units of a~$H_i$ is taken from the matching~$S_l$. Three types of lighting units are considered: incandescent, CFL, and LED.
Power ratings of lighting unit categories are taken from a study conducted by the Bonneville Power Administration (U.S.) where lighting fixtures were analyzed for a sample of~161 Northwest residences~\cite{BaselineLightingACEEE}.
For a given simulation day, we define an irradiance threshold ($\mathsf{Irr}^{\mathsf{i}}$) for a household~$H_i$. 
It indicates that occupants may consider switching on lights when outdoor lighting is less than $\mathsf{Irr}^{\mathsf{i}}$. $\mathsf{Irr}^{\mathsf{i}}$ is sampled from a normal distribution~\cite{RICHARDSON2009Light} $\mathbf{\mathsf{Normal}}(60,10)$.
All notations used in the model are described in Table~\ref{tab:notations}.
Annual lighting data for the U.S. is summarized for different household sizes from the RECS survey.

Literature shows that lighting usage increases by number of occupants in the household, however, the lighting usage does not double for every occupant added in the house.
In order to simulate shared lighting usage, the concept of effective occupancy~\cite{RICHARDSON2009Light} of a household $\langle \hat{O}_{i,0},\hat{O}_{i,t},\ldots,\hat{O}_{i,23}\rangle$ is introduced.  
Effective occupancy ($\hat{O}_{i,t}$) is defined as a function of active occupancy ($O_{i,t}$). The values for effective occupancy are
derived by scaling the annual lighting demand by household size such that the effective occupancy of a dwelling with one active occupant is one.
The next step is to obtain the details of lighting units in a household. The proportion of lighting unit types are obtained from a RECS household~$S_l$ that matches~$H_i$ (RECS Model). Power ratings are attached to each lighting unit.
In general, not all lighting units are used at the same frequency. This is observed in literature surveys such as DECADE report~\cite{DecadeReport2}.
The frequency of usage of lighting units in households can be roughly modeled as a natural log curve~\cite{RICHARDSON2009Light}, however, no formal methods have been presented in the literature due to lack of quantitative data.
We use the natural log curve presented in Richardson et al.~\cite{RICHARDSON2009Light} to model the relative usage of a lighting unit.
Once weights are assigned to lighting units, the probability of a switch-on event for every lighting unit is calculated at a regular time interval (in our case 1 hour).
The probability of a switch-on event $P^{\mathsf{on}}_b$ of lighting unit~$b$ at hour~$t$ is calculated as
\begin{equation}{\label{eq:light-switch-on}}
    \begin{array}{l}
    P^{\mathsf{on}}_b\;=\;\mathbb{I}_b\;\times\;b^{\mathsf{weight}}\;\times\;\hat{O}_{i,t}\;\times\;\gamma \;, \quad \text{where}   \\ 
    \mathbb{I}_b\;=\; \begin{cases}
      1 & \text{irradiance threshold condition is $\mathsf{True}$ for bulb $b$ at time $t$ $\mathsf{if}$} \;\; \mathsf{Irr}_t \le \mathsf{Irr}^{\mathsf{i}}\;,\\
      0 & \text{otherwise.}
    \end{cases}    
    \end{array}
\end{equation}
Here~$b^{\mathsf{weight}}$ is sampled from a natural logarithmic curve, $\gamma$ is a calibration constant used to achieve the appropriate annual lighting consumption for the U.S., and~$\hat{O}_{i,t}$ is the effective occupancy of~$H_i$ at time~$t$. If a switch-on event occurs, then energy consumption is calculated for the respective lighting unit~$b$. The lighting duration is picked randomly from the distribution described in Stokes et al.~\cite{STOKES2004103}.

\subsection*{Refrigeration $E^{\mathsf{refr}}$}
The energy consumed by a refrigerator depends upon its size, age, ambient temperature, and several other factors as described in literature. They consume 3\%--5\% of the total residential energy usage.
Shimoda et al.~\cite{SHIMODA20071617} show that the daily refrigerator consumption is affected by outside temperature, while Tsuji et al~\cite{Tsuji2004} show a linear relationship between outside temperature and annual refrigerator demand.
Both these work are done in context of refrigerators in Japan.
The Lawrence Berkeley National Laboratory in California uses field metered energy use data from~$\sim$1500 refrigerators and freezers to develop a model that predicts annual usage of different freezer and refrigerator categories~\cite{LBNLRefr2012}.
All of the above models collected relevant data from the field or utilized detailed surveys on refrigeration.

Our approach is to develop a regression model for predicting daily refrigerator usage (kWh/day) of a household (${E}^{\mathsf{refr}}_{i}$) as a function of outside environment temperature.
The model is trained with the metered refrigerator usage data from Pecan Street Inc, where~30\% of the total metered data is used for training and testing the model. The 30\% data is obtained by conducting stratified sampling based on climate zones and daily average temperature bins.
The dependent variable is the daily refrigerator usage ${E}^{\mathsf{refr}}_i$ in kWh/day for $H_i$. The independent variables are daily average temperature $\hat{T}^{\mathsf{out}}$ ($^{\circ}F$) and categorical attributes indicating three major climate zones.
The 24 hour load profile of a refrigerator $\langle E^{\mathsf{refr}}_{i,0},E^{\mathsf{refr}}_{i,1},\ldots,E^{\mathsf{\mathsf{refr}}}_{i,23} \rangle$ is constructed from the daily usage, and
the variation in the hourly usage of the refrigerator is modeled using a Guassian distribution.
The refrigerator operates in an automated/standby mode, that is, occupant presence does not influence the energy consumption of this activity~\cite{Tsuji2004,Subbiah2017}.
Thus, computing the~24 hour profile of the refrigerator by adding a small Gaussian noise to the hourly load can be considered acceptable.
The validation section shows that addition of this noise creates good match to real data.

\begin{table*}[!h]
\centering
  \caption{Summary of referenced end-use modeling methods, including how these models are extended in this paper.} \label{tab:method_summaries}
  
\begin{tabular}{l l l}
    \hline
    
    \textbf{End-use}  & \textbf{Relevant models}  & \textbf{Our approach} 
    \rule{0pt}{3ex} \rule[-1.5ex]{0pt}{0pt} \\
    
    \hline
    
    $\mathsf{HVAC}$
    & 
    \parbox{3.5cm} {
    Muratori et.al~\cite{MURATORI2013465},
    Subbiah et.al~\cite{Subbiah2017},
    Thorve et.al~\cite{Thorve2018},
    Tsuji et.al~\cite{Tsuji2004}
    }
    & 
    \parbox{10.5cm} {
    Our model is based on the approach adopted in Subbiah et.al~\cite{Subbiah2017} and Thorve et.al~\cite{Thorve2018}. These models were specific to Virginia state. The method employed in these works as well as ours is a physics model. This model is also documented in NREL Technical Reports. Additional details about thermostat settings, building characteristics such as insulation are obtained from RECS survey, EIA website, and NREL Technical Reports. 
    }
    \T\rule[-9ex]{0pt}{0pt} \\
    
    $\mathsf{DHW}$
    & 
    \parbox{3.5cm} {
    Maguire et.al~\cite{NRELhotwater129},
    Hendron et.al~\cite{Hendron2010H2o},
    Thorve et.al~\cite{Thorve2018}
    }
    & 
    \parbox{10.5cm} {
    Hendron et.al~\cite{Hendron2010H2o} and Maguire et.al~\cite{NRELhotwater129} present a general stochastic method to reproduce sample hot water draws based on two water usage surveys conducted in the U.S. The analyses concludes by reporting distributions related to hot water usage events such as showering, using dishwasher, and using clothes washer. Some of these results are summarized in Table~\ref{tab:dhw-specifics} and used in our model. Hot and cold water temperatures for specific end-uses are obtained from NREL surveys.
    The above model does not consider the setting of specific household schedules.
    This context of household occupancy and occurrence of events is added to an existing model in literature presented in Thorve et.al~\cite{Thorve2018} in order to schedule these hot water usage events.
    }
    \T\rule[-16ex]{0pt}{0pt} \\
    
    $\mathsf{light}$
    &
    \parbox{3.5cm}{
    Richardson et.al~\cite{RICHARDSON2009Light},
    Stokes et.al~\cite{STOKES2004103},
    Paatero \& Lund et.al~\cite{PaateroLund2006}
    }
    &
    \parbox{10.5cm}{
    We mainly improve upon the stochastic lighting model developed for U.K. household by Richardson et.al by adding context of U.S. households such as household size, household occupancy, annual lighting consumption in the U.S. for different household sizes, calibration of $\gamma$ for U.S. households, and proportion of light bulbs in the U.S. households and their power ratings.  The probability of switch-on event is modeled from Paatero \& Lund et.al~\cite{PaateroLund2006} and Richardson et.al~\cite{RICHARDSON2009Light}.
    Duration of switch-on event is taken from Stokes et.al~\cite{STOKES2004103}.
    Power ratings for different categories of lighting units in U.S. is obtained from a study conducted by Bonneville Power Administration~\cite{BaselineLightingACEEE}.
    Proportion of lighting units in U.S. households and annual lighting consumption by household size is derived from RECS survey. Irradiance data for the U.S. is obtained from NREL.
    }
    \T\rule[-3ex]{0pt}{0pt} \\
    
    $\mathsf{refr}$
    &
    \parbox{3.5cm}{       ---}
    &
    \parbox{10cm}{
    A linear regression model is developed to predict daily refrigerator usage for a household based on outside temperature and climate zones. 
    }
    \T\rule[-5ex]{0pt}{0pt} \\
    
    \parbox{1cm}{
    $\mathsf{misc}$, $\mathsf{act}$}
    &
    \parbox{3.5cm}{ Subbiah et.al~\cite{Subbiah2017}, Thorve et.al~\cite{Thorve2018}, Tsuji et.al~\cite{Tsuji2004}}
    &
    \parbox{10.5cm}{
    All the three referenced models have inspired the design of activity models involving use of appliances. The actual activity occurrence is obtained from the individual/household occupancy schedule. Duration and power usage distributions of appliances is modeled from NIST datasets~\cite{NIST_cook,NIST_dw_cw_cd,NIST_dw} and other datasets~\cite{energystar_computer_v5,energystar_TV_2021,energystar_vacuum_2011,electrolux_vacuum_2013}. The start time is chosen randomly within the duration reported by ATUS individuals and the power ratings and duration of the activity/appliance is selected from the above mentioned distributions.
    }
    
    \T\rule[-11ex]{0pt}{0pt} \\
    \hline
    
\end{tabular}
\end{table*}

\subsection*{Activity model $E^{\mathsf{appliances}}$}

The energy consumption in a households that is attributed to appliance usage and plug load is~20\%--26\%. This energy is a result of the occupants' desires to perform activities such as taking baths, making hot meals, using the dishwasher, doing laundry, charging electronics such as TVs and computers, or using any other appliances that consume electricity. Equations~\ref{eq:energy_eq}b and~\ref{eq:energy_eq}c  are used in this model.
Based on the aforementioned end-uses, appliance usage behavior is characterized by~\cite{Tsuji2004} through
operational mode of appliances, duration of operation, power consumption, limit on daily event occurrence, and saturation rate. Operational mode of appliances describes the functioning appliances and related behavior that can be categorized into three types: automatic (appliance use is independent of person), semi-automatic (appliance turned on by household member but turned off automatically), and manual (appliance turned off and on manually).
The saturation rate can be used to determine the presence and/or penetration of certain appliances in households.
Generally, the operational mode of appliances and saturation rate are deterministic in nature. 
However, parameters such as probability of activity occurrence, start time, duration, power consumption, and maximum occurrences vary from household to household and day to day. 
In general, some appliance usages can overlap and/or occur in parallel. These details are handled in this model.

Table~\ref{tab:activity-setup} outlines all the modeled activities and related appliances, their modes of operation, maximum allowed daily occurrences, activity duration, and power consumption. The distributions marked with an asterisk (*) denote that they are modeled by engineering judgement and/or other sources such as \href{https://energyusecalculator.com}{\textcolor{black}{Energy Calculator (energyusecalculator.com)}}.
Power rating distributions for dishwashers are obtained from a survey conducted by NIST~\cite{NIST_dw, NIST_dw_cw_cd}.
Power ratings and duration distributions for laundry appliances are derived from literature~\cite{Subbiah2017,Thorve2018} and surveys~\cite{NIST_dw_cw_cd}; 
power ratings for appliances in $\mathsf{cook}$ activity include electric ovens, microwaves, and electric cooktops (small- and large burners.) 
Power rating distributions for these appliances are derived from the NIST efficiency study~\cite{NIST_cook}, and
durations of appliance usage are obtained from ATUS data, where the maximum limit for cooking activities is capped to three.
Sample power ratings for TVs are observed from EnergyStar reports~\cite{energystar_TV_2021} and modeled using a normal distribution. 
The $\mathsf{tv}$ activity duration is modeled as a log-normal distribution after examining the ATUS survey data.
Power ratings for $\mathsf{computer}$ use activity are derived from a small study conducted by EnergyStar~\cite{energystar_computer_v5}.
Standard values for charging duration are used from reputed laptop manufacturers.
Vacuum related data are obtained from EnergyStar vacuum report and a survey conducted by Electrolux covering 28,000 consumers from 23 countries including U.S.~\cite{energystar_vacuum_2011,electrolux_vacuum_2013}. 
We assume that all households have vacuum cleaners. The usage frequency of vacuuming is 1-5 times per week~\cite{electrolux_vacuum_2013} and the maximum number of daily occurrences is 1.
Assuming $\mathsf{Normal}$ distribution for power ratings and duration of appliance usage is reasonable after examining rudimentary results from surveys/reports. 
The results of the hot water usage study conducted by NREL~\cite{Hendron2008Building,Hendron2010H2o} as summarized in Table~\ref{tab:dhw-specifics} show that most of the processes can be modeled as a $\mathsf{Normal}$ distribution.

\begin{table}[!h]
\centering
\caption{Modeled activity and appliance usage behaviors}
\begin{tabular}{l l l c l l c}
	\hline
	\textbf{Activity} & \textbf{Appliance} & \textbf{Mode} & \begin{tabular}{@{}c@{}}    \textbf{Max} \\ \textbf{occ.} \end{tabular} &  \begin{tabular}{@{}c@{}}    \textbf{Duration} \\ \textbf{(minutes)} \end{tabular}& \textbf{Power (W)}&\textbf{Hot Water} \\ 
	\hline
	
	$\mathsf{dwasher}$
    & 
    dishwasher
    &
    Semi-automatic
    & 
    2
    & 
    $\mathsf{Normal}(90,30)^{*}$
    &
    $\mathsf{Normal}(900,100)$
    &
    Yes
    
	\rule{0pt}{3ex} \rule[-1.5ex]{0pt}{0pt} \\
	
	$\mathsf{cwasher}$
    & 
    clothes washer
    & 
    Semi-automatic
    & 
    2
    & 
    $\mathsf{Normal}(45,20)^{*}$
    &
    $\mathsf{Normal}(400,50)^{*}$
    &
    Yes
    
	\rule{0pt}{3ex} \rule[-1.5ex]{0pt}{0pt} \\
	
	$\mathsf{cdryer}$
    & 
    clothes dryer
    &
    Semi-automatic
    & 
    2
    & 
    $\mathsf{Normal}(45,20)^{*}$
    &
    $\mathsf{Normal}(2500,200)^{*}$
    &
    No
    
	\rule{0pt}{3ex} \rule[-2.0ex]{0pt}{0pt} \\
	
	$\mathsf{cook}$
    & 
    \begin{tabular}{@{}l@{}} 
    oven\\
    microwave\\
    cooktop (large)\\
    cooktop (small)
    \end{tabular}
    &
    \begin{tabular}{@{}l@{}} 
    Manual/ \\ Semi-automatic
    \end{tabular}
    & 
    3
    & 
    
    $\mathsf{LogNormal}(3,0.96)$
    &
    \begin{tabular}{@{}l@{}} 
	$\mathsf{Normal}(1426,13.3)$\\
	$\mathsf{Normal}(880,14)$\\
	$\mathsf{Normal}(213,1.2)$\\
	$\mathsf{Normal}(393,3.1)$
	\end{tabular}
    &
    No
    
	\rule{0pt}{4ex} \rule[-1.5ex]{0pt}{0pt} \\
	
	$\mathsf{tv}$
	&
	television
    & 
    Manual
    & 
    --
    & 
    $\mathsf{LogNormal}(4.24,0.79)$
    &
    $\mathsf{Normal}(120,20)^{*}$
    &
    No
    
	\rule{0pt}{3ex} \rule[-1.5ex]{0pt}{0pt} \\
	
	$\mathsf{computer}$
    & 
    \begin{tabular}{@{}l@{}} 
    desktop\\
    notebooks
    \end{tabular}
    &
    Manual
    & 
    --
    & 
    $\mathsf{Normal}(90,30)^{*}$    
    &
	\begin{tabular}{@{}l@{}} 
	$\mathsf{Normal}(191.5,32.7)$\\$\mathsf{Normal}(60.5,20.5)$
	\end{tabular}
    &
    No
    
	\rule{0pt}{5ex} \rule[-2.5ex]{0pt}{0pt} \\
	
	$\mathsf{cleaning}$
	&
	vacuum
    & 
    Manual
    & 
    1
    & 
    $\mathsf{Normal}(30,15)$
    &
    $\mathsf{Normal}(1200,300)$
    &
    No
    
	\rule{0pt}{3ex} \rule[-1.5ex]{0pt}{0pt} \\
	\hline
	
\end{tabular}
\label{tab:activity-setup}
\end{table}

The activity model simulates appliance usage based on activity indicators provided by ATUS when the occupant is present in the house.
Considering the presence of appliance in each household (from matching RECS household)
The time use diaries of adults in the synthetic population and frequency of occurrence of appliance usage such as dishwasher and laundry, and activities such as cooking are taken from RECS household.
The activity model focuses on activities performed by an individual when at home.
Similar to lighting, activities such as cooking, vacuuming, and leisure activities such as watching TV are shared by household members.
A procedure is outlined below for generating household level activity sequence $\mathsf{ActSeq}_i$.
\noindent
Let $M$ be the number of adult members in the synthetic household. Then each household member $P_{i,j}$ has an activity sequence $\mathsf{ActSeq}_{i,j}$.
The goal is to find one household level activity sequence $\mathsf{ActSeq}_i$ composed of $n$ activities (individual + shared appliance usage related activities) such that the sequence satisfies following constraints:
\begin{enumerate}
\item Each activity is performed when at least one occupant is home. 
\item The limit on repeated usage is respected for each activity type.
\item Presence of appliance is considered for activities such as dishwasher, and laundry appliances.
\end{enumerate}
Once the above constraints are satisfied, a start time is randomly selected for each activity from the activity duration reported by ATUS. 
The actual duration and power ratings for appliances used in different activities is chosen from Table~\ref{tab:activity-setup}.

\begin{figure}[!h]
    \centering
    \includegraphics[width=14cm, height=8.5cm]{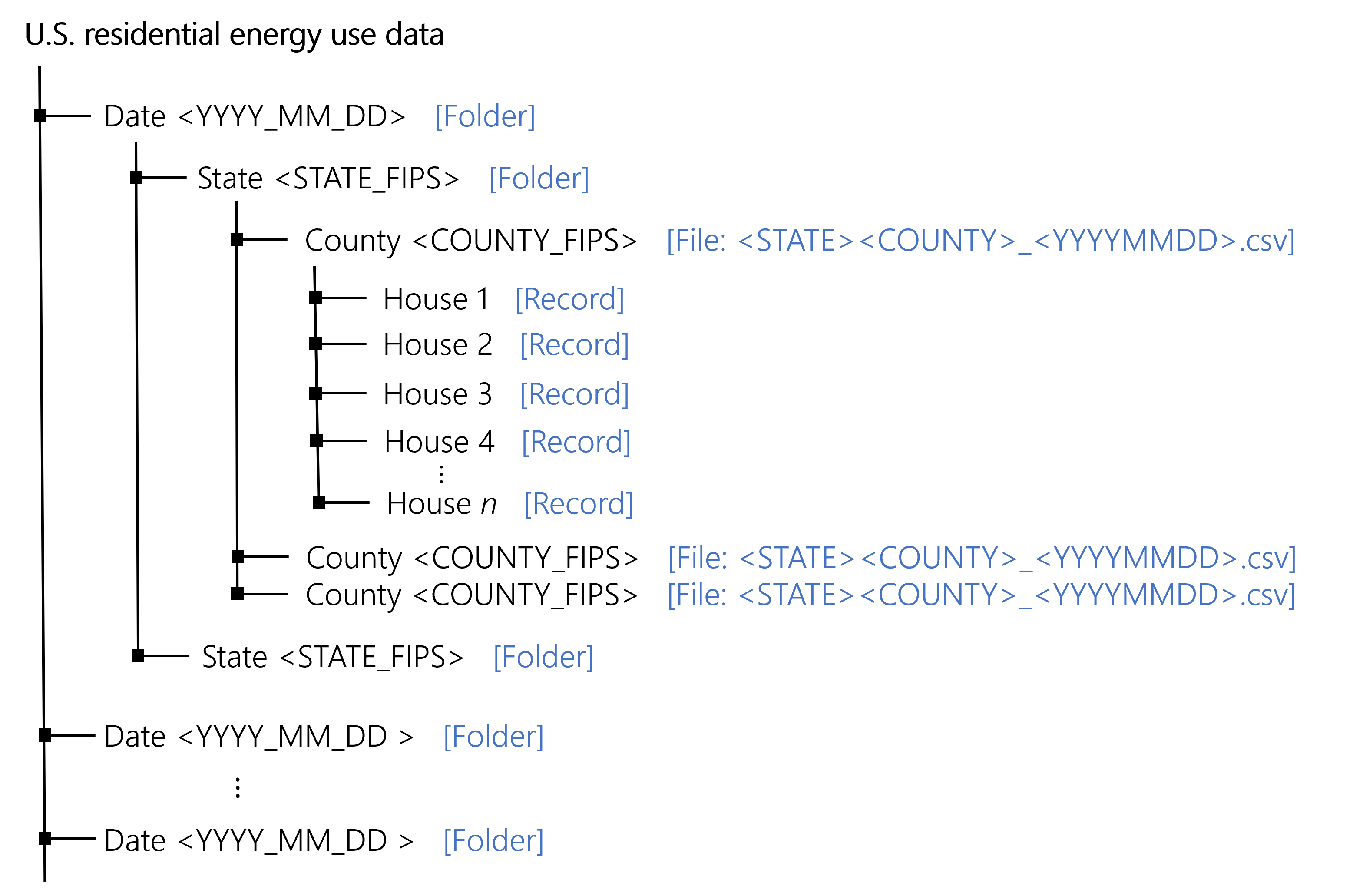}
    \caption{\textbf{Data organization.} Dataset is available in the form of csv files. The files are organized by dates (temporal) and states (spatial). The blue text indicates the type (e.g. folder, file, record). The text within angular brackets denotes nomenclature templates of folders and files. A record csv file contains energy use data and metadata for a synthetic household in the SPEW population. There will be one file per county and date. One day generates several GBs of data.}
    \label{fig:datarecords}
\end{figure}

\section*{Data Records}

The dataset for the entire year of 2014 for U.S. households is publicly available for download from the net.science repository through University of Virginia Dataverse~\cite{thorve_scidata2022_doi}.
The dataset is available in the form of csv files.
It  is organized in folders according to date and state. Figure~\ref{fig:datarecords} shows the hierarchy of data organization and file name templates. Each file corresponds to a U.S. county identifier and date. A county identifier is a FIPS code. FIPS codes are numbers which uniquely identify geographic areas by the U.S. census.
A record in the file corresponds to a synthetic household. The record includes synthetic household metadata and energy data for that particular date. All energy related data is in kWh. All the energy data is timestamped by local timezones in the country. A data header codebook is also included in the downloads.
Note that, this work was reviewed by the University of Virginia's Institutional Review Board (IRB) and was determined to be exempt from board IRB approval, as this research project did not involve human subject research.

\begin{figure}[!h]
    \centering
    \includegraphics[width=.98\textwidth]{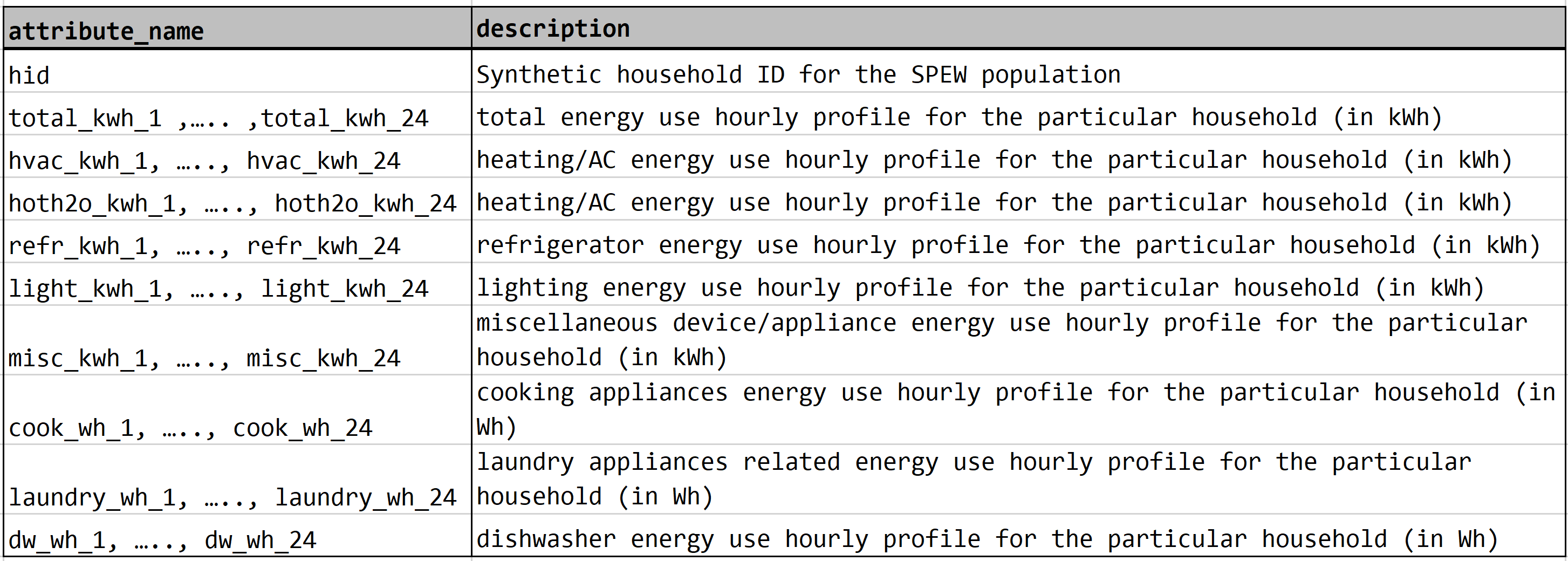}
    \caption{\textbf{Data Attributes.} 24-hour dis-aggregated hourly household energy demand profiles are made available. 1-24 indicates the hour starting midnight. Eight end-use profiles are described (lines 3-10).}
    \label{fig:datafeatures}
\end{figure}

\begin{table}[!h]
	\centering
	\caption{Datasets used for validation}
	\begin{tabular}{l l l l c c c c c c}
		\hline
		\textbf{Climate} & \textbf{Location} & \textbf{Source}& \textbf{Year}& \begin{tabular}{@{}c@{}} \textbf{Sample}\\\textbf{size} \end{tabular} &\begin{tabular}{@{}c@{}} \textbf{Area}\\\textbf{type} \end{tabular} &
		 \textbf{Resolution} &\begin{tabular}{@{}c@{}} \textbf{Is open-}\\\textbf{source} \end{tabular} & \begin{tabular}{@{}c@{}} \textbf{Is data}\\\textbf{complete?} \end{tabular}  & \begin{tabular}{@{}c@{}} \textbf{Is data}\\\textbf{dis-}\\\textbf{aggregated?} \end{tabular} \\ 
		\hline
		\begin{tabular}{@{}c@{}}
        Hot-\\Humid \end{tabular}
        & 
        Austin,TX
        & 
        \begin{tabular}{@{}c@{}} Pecan\\Street \end{tabular} 
        &
        2018
        &
        25
        & 
        Urban
        &
        \begin{tabular}{@{}l@{}} 
        15-min
        \end{tabular}
         & Yes & Yes & Yes
		\rule{0pt}{3ex} \rule[-1.5ex]{0pt}{0pt} \\
		\hline
		
	    \begin{tabular}{@{}c@{}}
        Hot-\\Humid \end{tabular}
        & 
        Horry,SC
        & 
        \begin{tabular}{@{}c@{}} NRECA \end{tabular} 
        &
        2017
        &
        ~56000
        & 
        \begin{tabular}{@{}c@{}} 
        Rural\\Semi-\\urban
        \end{tabular}
        &
        \begin{tabular}{@{}l@{}} 
        Hourly
        \end{tabular}
         & No & Yes & No
		\rule{0pt}{3ex} \rule[-1.5ex]{0pt}{0pt} \\
		\hline
		
		\begin{tabular}{@{}c@{}}
        Mixed-\\Humid \end{tabular}
        & 
        \begin{tabular}{@{}l@{}} 
        Rappaha-\\nnock \\ in VA
        \end{tabular}
        & 
        \begin{tabular}{@{}c@{}} NRECA \end{tabular} 
        &
        2016
        &
        100
        & 
        Rural
        &
        \begin{tabular}{@{}l@{}} 
        Hourly
        \end{tabular}
         & No & Yes & No
		\rule{0pt}{3ex} \rule[-1.5ex]{0pt}{0pt} \\
		\hline
		
		Cold
        & 
        \begin{tabular}{@{}l@{}} 
        Tompkins\\
        Cayuga\\
        in NY
        \end{tabular}
        & 
        \begin{tabular}{@{}c@{}} Pecan\\Street \end{tabular} 
        &
        2019
        &
        25
        &
        Urban
        &
        \begin{tabular}{@{}l@{}} 
        15-min
        \end{tabular} 
        & Yes & No & Yes
        
		\rule{0pt}{3ex} \rule[-1.5ex]{0pt}{0pt} \\
		\hline
		
		Cold
        & 
        \begin{tabular}{@{}l@{}} 
        Los Alamos\\
        in NM
        \end{tabular}
        & 
        \begin{tabular}{@{}c@{}} Open data\\Dryad\\repository \end{tabular} 
        &
        \begin{tabular}{@{}c@{}} 2014 \end{tabular} 
        &
        1600
        &
        \begin{tabular}{@{}c@{}}
        Semi-\\urban\end{tabular}
        &
        \begin{tabular}{@{}l@{}} 
        Hourly
        \end{tabular} 
        & No & Yes & No
        
		\rule{0pt}{3ex} \rule[-1.5ex]{0pt}{0pt} \\
		\hline
		
		Cold
        & 
        \begin{tabular}{@{}l@{}} 
        MT
        \end{tabular}
        & 
        \begin{tabular}{@{}c@{}} NEEA \end{tabular} 
        &
        2019
        &
        9
        &
        -
        &
        \begin{tabular}{@{}l@{}} 
        Hourly
        \end{tabular} 
        & Yes & No & Yes
        
		\rule{0pt}{3ex} \rule[-1.5ex]{0pt}{0pt} \\
		\hline
		
	    Cold
        & 
        \begin{tabular}{@{}l@{}} 
        ID
        \end{tabular}
        & 
        \begin{tabular}{@{}c@{}} NEEA \end{tabular} 
        &
        2019
        &
        19
        &
        -
        &
        \begin{tabular}{@{}l@{}} 
        Hourly
        \end{tabular} 
        & Yes & No & Yes
        
		\rule{0pt}{3ex} \rule[-1.5ex]{0pt}{0pt} \\
		\hline
        
		\begin{tabular}{@{}c@{}}
		Cold\\
		Marine
		\end{tabular}
        & 
        \begin{tabular}{@{}l@{}} 
        OR
        \end{tabular}
        & 
        \begin{tabular}{@{}c@{}} NEEA \end{tabular} 
        &
        2019
        &
        102
        &
        -
        &
        \begin{tabular}{@{}l@{}} 
        Hourly
        \end{tabular} 
        & Yes & No & Yes
        
		\rule{0pt}{3ex} \rule[-1.5ex]{0pt}{0pt} \\
		\hline
		
		\begin{tabular}{@{}c@{}} 
		Cold\\Marine
		\end{tabular}
        & 
        \begin{tabular}{@{}l@{}} 
        WA
        \end{tabular}
        & 
        \begin{tabular}{@{}c@{}} NEEA \end{tabular} 
        &
        2019
        &
        78
        &
        -
        &
        \begin{tabular}{@{}l@{}} 
        15-min
        \end{tabular} 
        & Yes & No & Yes
        
		\rule{0pt}{3ex} \rule[-1.5ex]{0pt}{0pt} \\
		\hline
		
		\begin{tabular}{@{}l@{}}
		Hot-Dry/\\Mixed-\\Dry
		\end{tabular}
        & 
        \begin{tabular}{@{}l@{}} 
        San Diego\\ in CA
        \end{tabular}
        & 
        \begin{tabular}{@{}c@{}} Pecan\\Street \end{tabular} 
        &
        \begin{tabular}{@{}l@{}} 
        2014\\
        2015\\
        2016
        \end{tabular}
        &
        25
        &
        Urban
        &
        \begin{tabular}{@{}l@{}} 
        15-min
        \end{tabular} 
         & Yes & No & Yes
		
		\rule{0pt}{3ex} \rule[-1.5ex]{0pt}{0pt} \\
		\hline
		 
	\end{tabular}
	\label{tab:validation-datasets}
\end{table}

\section*{Technical Validation}

Three studies are presented for validating the synthetic energy profiles.
The first study quantifies the similarity between the real and synthetic energy use probability distributions using Jensen-Shannon and Hellinger distance.
Comparisons are performed by end-use for real and synthetic data in all representative locations of the U.S.
Strong similarities are observed for appliance use distributions between real and synthetic data as well as across spatial locations.
TCL loads show differences in distributions across locations.
The second study examines variations in the 24-hour energy use timeseries in real and synthetic data in all representative locations in the U.S.
We uncover unique energy use patterns in the real and synthetic datasets and study similarities in patterns using unsupervised learning. 
We introduce two metrics in the process -- coverage and closeness.
The synthetic data has patterns similar to that of real data.
The last study is focused on observing trends in the synthetic energy use in different representative locations in the U.S. 
We notice that the synthetic data is able to incorporate the effects of mixture of variables such as weather, irradiance, building attributes and demographic characteristics on household level energy usage. 
The study is a quick demonstration of energy use variability at multiple spatio-temporal levels in different end-uses. 

The remaining V\&V section is outlined as follows. First, we describe challenges in validating a large synthetic dataset for energy use.
Then, we highlight the temporal and spatial resolutions of the data that are considered in the validation experiments.
Next, ground truth datasets (real/recorded/actual data) used for evaluation are briefly described.
This is followed by description of the experimental setup and results.

Validating the quality of the large-scale synthetic timeseries data for a sizeable region such as the U.S. is challenging, owing to the vast extent, diversity, and contrasting climates in the country.
One of the challenges of validating an energy consumption timeseries at household level is the large variety and variability of the load patterns within and between households.
In addition to external elements such as weather and building characteristics, consumer lifestyles and affordances play a vital role in shaping the demand such as a curve with morning peak, or a curve with a small afternoon peak and sharp evening peak.
This leads to a big spectrum of variations and patterns in energy use.
Thus, in-depth comparative analyses of synthetic data to actual data is required.
However, it is conditioned on the availability of a reasonable amount of representative real data.
Here, we employ real/recorded data such as load research data, end-use metering data, and smart meter data from ten locations in the country that are representative of the U.S. climate zones (Table~\ref{tab:validation-datasets}).
The availability of public smart meter data in the U.S. is limited, which may cause a potential skew towards the selected sample of households and may not be spatially representative. 
Thus, framing our understanding of validation in this context is important.

We address the quality of the synthetic energy consumption data on two intrinsic qualities of energy use data : magnitude (usage over 24 hours) and load shape (pattern of consumption). 
Magnitude and load shape can be examined across the temporal (hour/day/month/year) and spatial (household/census tract/city/county/state/climate zones) axes.
Thus, the verification and validation (V\&V) process covers:
 \begin{itemize}
    \item \emph{Spatial representativeness and resolutions.} Due to limited availability of real data, we define spatial representativeness by choosing atleast one location in each climate zone in the U.S. to carry out validation experiments.
    The major climate zones~\cite{BAClimateZones} in the contiguous United States are as follows: ($i$) marine, ($ii$) hot-dry/mixed-dry, ($iii$) hot-humid, ($iv$) mixed-humid, and ($v$) cold/very-cold. 
    Comparisons are then performed at household and city/county resolutions.
    \item \emph{Temporal representativeness and resolutions.} Temporal representativeness is studied by observing similarities between real and synthetic hourly demand profiles.
    Furthermore, daily and seasonal energy usage is studied for different locations.
    \item \emph{Dis-aggregate energy use.} Note that we publish dis-aggregated energy use data at household level. Thus, a finer level of evaluation such as an energy use sub-type (e.g. HVAC, cooking, etc,.) is possible at various temporal and spatial levels. 
\end{itemize}

All the real datasets used in the V\&V process are listed in Table~\ref{tab:validation-datasets}. Recorded datasets are obtained from Pecan Street Dataport~\cite{pecanstreet}, Northwest Energy Efficiency Alliance (NEEA)~\cite{NEEAData}, National Rural Electric Cooperative Association (NRECA). The Los Alamos dataset is obtained from a public data sharing repository Dryad~\cite{LADPU_dryad}.
Unfortunately, we do not have any metadata about households (e.g. household size, dwelling type, etc) in these datasets.
The datasets only have energy use timeseries.

\noindent
Three studies are presented to cover temporal, spatial, and dis-aggregate nature of the synthetic time-series:

\noindent
I. Comparing real and synthetic end-use energy usage (magnitude)

\noindent
II. Comparing real and synthetic energy use patterns (shape/structure)

\noindent
III. Observing differences and similarities in synthetic energy use data in spatially representative locations

\subsection*{I. Comparing real and synthetic end-use energy usage (magnitude)}

\begin{figure*}[!h]    
    \begin{subfigure}{.48\textwidth}
    \centering
    \includegraphics[width=6.6cm,height=6.1cm]{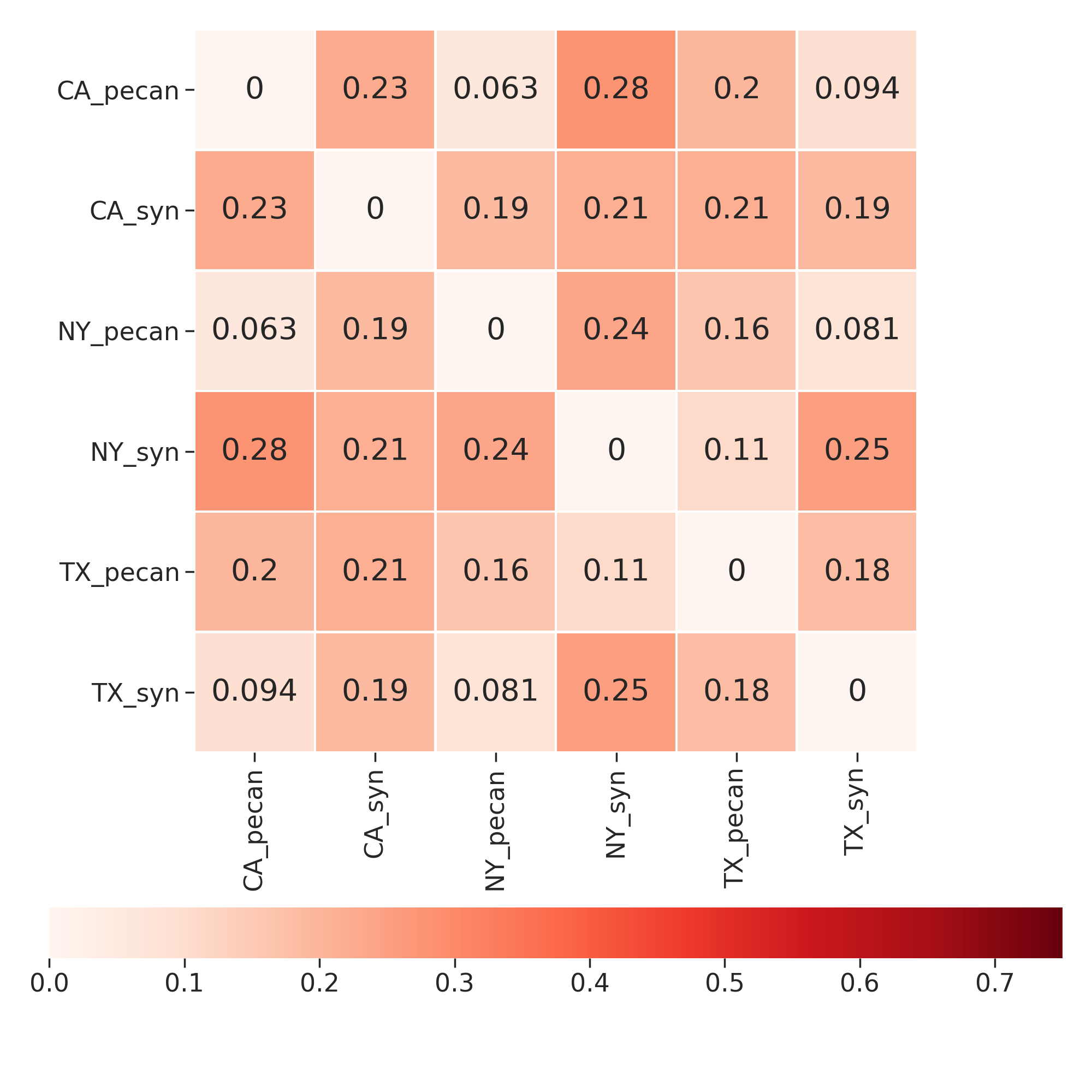}
    \caption{HVAC (Jensen-Shannon)}
      \label{fig:}
    \end{subfigure}
    \begin{subfigure}{.48\textwidth}
    \centering
    \includegraphics[width=6.6cm,height=6.1cm]{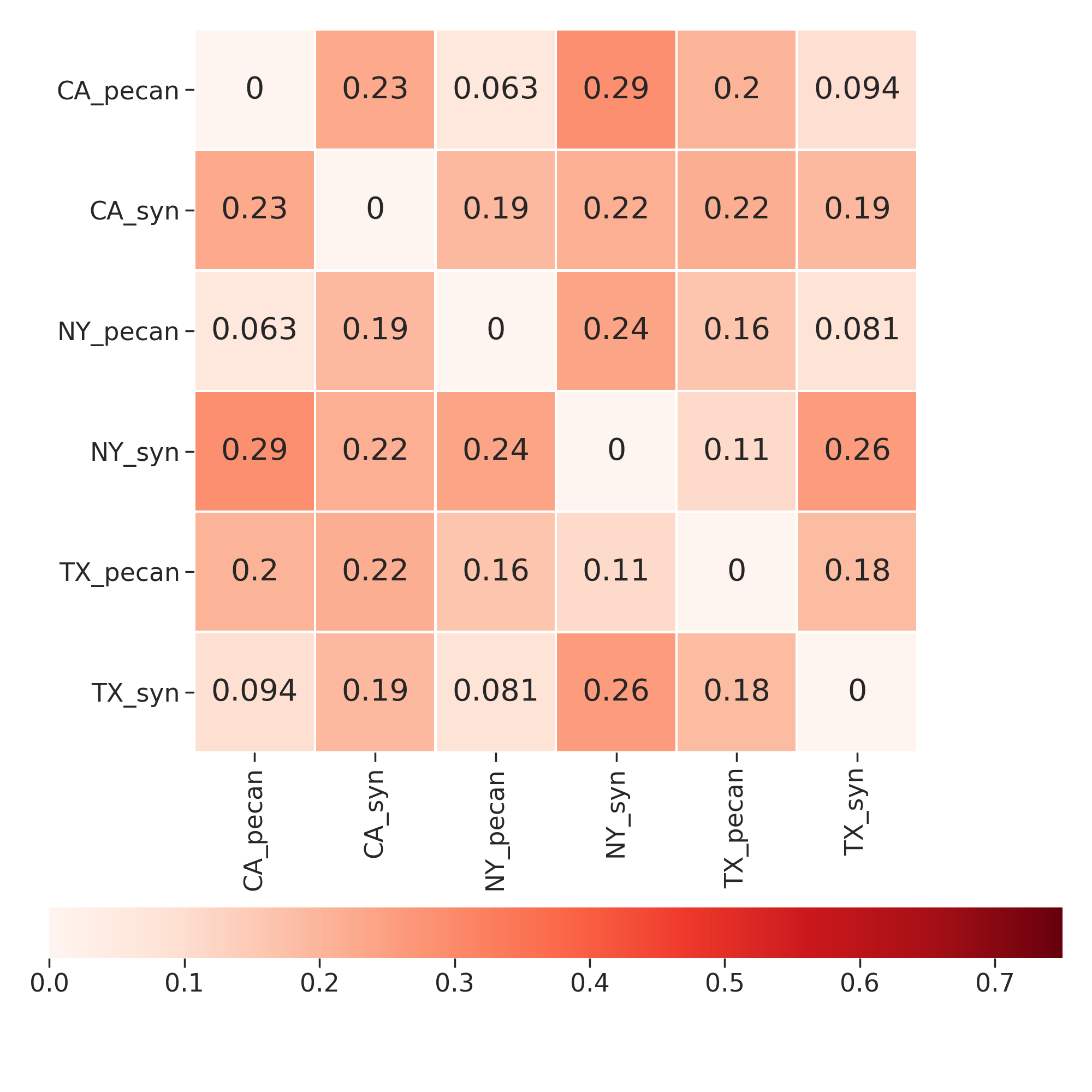}
    \caption{HVAC  (Hellinger)}
      \label{fig:}
    \end{subfigure}

    \begin{subfigure}{.48\textwidth}
    \centering
    \includegraphics[width=6.6cm,height=6.1cm]{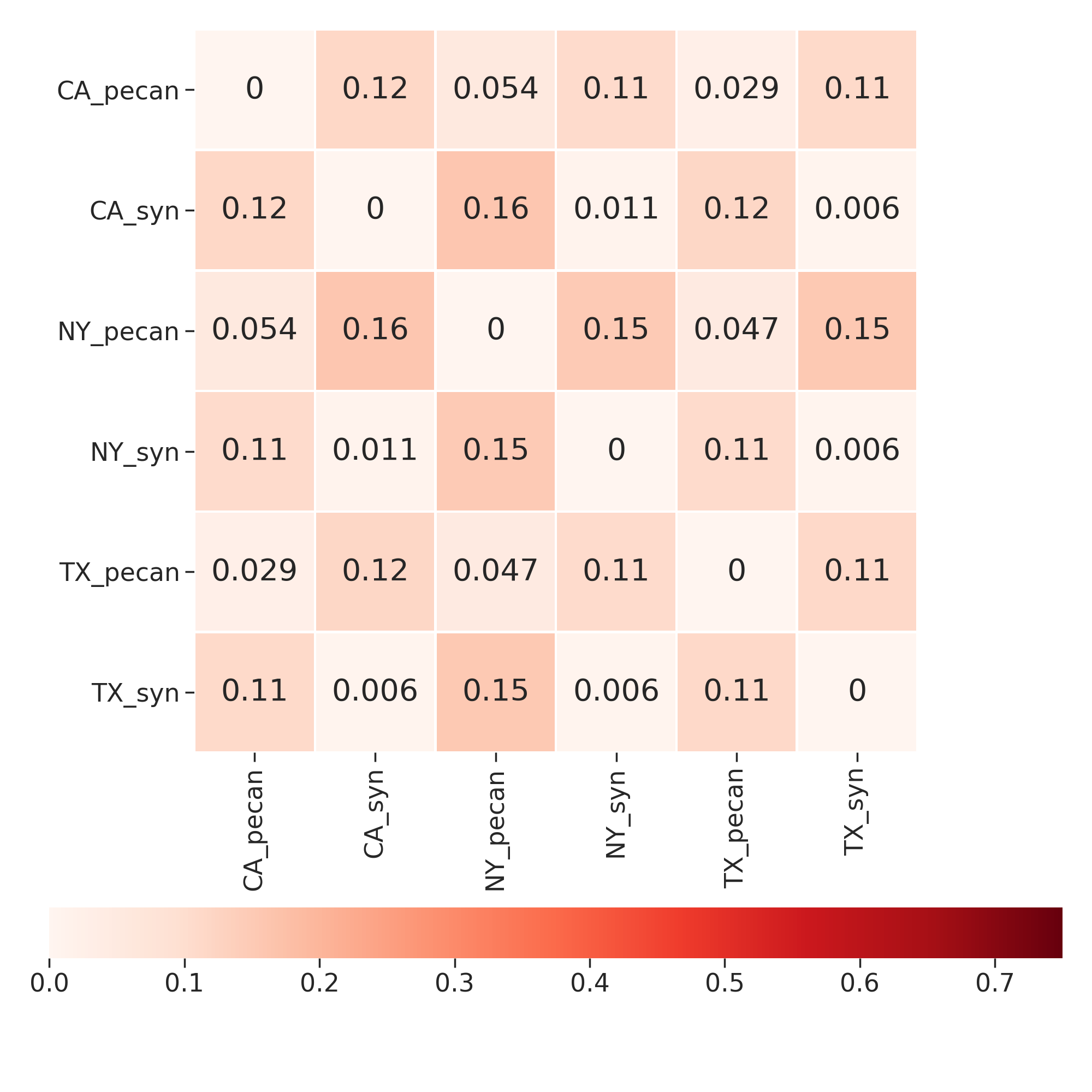}
    \caption{Refrigerator  (Jensen-Shannon)}
      \label{fig:}
    \end{subfigure}
    \begin{subfigure}{.48\textwidth}
    \centering
    \includegraphics[width=6.6cm,height=6.1cm]{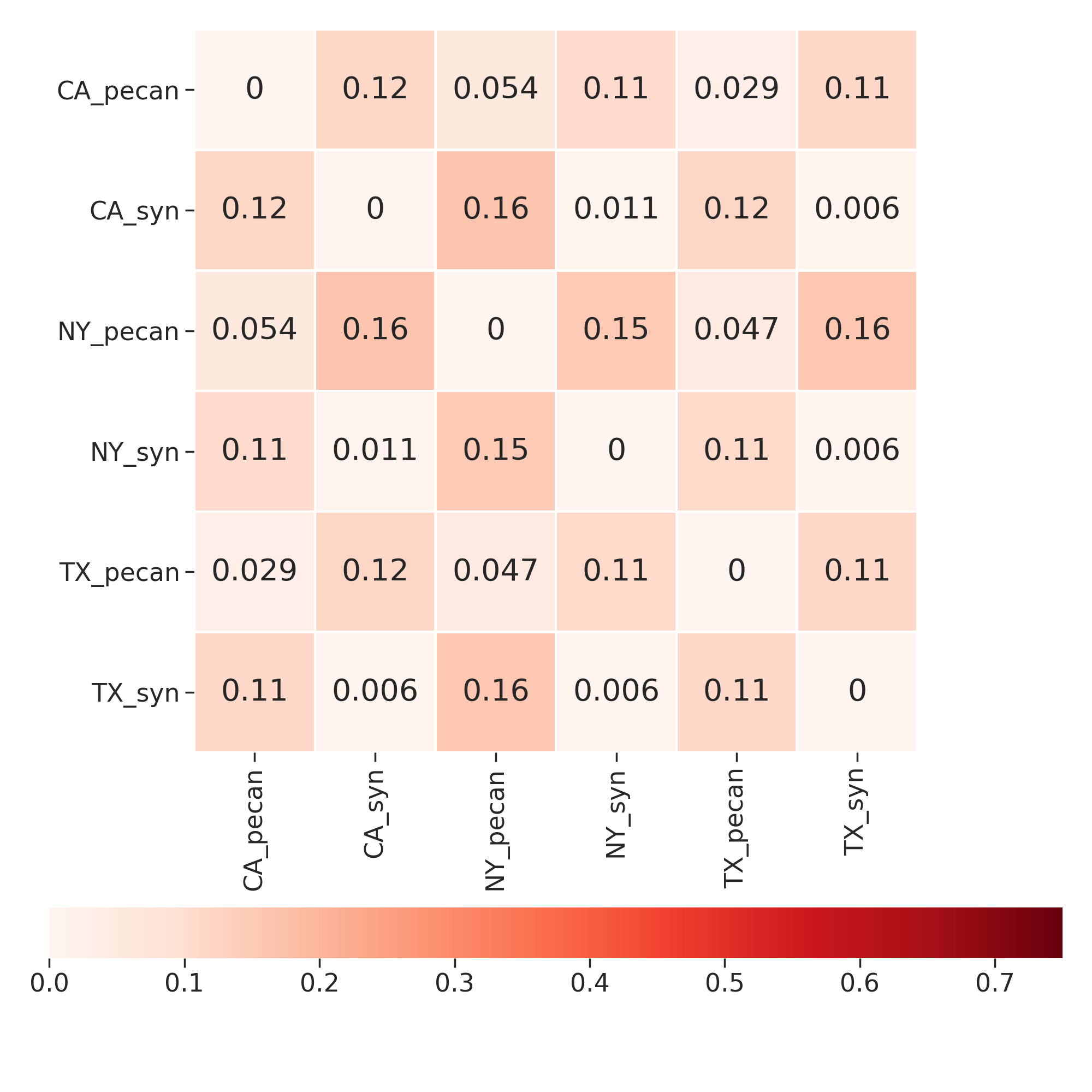}
    \caption{Refrigerator (Hellinger)}
      \label{fig:}
    \end{subfigure}
    
    \begin{subfigure}{.48\textwidth}
    \centering
    \includegraphics[width=6.6cm,height=6.1cm]{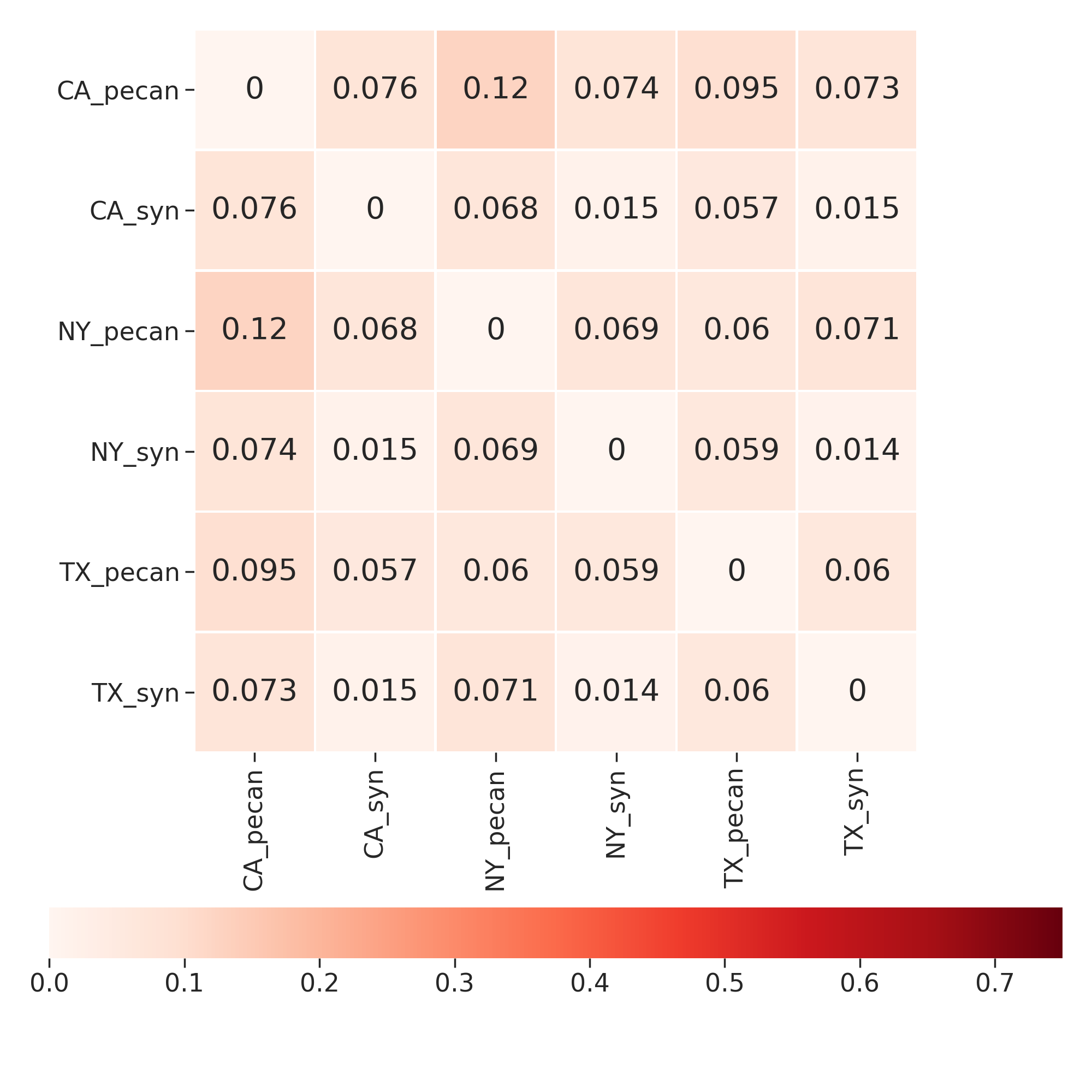}
    \caption{Cooking appliances  (Jensen-Shannon)}
      \label{fig:}
    \end{subfigure}
    \begin{subfigure}{.48\textwidth}
    \centering
    \includegraphics[width=6.6cm,height=6.1cm]{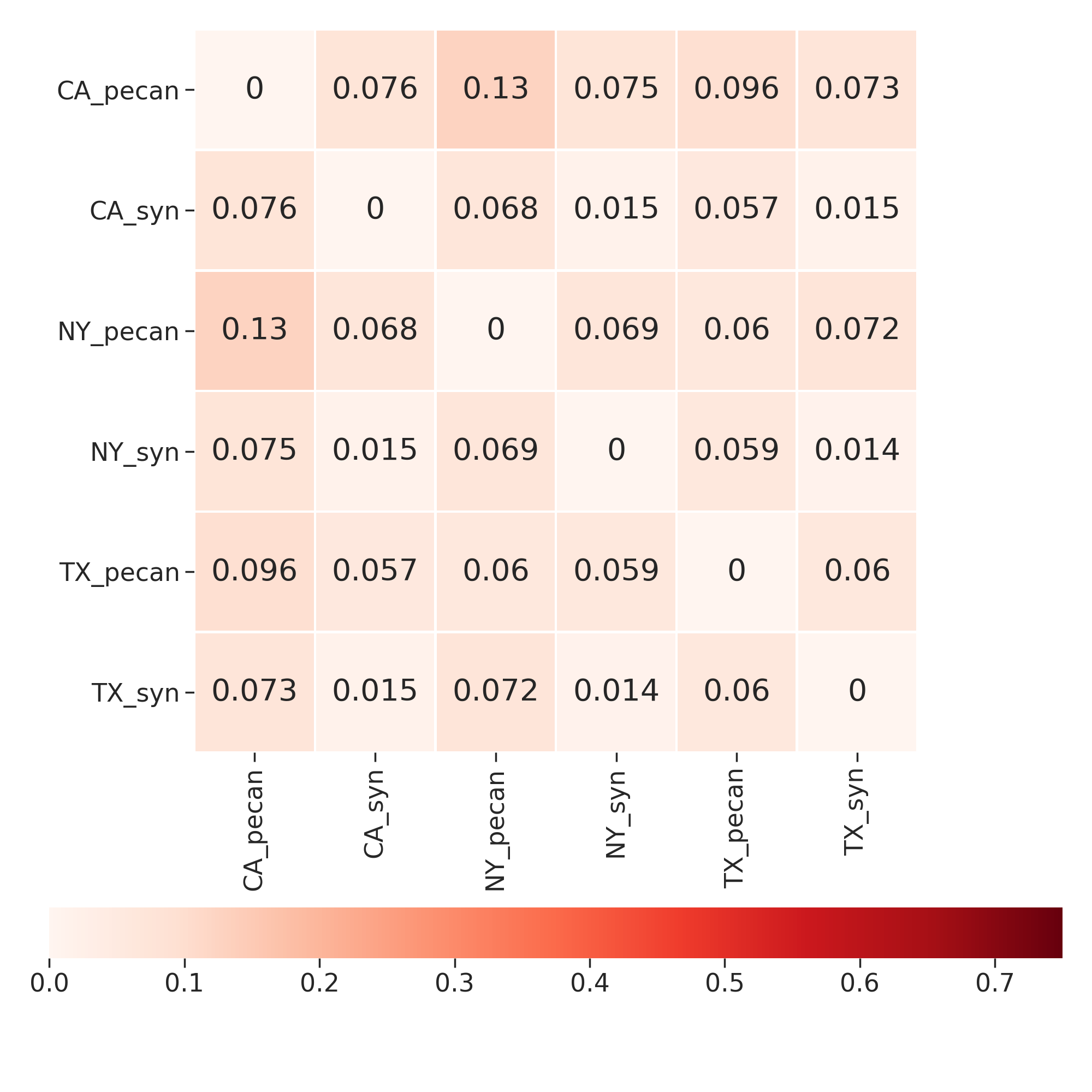}
    \caption{Cooking appliances (Hellinger)}
      \label{fig:}
    \end{subfigure}
    \caption{\textbf{Left column: Jensen-Shannon distance matrices, Right column: Hellinger distance matrices}. Each of the column shows Jensen-Shannon distance and Hellinger distance matrices between end-use probability distributions. Each matrix represents distances between two energy usage distributions for a particular enduse (e.g. HVAC, refrigerator, cooking). The row and column headers of the matrix represent different data-sources and different regions and each cell represents the probability distribution similarity/distance value in the form of heatmap, where the bar shows the range of the values on a continuous scale. }
    \label{fig:jensen-hellinger}
\end{figure*}

In this experiment, distributions of synthetic and real daily end-use data are compared using statistical metrics.
One way of comparing these distributions is by measuring distance between the real and synthetic end-use distributions.
Many metrics can be used to perform this task (e.g., Kullback–Leibler divergence (KL), the Hellinger distance, total variation distance (TVD), the Wasserstein metric, the Jensen-Shannon divergence (JS), and the Kolmogorov–Smirnov statistic (KS)).
Klemenjak et al.~\cite{Klemenjak2020} use JS distance and Hellinger distance as examples to compare distributions of appliance energy use between different datasets.
A similar method is implemented in this section using the JS distance and the Hellinger distance metric.
In our case, computing the distances between daily end use distributions allows us to perform regional comparisons as well as comparisons between real and synthetic datasets.

The Jensen-Shannon distance is the square root of the Jensen-Shannon divergence~\cite{JS1991}. The range of this metric ranges between $[0,1]$ where~0 implies the distributions are similar. 
We prefer JS divergence over KL divergence since it is a symmetric measure.
If~$P$ and~$Q$ are two probability vectors, then the JS distance~$\mathsf{JS}(P,Q)$ is given by 
\begin{equation}
    \mathsf{JS}(P,Q)\;=\;\sqrt{\frac{\mathsf{KL}(P||M)+\mathsf{KL}(Q||M)}{2}} \;,
\end{equation}
\noindent
where~$M$ is the pointwise mean of~$P$ and~$Q$ and~$\mathsf{KL}$  is the Kullback-Leibler divergence.
To supplement our study, we use Hellinger distance as a second metric to quantify the similarity between two probability distributions.
Hellinger distance is also a symmetric measure. Its range of values is $[0,1]$ with~0 encoding that the distributions are similar.
The Hellinger distance of two probability vectors~$P$ and~$Q$ is denoted by~$\mathsf{H}(P,Q)$ and defined as
\begin{equation}\label{eq:hellinger}
    \mathsf{H}(P,Q)\;=\;\frac{1}{\sqrt{2}}\sqrt{\sum^{k}_{i=1} (\sqrt{p_i} - \sqrt{q_i})^{2}} \;,
\end{equation}
\noindent
where~$k$ is the length of the vectors, and~$p_i$,~$q_i$ are the~$i^{\text{th}}$ elements of the vectors~$P$ and~$Q$, respectively.

Daily end-use energy usage (e.g. ${E}^{\mathsf{hvac}}_i$) at household level are compared in the real and synthetic data for every location specified in Table~\ref{tab:validation-datasets}.
Vectors $P$ and $Q$ denote values in a single end-use for two datasets.
Tables~\ref{fig:jensen-hellinger}(a)(b)(c) list JS distances and Tables~\ref{fig:jensen-hellinger}(d)(e)(f) list Hellinger distances for selected end-uses (HVAC, refrigerator, cooking appliances). 
Each matrix represents distances between two energy usage distributions for an end-use.
The row and column headers represent different data-sources and different regions and each cell represents the probability distribution similarity/distance value in the form of heatmap where the bar shows the range of the values on a continuous scale. 

The JS and Hellinger distance tables for end-uses show strong similarities (the distance is close to zero).
Furthermore, within each matrix three types of comparisons are performed. We compute similarity between end-use distributions for different regions within synthetic data, different regions within real data, and different regions in different data sources (namely real and synthetic data).
For appliance usage (e.g. cooking), the distributions are quite  similar across regions and data-sources.
This supports findings from Figure~\ref{fig:regional-linechart-activities} that there exists significant similarities between different regions for synthetic daily energy consumption of different appliances. 
For HVAC end-use, it is observed that the distributions grow apart between regions for both -- synthetic and real data sources. This is particularly true due to the strong association of HVAC with outdoor/environment temperature conditions and the time span for which these temperature conditions prevail (e.g., warmer temperatures are observed for a longer time in Texas (TX))

\subsection*{II. Comparing energy use patterns (load shape/structural similarity)}

\begin{figure*}[!h]    
    \begin{subfigure}{.495\textwidth}
    \centering
    \includegraphics[width=8.25cm,height=11cm]{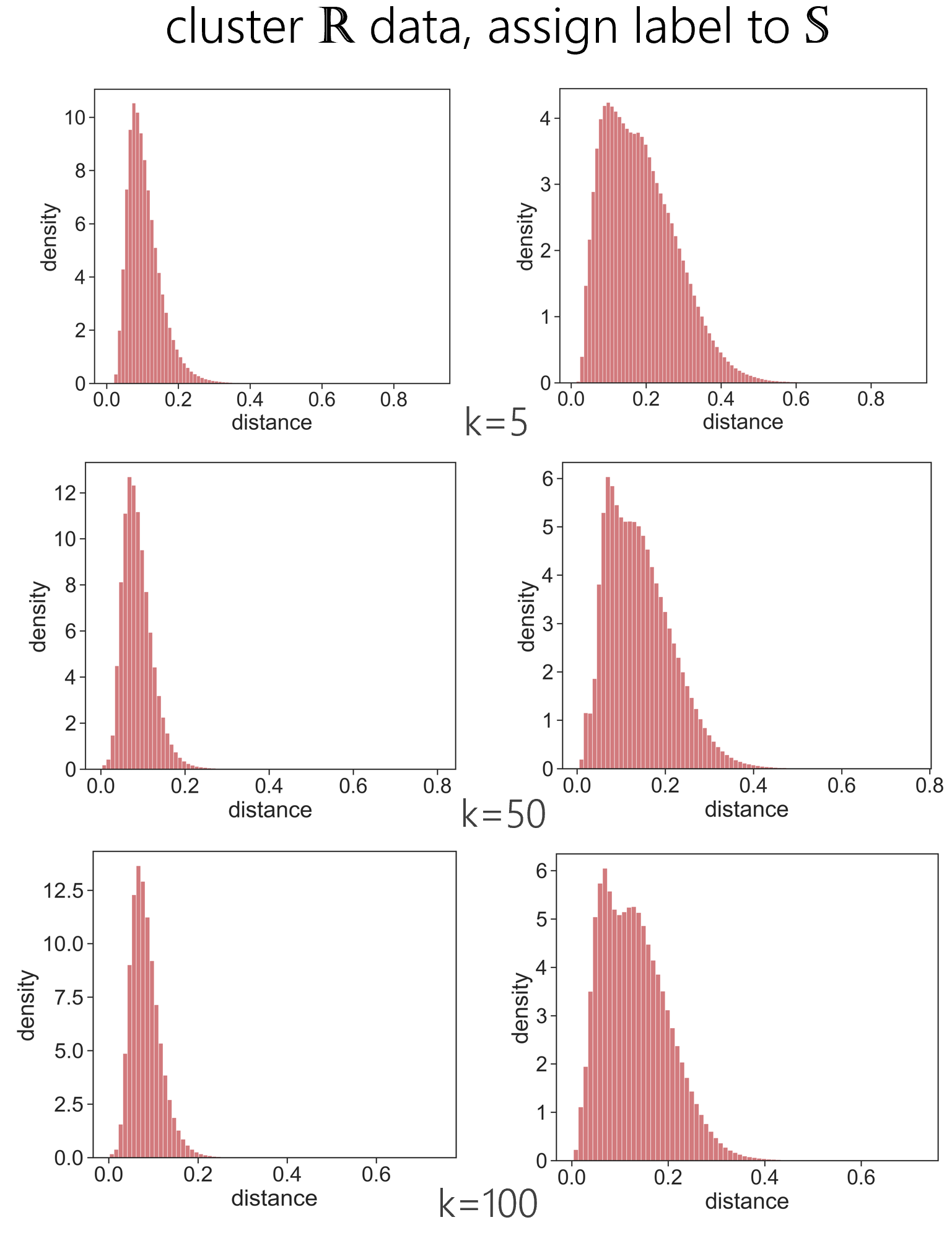}
    \caption{Case 1: $closeness(\mathscr{R},\mathscr{S})$}
      \label{fig:}
    \end{subfigure}
    \begin{subfigure}{.49\textwidth}
    \centering
    \includegraphics[width=8.25cm,height=11cm]{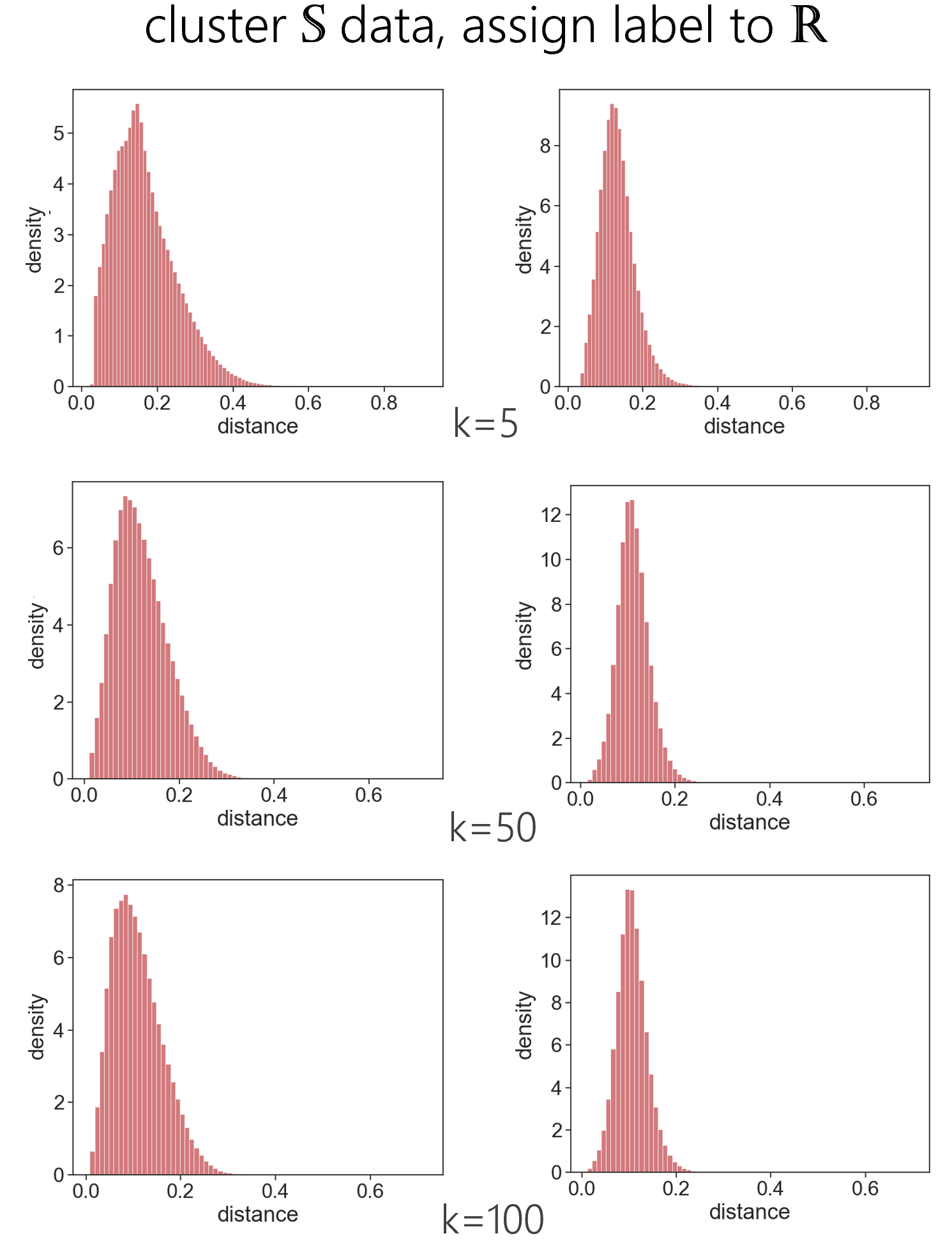}
    \caption{Case 2: $closeness(\mathscr{S},\mathscr{R})$}
      \label{fig:}
    \end{subfigure}
    
    \caption{\textbf{Example of \emph{closeness} in different cases with varying $k$.} Figures show the distances of data points from sets  $\mathscr{R}$ and $\mathscr{S}$ to their respective cluster center. (a) demonstrates histograms of distances for different $k$. The plot on left is for real data points and on right is for synthetic data points. Then, we calculate $closeness(\mathscr{R},\mathscr{S})$ using Hellinger distance (corresponds blue line in Figure~\ref{fig:overall-vv-results}(c)). For $k=5$ a bimodal pattern is observed in distances for synthetic data points which tends to diminish as the number of clusters $k$ increases.  Figure (b) shows histograms of distances for different $k$ for case 2. The plot on left is for synthetic data points and on right is for real data points.  $closeness(\mathscr{S},\mathscr{R})$ is calculated using Hellinger distance (corresponds to orange line in Figure~\ref{fig:overall-vv-results}(c)).
    }
    \label{fig:closeness_proximity}
\end{figure*}

\begin{figure*}[!h]    
    \begin{subfigure}{.48\textwidth}
    \centering
    \includegraphics[width=7.5cm,height=5.5cm]{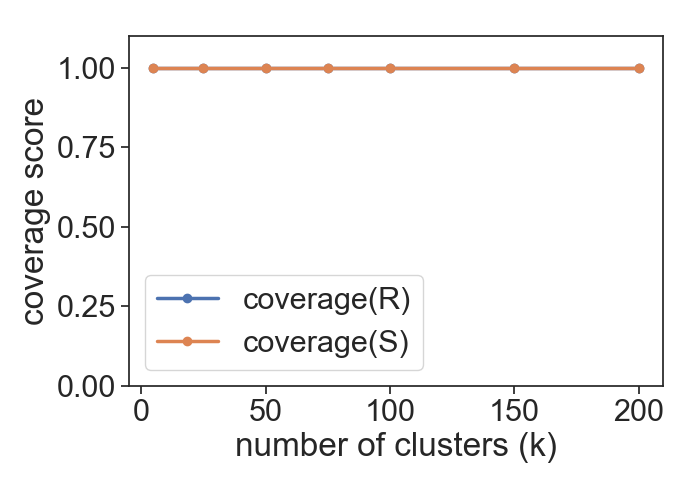}
    \caption{Coverage}
      \label{fig:}
    \end{subfigure}
    \begin{subfigure}{.48\textwidth}
    \centering
    \includegraphics[width=7.5cm,height=5.5cm]{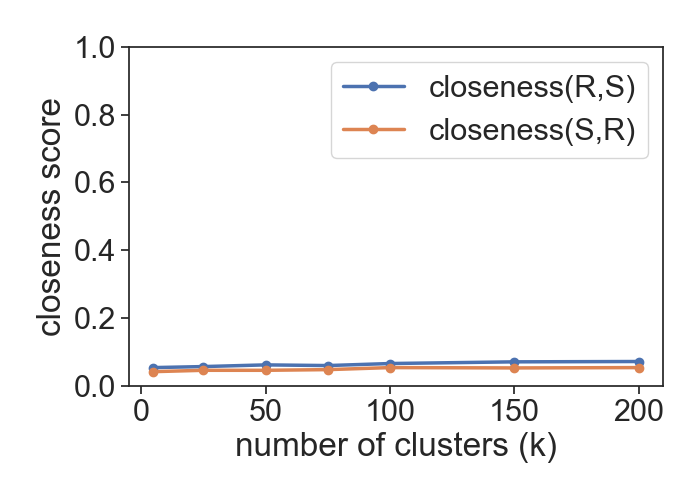}
    \caption{Closeness}
      \label{fig:}
    \end{subfigure}
    \caption{\textbf{Summary of the two case scenarios.} Orange color is denoted for findings of case 1 where we cluster real data set $\mathscr{R}$ and assign a cluster label to synthetic data set $\mathscr{S}$. Blue color is denoted for findings of case 1 where we cluster synthetic data set $\mathscr{S}$ and assign a cluster label to real data set $\mathscr{R}$. (a) illustrates 100\% coverage in both cases even as $k$ varies. This means that, in each case at least one data point belongs to every cluster for a given $k$. (b) shows the closeness between the two distance vectors : distance of real data points in a cluster to its respective centroid and  distance of synthetic data points in a cluster to its respective centroid. Closeness is given by the Hellinger distance which suggests that a value of 0 signifies that the two distributions are similar. The value of distances is close to 0 for all values of $k$ in both the cases. However, an upward trend is observed as $k$ increases. Overall we see the robustness of results w.r.t. $k$.} 
    \label{fig:overall-vv-results}
\end{figure*}

In this section, the synthetic energy use timeseries are evaluated using the concepts of diversity, coverage, and closeness. 
The diversity in energy use patterns is captured by segmenting the normalized timeseries $\langle \overline{e}_0,\ldots,\overline{e}_{23}\rangle$  using unsupervised learning techniques such as clustering.
This is followed by studying \emph{coverage} in terms of what percentage of synthetic timeseries population is represented in the real timeseries population and vice versa. Thus, coverage is used to measure  diversity.
However, learning only coverage is not sufficient. It is necessary to measure the accuracy of the matches found. Hence, we introduce the \emph{closeness} metric. It studies how close (e.g. $dist(i,j)$) are the synthetic and real data points.

Let $\mathscr{R}$ and $\mathscr{S}$ be the set of load shapes of real and synthetic energy use timeseries. 
Let $K_\mathscr{R}$ be the number of unique load shapes (segments/patterns/clusters) found in set $\mathscr{R}$.
Then, we define the ${coverage(\mathscr{S})}$ as a ratio

\begin{equation}
\begin{array}{l}
    coverage(\mathscr{S})\;=\;\frac{\textrm{Number of unique shapes in $\mathscr{R}$ that contain atleast one data point from $\mathscr{S}$}}{\textrm{Number of unique shapes in $\mathscr{R}$}} \\ \\
   \quad \quad \quad \quad \quad \; \; =\;\frac{1}{K_\mathscr{R}} \times \sum_{b=1}^{K_\mathscr{R}} \mathbb{I}_b
    \quad \text{where}   \\ \\
    \mathbb{I}_b\;=\; \begin{cases}
      1 & \text{if cluster $b$ contains atleast one timeseries $j\in\mathscr{S}$ }\;,\\
      0 & \text{otherwise.}
    \end{cases}    
    \end{array}
\end{equation}

\noindent
Thus, $coverage(\mathscr{S})$ reflects the degree to which samples from set $\mathscr{S}$ cover the patterns in set $\mathscr{R}$.
Similarly, if $K_\mathscr{S}$ is the number of unique segments in set $\mathscr{S}$, then,  $coverage(\mathscr{R})$ reflects the the percentage of unique patterns in set $\mathscr{S}$ covered by data points in set $\mathscr{R}$.
Coverage is bounded between 0 and 1.
Figure~\ref{fig:} shows $coverage(\mathscr{S})$ and $coverage(\mathscr{R})$ as $K$ varies.

To measure closeness we calculate distance of individual timeseries to it's respective cluster center/representative.
If $K_\mathscr{R}$ is the number of clusters in set $\mathscr{R}$, then, the \emph{closeness($\mathscr{S}$,$\mathscr{R}$)} of set $\mathscr{S}$ to set $\mathscr{R}$ is measured by comparing the distributions of distances of individual timeseries $i\in\mathscr{R}$ and $j\in\mathscr{S}$ in each cluster $c\in K_\mathscr{R}$ to the respective center/representative timeseries of the cluster.
Figure~\ref{fig:} illustrates the schematic of building the distance distributions.
Let $P_\mathscr{R}$ and $P_\mathscr{S}$ denote the probability vectors of distances of sets $\mathscr{R}$ and $\mathscr{S}$ respectively. To measure the degree of closeness, we compare the two probability distributions using Hellinger distance $\mathsf{H}(P_\mathscr{R},P_\mathscr{S})$ (Equation~\ref{eq:hellinger}).
If distributions $P_\mathscr{R}$ and $P_\mathscr{R}$ are similar, then we say that set $\mathscr{S}$ is close to set $\mathscr{R}$.
\begin{equation}
   closeness(\mathscr{S},\mathscr{R}) \;=\; \mathsf{H}(P_\mathscr{R},P_\mathscr{S})
\end{equation}
Closeness is bounded between 0 and 1. 0 implies that the two sets are close. 
Note that closeness is not a symmetric metric i.e. 
$closeness(\mathscr{S},\mathscr{R}) \neq closeness(\mathscr{R},\mathscr{S})$.
Figure~\ref{fig:} describes the variation in similarity score of the probability with different number of segments $K$.

Now, we briefly describe the experimental setup. Two cases are considered to examine coverage, closeness and robustness of cluster groupings ($k$). 
For each case the energy use timeseries is normalized resulting in a \emph{load shape} $\langle \overline{e}_0,\ldots,\overline{e}_{23}\rangle$.
We choose normalization by total consumption~(Equation~\ref{eq:total-normalization}) in order to consider pronounced effects of peak-load in the profile.
Household preferences or lifestyles can be typically captured by one or more load shapes~\cite{kwac2014}, hence we choose this representation for uncovering patterns in the data. 
Thus, every $i \in \mathscr{R}$ and $j \in \mathscr{S}$ are normalized energy use vectors of length 24.

\begin{equation}\label{eq:total-normalization}
    \overline{e}_t\;=\;\frac{e_t}{E^\mathsf{total}}\;,\quad \textrm{where} \; \; E^\mathsf{total} = \sum^{23}_{t=0} e_t
\end{equation}

In the first case (Case 1), we generate $K_\mathscr{R}$ patterns from set $\mathscr{R}$ by clustering the real normalized energy use vectors using k-means clustering algorithm with Euclidean distance.
This is followed by assigning a cluster label $k\in K_\mathscr{R}$ to each synthetic energy use timeseries $j\in \mathscr{S}$. Let $c_k$ be the center/representation vector of group $k$.
Then, $j \in \mathscr{S}$ is assigned to the cluster whose cluster center distance is minimum from $j$ and is given by 
$min(dist(j,c_0),\ldots, dist(j,c_{K_\mathscr{R}}))$.
Then, we calculate the coverage of synthetic data $coverage(\mathscr{S})$ and closeness of synthetic data to real data among all clusters as $closeness(\mathscr{S},\mathscr{R})$.
In Case 2, we generate $K_\mathscr{S}$ clusters from set $\mathscr{S}$ (synthetic data) by segmenting the normalized energy use vectors using k-means clustering algorithm with Euclidean distance.
This is followed by assigning a cluster label $k\in K_\mathscr{S}$ to each real energy use timeseries $i\in \mathscr{R}$. $i$ is assigned to the cluster whose cluster center distance is minimum from $i$ and is given by $min_{\forall k \in K_\mathscr{S}}  dist(i,c_k)$.
Then, we calculate the coverage of real data in synthetic groups $coverage(\mathscr{R})$ and closeness of real data and synthetic data among all synthetic clusters as $closeness(\mathscr{R},\mathscr{S})$.

Results of both the cases are summarized in Figures~\ref{fig:overall-vv-results}.
A 100\% \emph{coverage} is observed in both the cases for different values of $k$.
Observations for \emph{closeness} metric are interesting. The Hellinger distance is close to zero in all the scenarios, however there is a slight uptake in the value as $k$ increases. 
We inspect this further in Figure~\ref{fig:closeness_proximity}.
Figure~\ref{fig:closeness_proximity} shows histograms of distances of real data points and synthetic data points from their assigned cluster center.
In case 1, the distribution of distances of synthetic data points is slightly broader than the distribution of distances of real data points for all $k$. Thus, we see a distance for \emph{closeness($\mathscr{R},\mathscr{S}$)} in Figure~\ref{fig:overall-vv-results}(c).
As $k$ increases it is observed that some individual clusters have a broad and/or bimodal distance distribution indicating that there are data points that are very close to the cluster center while a few are far away. This difference is apparent as the number of clusters increases.

The goal of this V\&V exercise was to verify if the diversity and trends of the real energy use profiles are replicated in the synthetic energy use profiles.
Due to a biased and skewed sample of the real energy use data, it is challenging to perform validation of synthetic data.
Some of the characteristics of the real datasets that hinder the  implementation of using existing evaluation metrics \emph{as is} are mentioned below.
No supporting information of the real households is available (e.g. household size, dwelling type, square footage, indoor thermostat setting).
We have shown that all of these factors are extremely important in the generation of household demand at a given time.
Some of households in the real data may also be participants in demand-response programs resulting in unique load shapes due to shifting demand/reducing peak demand that may not be found in households not participating in DR programs (e.g. synthetic data).
The real datasets are collected for different years for each region.
The data are incomplete for some regions (e.g. San Diego samples do not have lighting data).
The sample size (number of households) is highly skewed. It varies from 9 households in Montana to 56000 households for Horry,SC.
Thus, it is important to note that $|\mathscr{R}|<<|\mathscr{S}|$ (e.g. the number of households simulated in our framework for Washington state is far greater than that of 78 households in real data for Washington state.) 
All of these observations are summarized in Table~\ref{tab:validation-datasets}.

\subsection*{III. Observing differences and similarities in synthetic energy use data in spatially representative locations}

\begin{figure*}[!h]    
    \begin{subfigure}{.17\textwidth}
    \centering
    \includegraphics[width=3cm,height=3cm]{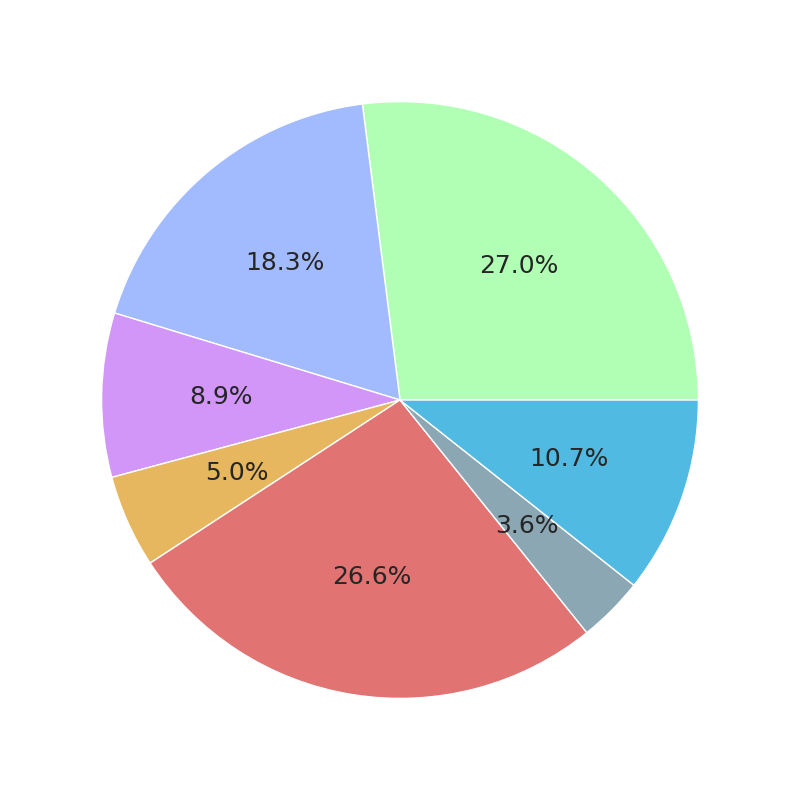}
    \caption{Arlington, VA}
      \label{fig:}
    \end{subfigure}
    \begin{subfigure}{.17\textwidth}
    \centering
    \includegraphics[width=3cm,height=3cm]{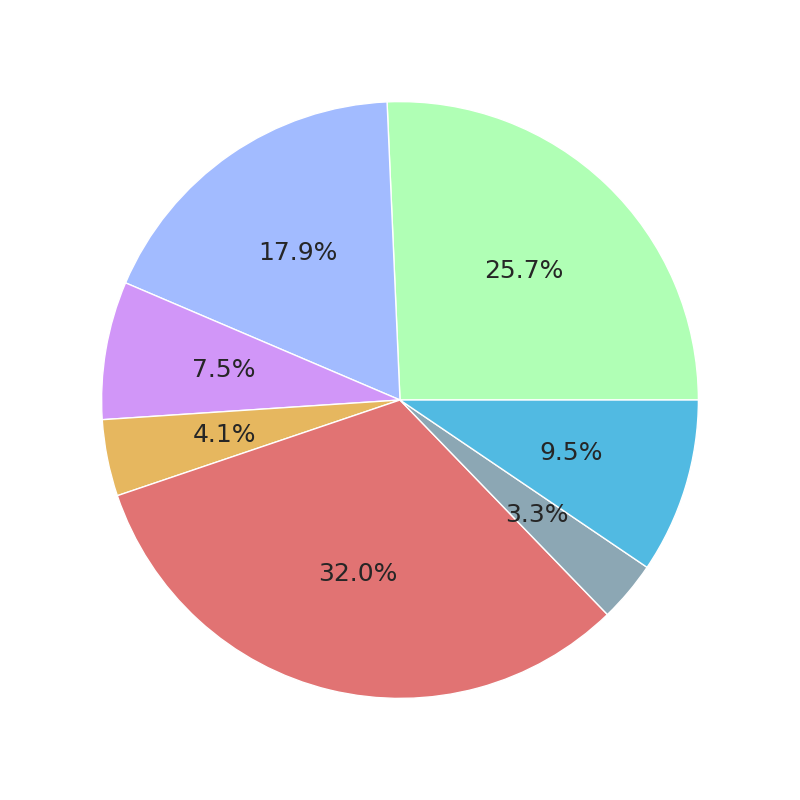}
    \caption{Cook, IL}
      \label{fig:}
    \end{subfigure}
    \begin{subfigure}{.17\textwidth}
    \centering
    \includegraphics[width=3cm,height=3cm]{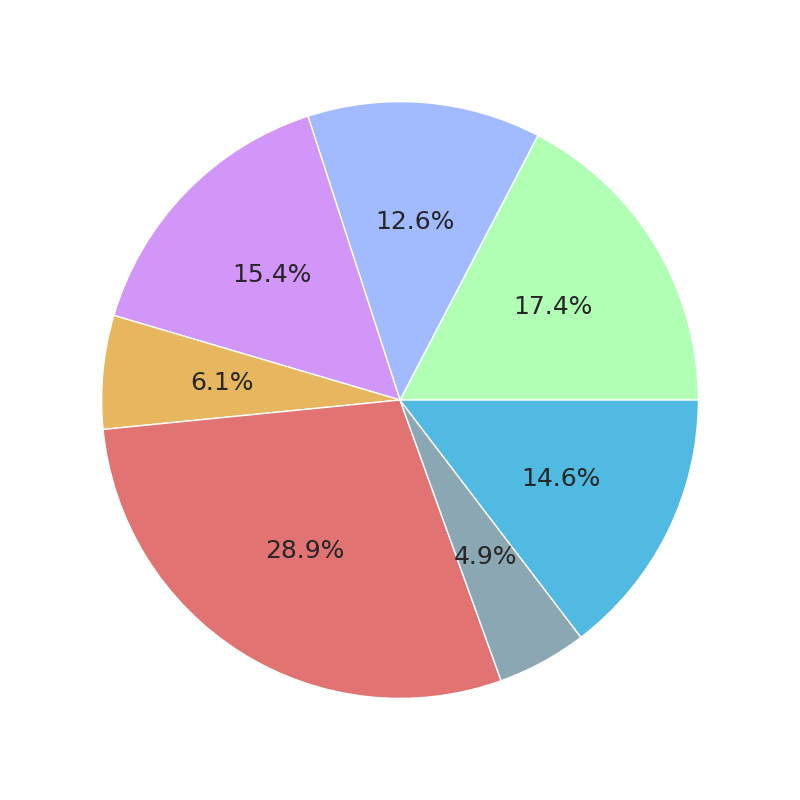}
    \caption{Houston, TX}
      \label{fig:}
    \end{subfigure}
    \begin{subfigure}{.17\textwidth}
    \centering
    \includegraphics[width=3cm,height=3cm]{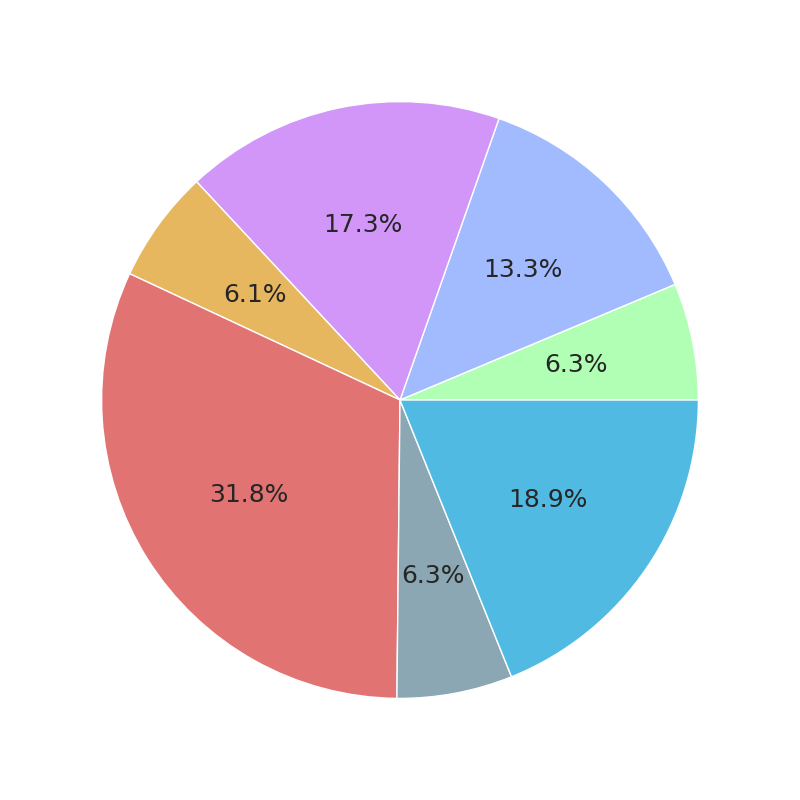}
    \caption{Maricopa, AZ}
      \label{fig:}
    \end{subfigure}
    \begin{subfigure}{.17\textwidth}
    \centering
    \includegraphics[width=3cm,height=3cm]{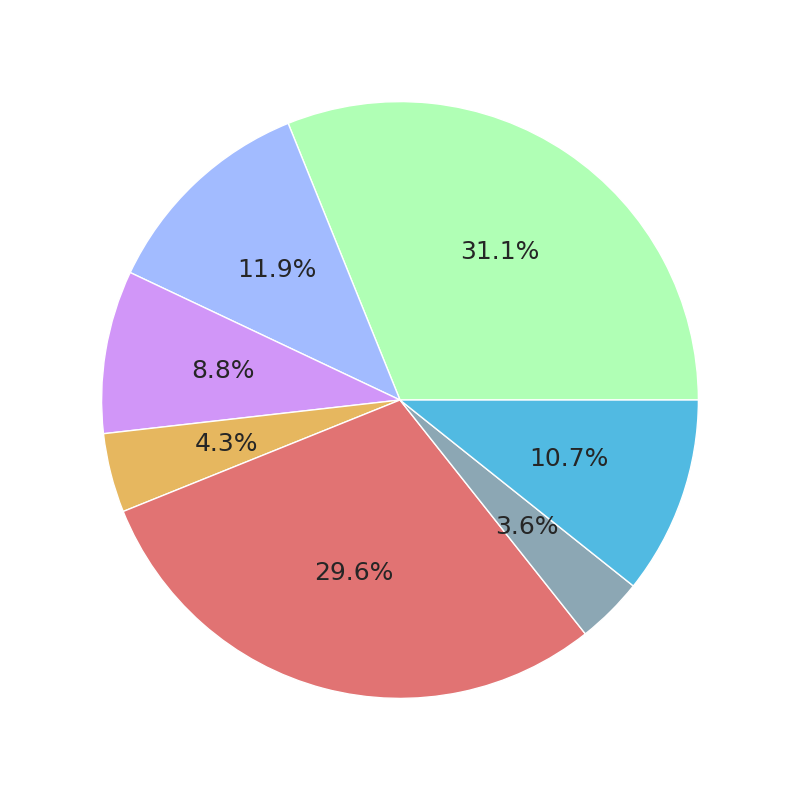}
    \caption{King, WA}
      \label{fig:}
    \end{subfigure}
    \begin{subfigure}{.13\textwidth}
    \centering
    \includegraphics[width=2cm,height=2.9cm]{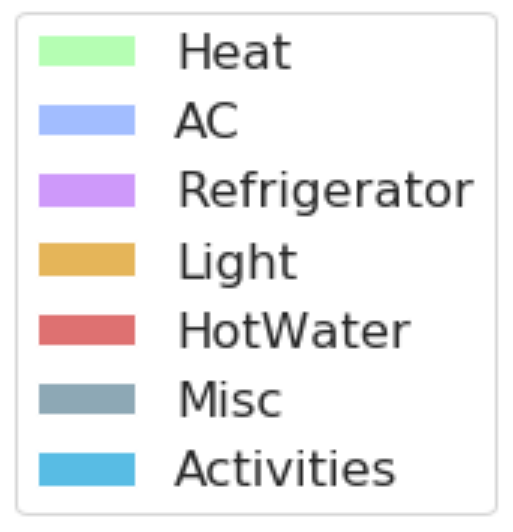}
    \caption{Legend}
      \label{fig:}
    \end{subfigure}
    
    \caption{\textbf{Composition of synthetic electric consumption in the representative target locations.} Heating and cooling constitute the majority part of the residential electric consumption. Refrigerators consume slightly higher energy in hotter regions such as Maricopa and Houston. Activities such as dishwashing, laundry, and cooking represents between 8-17\% for different regions. Lighting and water heating have a consistent proportion of consumption across all locations. The proportions bear similarities with data published by EIA.}
    \label{fig:pie-chart-regional}
\end{figure*}

This empirical study uses only the synthetic data to conduct a comparative regional analyses to examine similarities and dissimilarities between energy use for different end-uses.
We observe the spatio-temporal patterns and variations in different end-uses with respect to environmental elements such as irradiance and temperature as well as demographic and structural characteristics of the households.
The selected target locations are spatially representative of different climate zones of the U.S.:

\begin{center}
\emph{Arlington, VA}; \emph{Cook County, IL}; \emph{Houston County, TX}; \emph{Maricopa County, AZ}; \emph{King County, WA} \end{center}

The composition of electric consumption by end-uses is shown in the form of pie diagrams in Figure~\ref{fig:pie-chart-regional}.
EIA reports the shares of the major end-uses as follows: DHW 17-32\%, lighting 5-10\%, refrigerator 3-5\%, activities/appliances 20-26\%, space heating 25-47\%, and air conditioning 5-10\%.
In general, the percentages of major end-use categories lie in the ranges similar to those reported by EIA.
HVAC has a dominant share in the energy consumption in households as compared to usage of appliances and/or other activities.

\begin{figure*}[!h]    
    \begin{subfigure}{.24\textwidth}
    \centering
    \includegraphics[width=4.2cm,height=2.8cm]{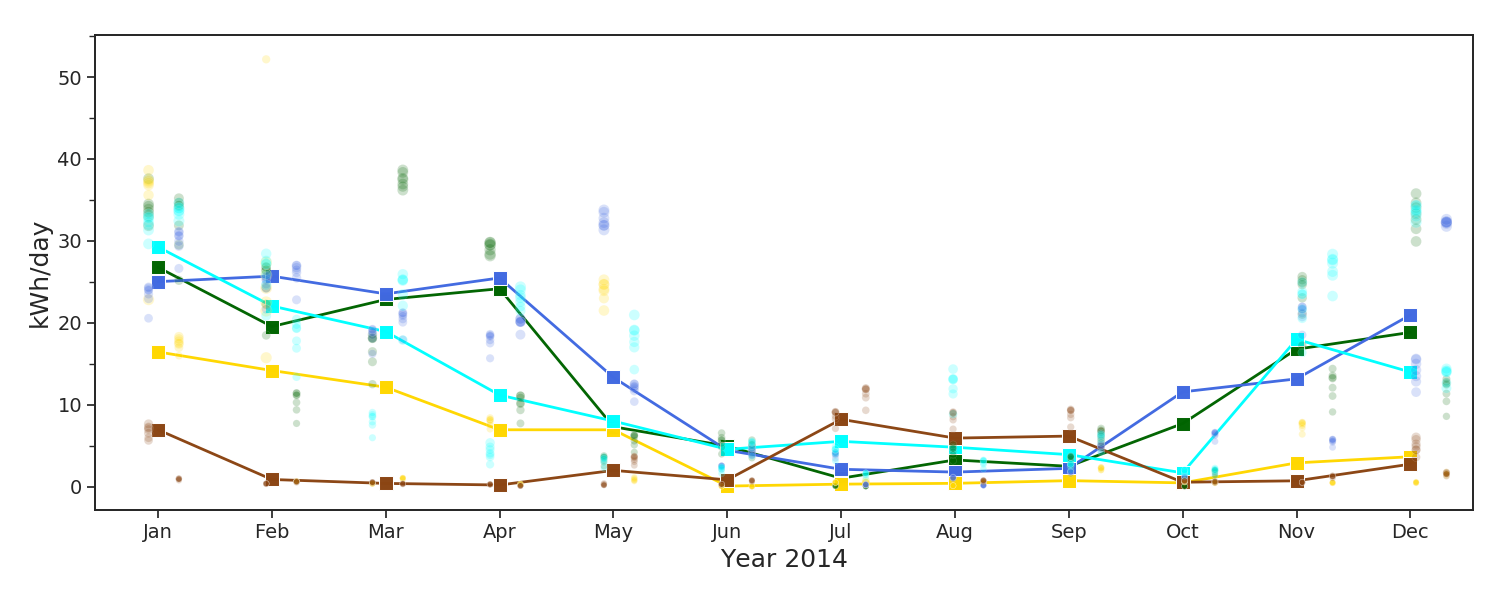}
    \caption{HVAC energy use}
      \label{fig:}
    \end{subfigure}
    \begin{subfigure}{.24\textwidth}
    \centering
    \includegraphics[width=4.2cm,height=2.8cm]{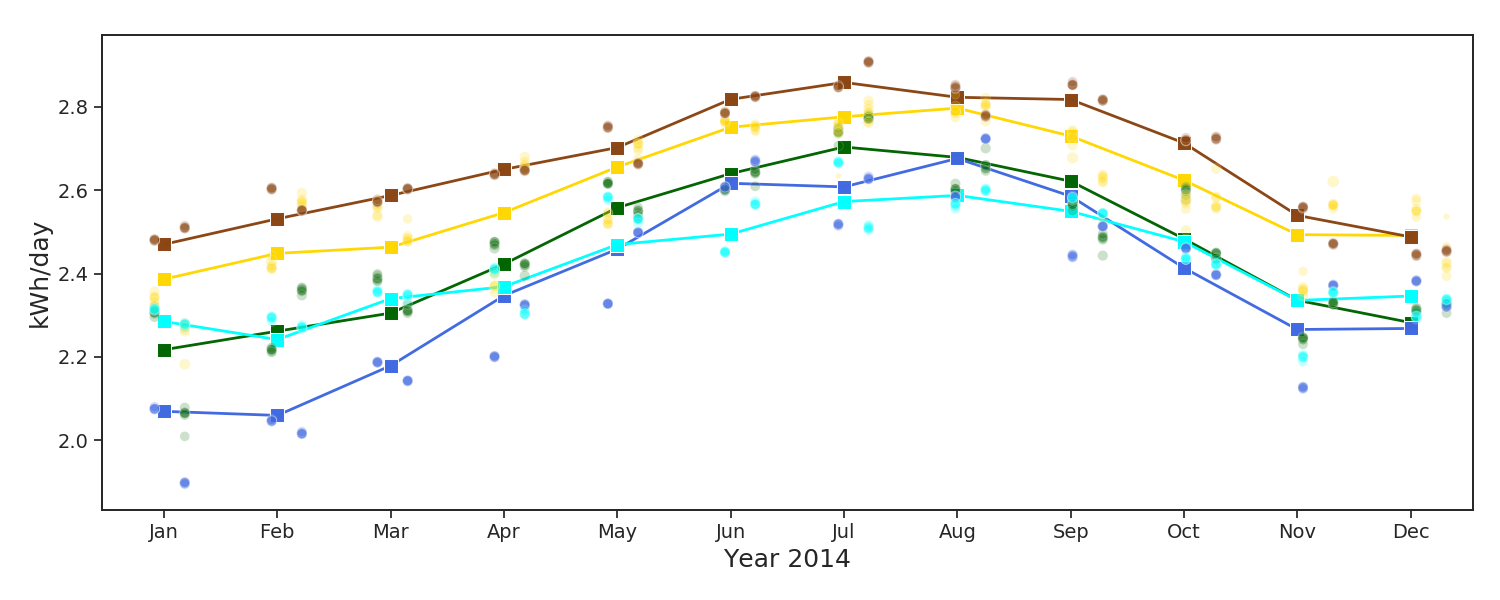}
    \caption{Refrigerator energy use}
      \label{fig:}
    \end{subfigure}
    \begin{subfigure}{.24\textwidth}
    \centering
    \includegraphics[width=4.2cm,height=2.8cm]{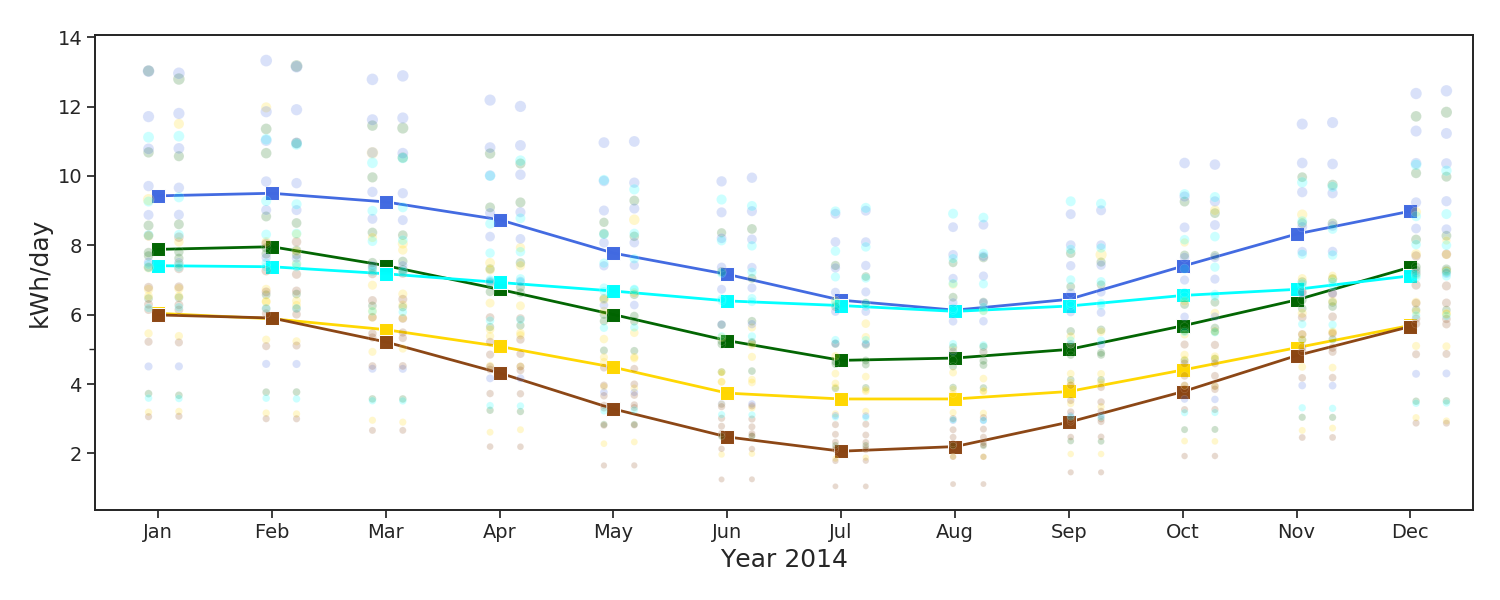}
    \caption{Hot water energy use}
      \label{fig:}
    \end{subfigure}
    \begin{subfigure}{.24\textwidth}
    \centering
    \includegraphics[width=4.2cm,height=2.8cm]{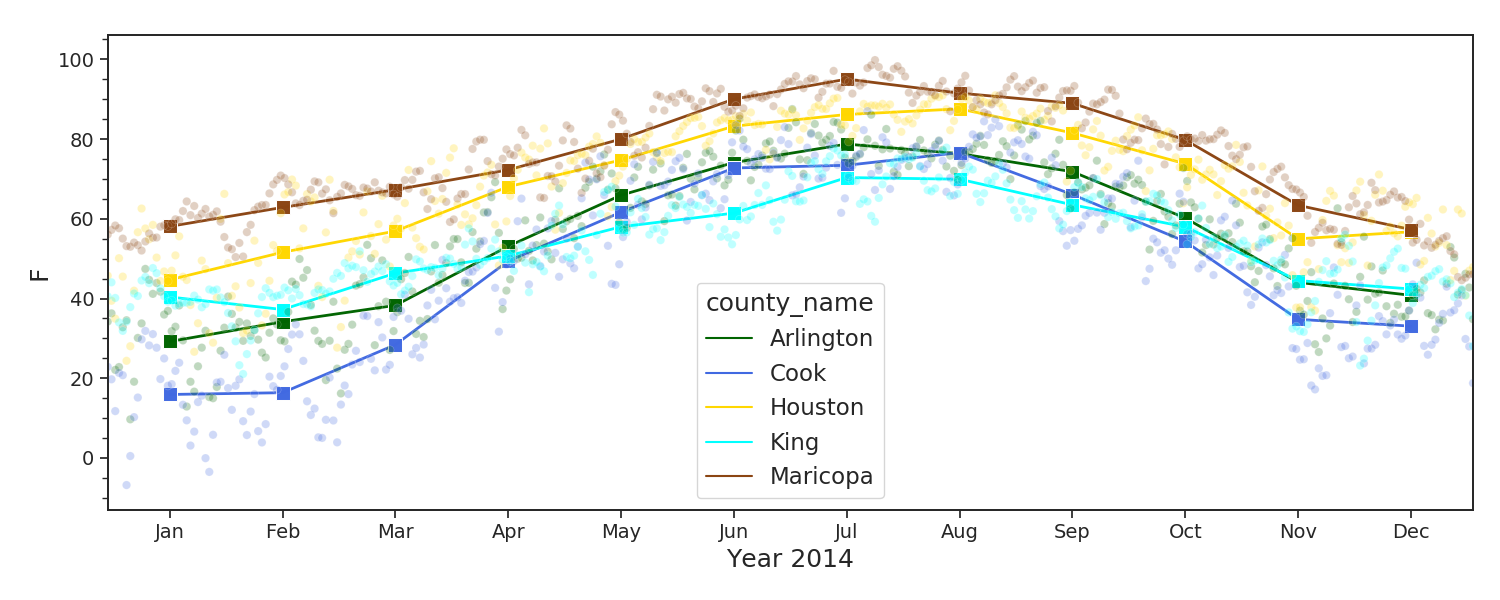}
    \caption{Outside temperature}
      \label{fig:}
    \end{subfigure}
    \caption{\textbf{Monthly synthetic energy use changes in end-uses such as HVAC, refrigerator, domestic hot water w.r.t. temperature.} The above line charts monthly energy use changes in end-uses such as HVAC, refrigerator, domestic hot water w.r.t. outside temperature.
    The line chart shows average daily consumption over all households in the target regions. The scatter plot in the background describes average daily consumption for an end-use for sampled days color coded by location. The size of the markers denotes the standard deviation of the end-use consumption. Legend: Arlington, VA (green); Cook County, IL (blue); Houston County, TX (yellow); Maricopa County, AZ (brown); King County, WA (cyan) }
    \label{fig:regional-linechart-major}
\end{figure*}

\begin{figure*}[!h]    
    \begin{subfigure}{.32\textwidth}
    \centering
    \includegraphics[width=5.5cm,height=2.5cm]{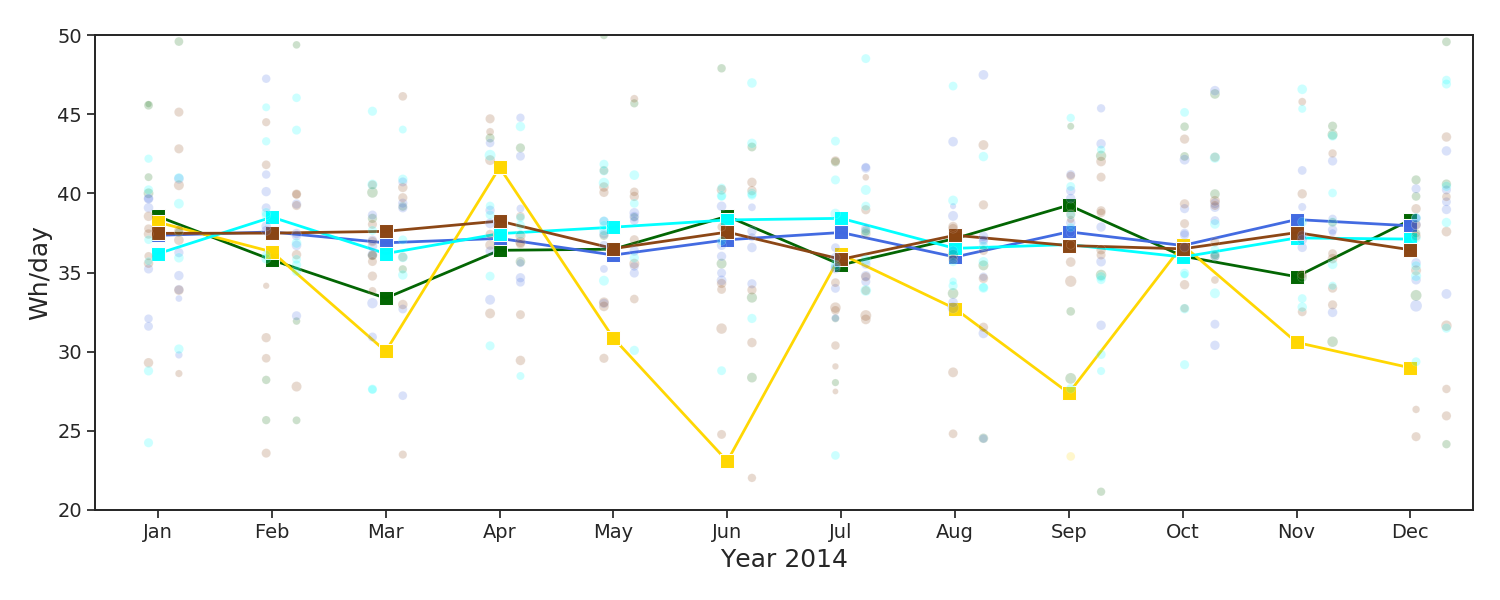}
    \caption{Dishwasher}
      \label{fig:}
    \end{subfigure}
    \begin{subfigure}{.32\textwidth}
    \centering
    \includegraphics[width=5.5cm,height=2.5cm]{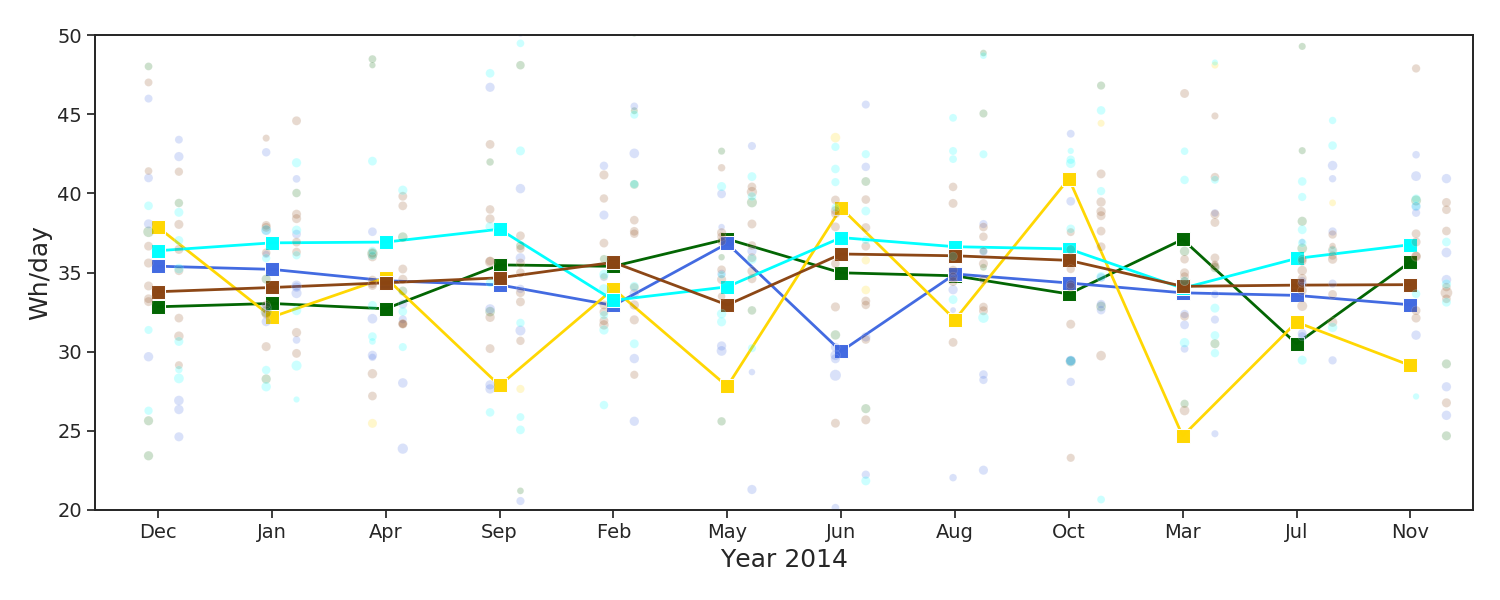}
    \caption{Laundry}
      \label{fig:}
    \end{subfigure}
    \begin{subfigure}{.32\textwidth}
    \centering
    \includegraphics[width=5.5cm,height=2.5cm]{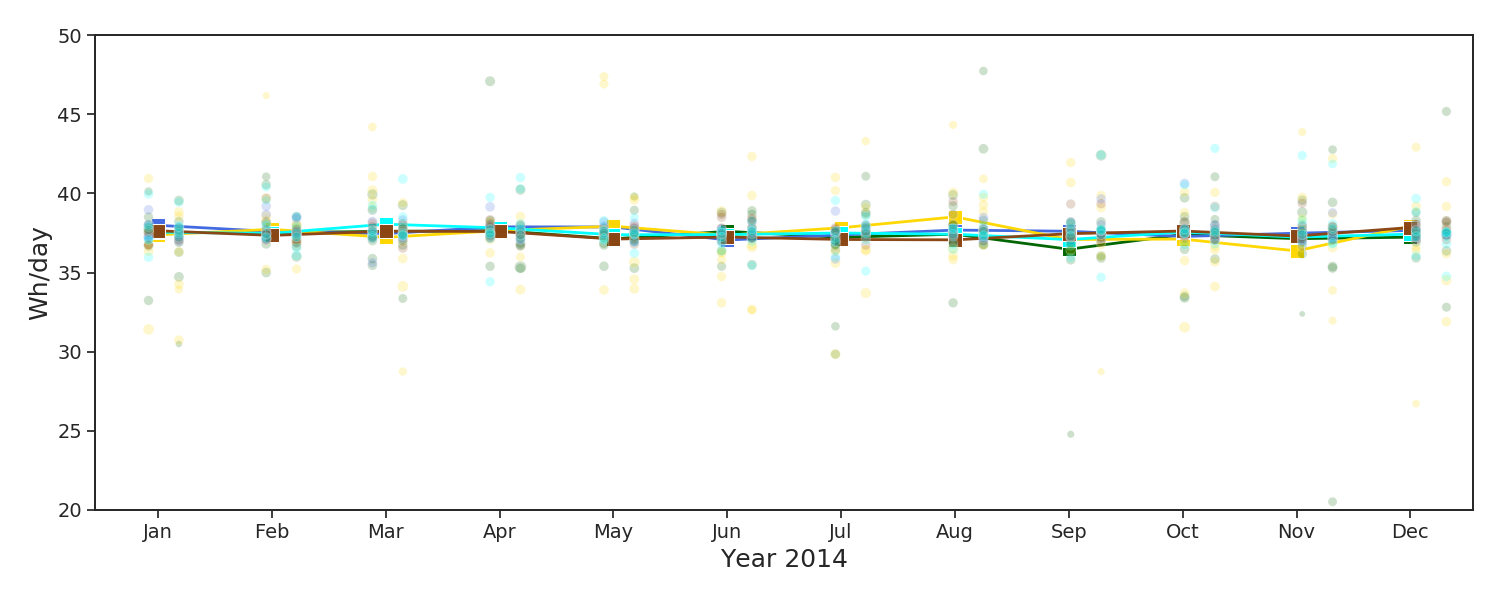}
    \caption{Cook}
      \label{fig:}
    \end{subfigure}
    
    \begin{subfigure}{.32\textwidth}
    \centering
    \includegraphics[width=5.5cm,height=2.5cm]{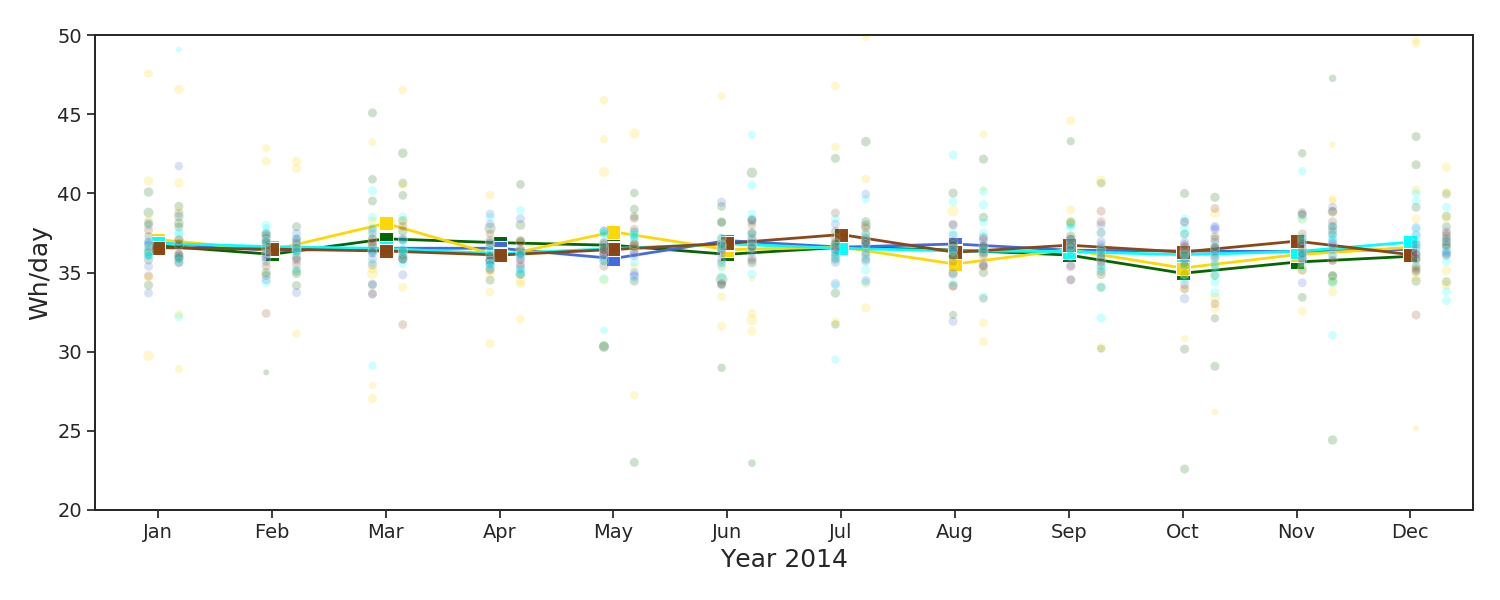}
    \caption{Cleaning}
      \label{fig:}
    \end{subfigure}
    \begin{subfigure}{.32\textwidth}
    \centering
    \includegraphics[width=5.5cm,height=2.5cm]{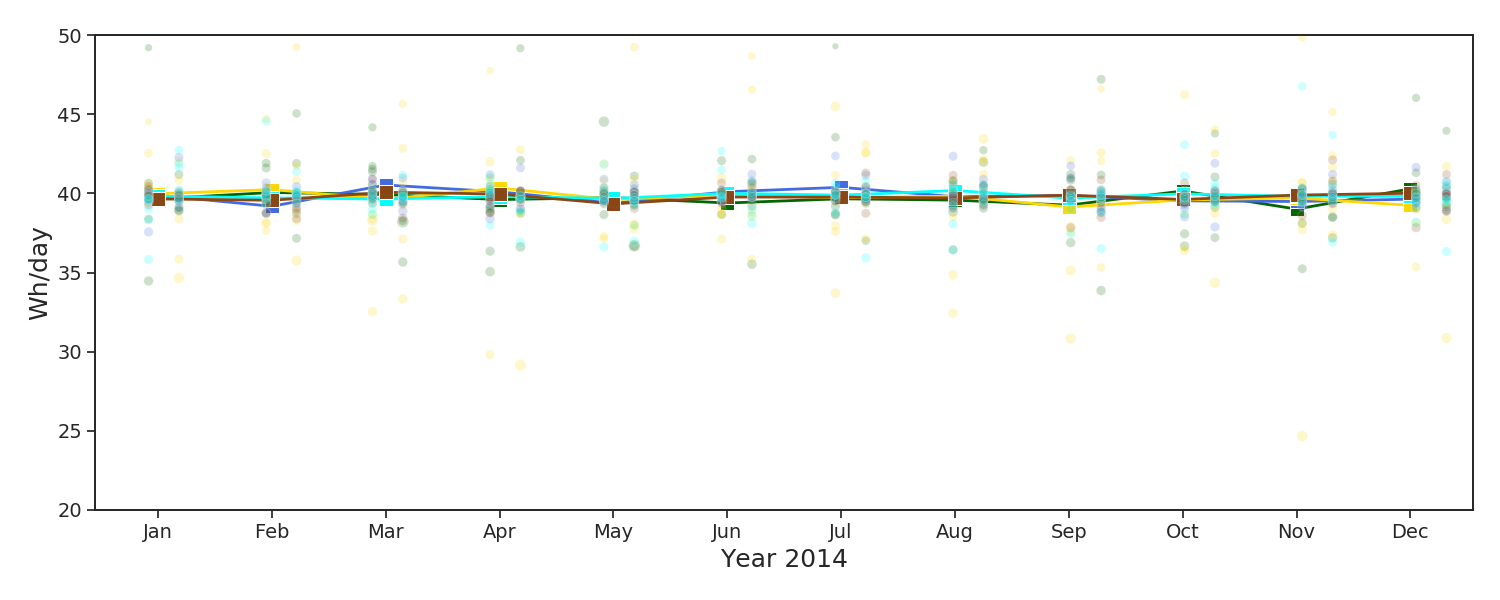}
    \caption{TV use}
      \label{fig:}
    \end{subfigure}
    \begin{subfigure}{.32\textwidth}
    \centering
    \includegraphics[width=5.5cm,height=3cm]{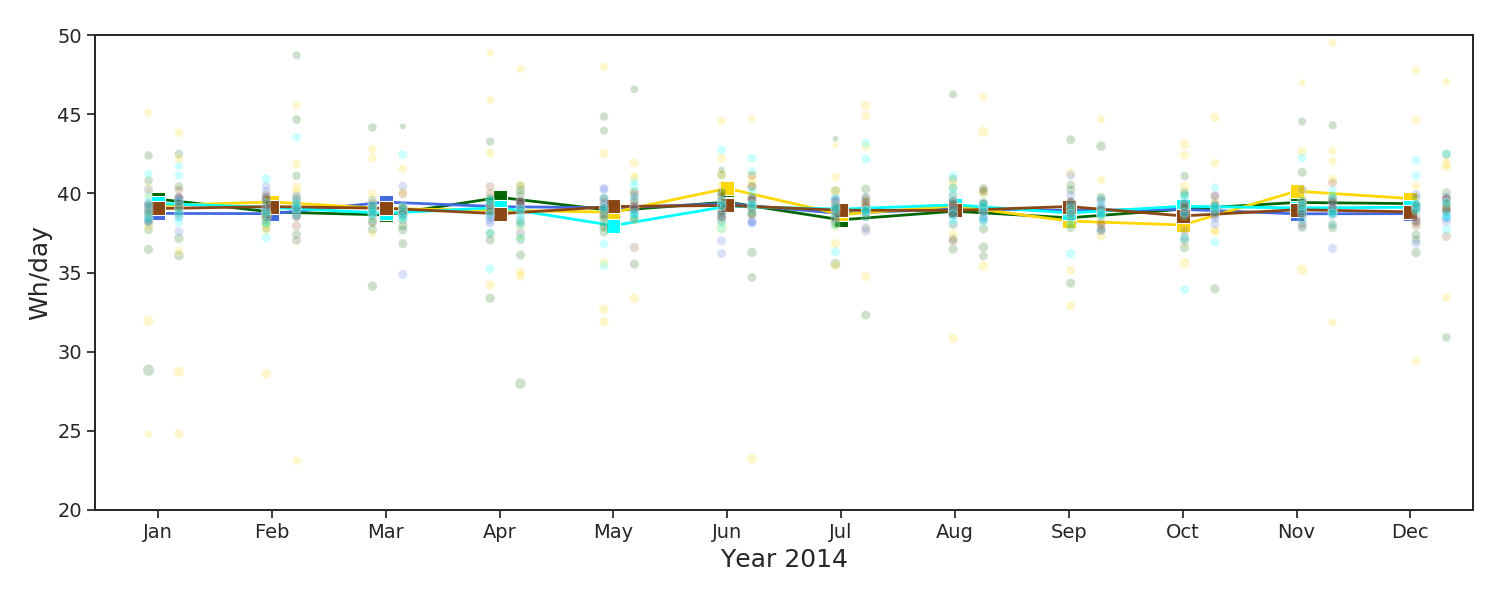}
    \caption{Computer use}
      \label{fig:}
    \end{subfigure}
    \caption{\textbf{Synthetic appliance energy use variation in target locations throughout the year.} The line charts show variation in daily energy consumption for different appliance energy use throughout the year averaged by month. The lines depict average daily consumption over all households in the target region. The scatter plot in the background describes average daily consumption for an end-use for sampled days color coded by location. The size of the markers denotes the standard deviation of the end-use consumption. There are noticeable similarities in appliance-usage throughout all locations indicating that people in different parts of the country use appliances in a similar style. This is a reasonable observation since day-to-day activities such as cooking and cleaning will occur in all households. Their usage pattern may change during the day, but the total energy consumed by the appliance at the end of the day is similar. Arlington, VA (green); Cook County, IL (blue); Houston County, TX (yellow); Maricopa County, AZ (brown); King County, WA (cyan)}
    \label{fig:regional-linechart-activities}
\end{figure*}

Seasonal energy use variations for HVAC, refrigerator, and hot water is captured in Figure~\ref{fig:regional-linechart-major}.
The plot shows variation in daily average energy use of the four end-uses on a monthly basis alongwith temperature across the year 2014.
Refrigerator energy use increases slightly with temperature while energy used to heat water decreases with increase in temperature.

Electricity usage for heating water is the lowest during summer months for all locations (Figure~\ref{fig:regional-linechart-major}c). 
In particular, regions from hot-humid and hot-dry climate zones consume the least amount of energy.
This observation stems from the relation between $E^{\mathsf{h2o,v}}$ and $T^{\mathsf{cold}}_{m,z}$ described in Equation~\ref{eq:dhw}.
The water inlet temperature ( $T^{\mathsf{cold}}_{m,z}$) differs across temporal as well as spatial scale and is dependent on outside environment temperatures~\cite{NRELhotwater129} (Details in Appendix). 
Figure~\ref{fig:regional-h2o-charts} shows plots describing relation between household size and the number of gallons of hot water consumed and energy required to heat water.
Note that, we consider only electric water heaters in this work.

\begin{figure*}[!h]    
    \begin{subfigure}{.49\textwidth}
    \centering
    \includegraphics[width=8.5cm,height=4cm]{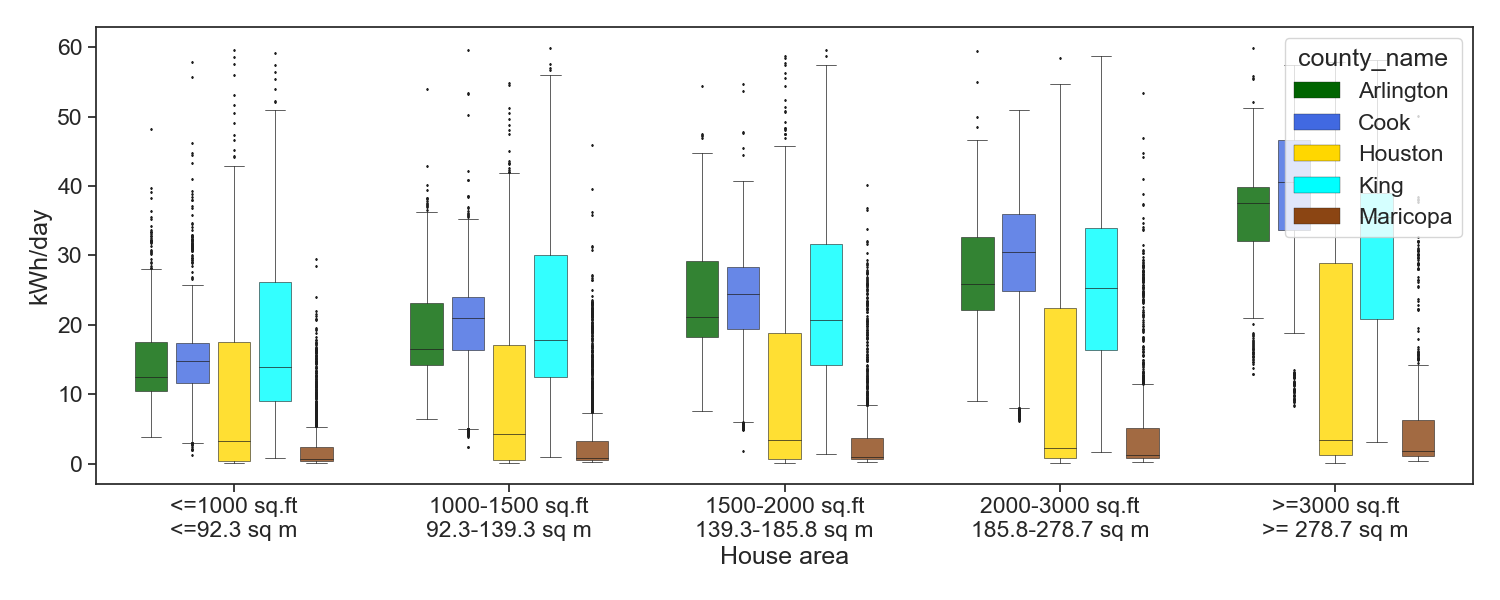}
    \caption{HVAC vs. house area (i.e. floor area)}
      \label{fig:}
    \end{subfigure}
    \begin{subfigure}{.49\textwidth}
    \centering
    \includegraphics[width=8.5cm,height=4cm]{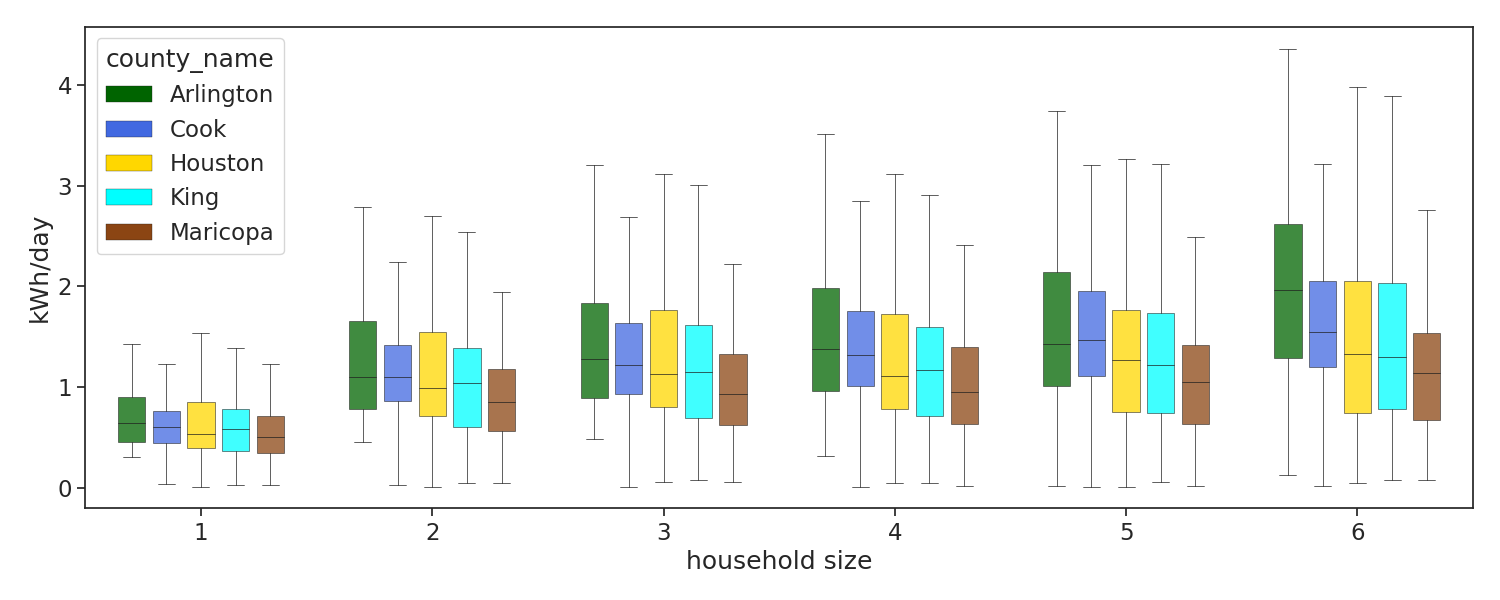}
    \caption{Lighting vs. household size}
      \label{fig:}
    \end{subfigure}
    \caption{\textbf{(a) Synthetic HVAC use and house area (i.e. floor area).} Boxplot comparing daily HVAC consumption in a winter day for the selected target locations by house area (i.e. floor area). The x-axis groups floor area of houses in five bins denoted in two units sq. ft ($\text{ft}^2$)  and sq m ($\text{m}^2$). The bins are as follows : $\leq$ 1000 $\text{ft}^2$, 1000 - 1500 $\text{ft}^2$, 1500 - 2000 $\text{ft}^2$, 2000 - 3000 $\text{ft}^2$, $\geq$ 3000 $\text{ft}^2$. It is observed that as floor area of the house increases HVAC consumption increases in all regions. Winter temperatures are relatively moderate in AZ and TX, thus, the HVAC consumption is less as compared to other regions. \textbf{(b) Synthetic lighting use and household size.} Lighting consumption increases as household size increases. Household size indicates number of members in a household.
    }
    \label{fig:hvac-sqft-light-hhsize}
\end{figure*}

Figure~\ref{fig:regional-linechart-major}(a) shows that the HVAC consumption varies significantly throughout the year. HVAC use is higher in hot-dry areas in summer as compared to other regions possibly due to higher temperatures. 
Structural characteristics such as dwelling size (square footage), insulation quality, age and efficiency of HVAC equipment also affect household HVAC consumption. 
Another important variable that drives HVAC consumption is indoor thermostat behavior which is related to household occupants' behavior/actions. In this work, indoor thermostat temperatures are set constant throughout the day. 
Insulation quality is not monitored in households (due to lack of data). We assume that the dwelling is well-insulated and the insulation values are implemented according to the DOE standards for the respective climate zones. 
In Figure~\ref{fig:hvac-sqft-light-hhsize}a we show effect of square footage (conditioned space) of a dwelling on hvac energy use.
In general, we observe that as the conditioned space in the dwelling increases, the HVAC consumption increases. 

\begin{figure*}[!h]    
    \begin{subfigure}{.49\textwidth}
    \centering
    \includegraphics[width=8.6cm,height=3.7cm]{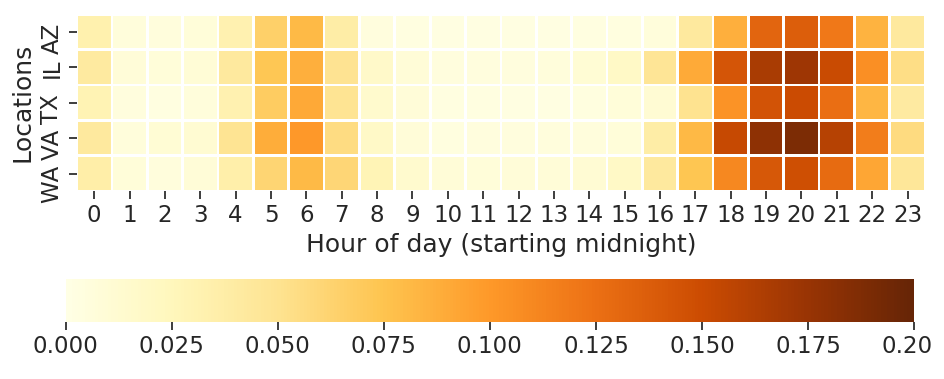}
    \caption{Lighting profile}
    \end{subfigure}
    \begin{subfigure}{.49\textwidth}
    \centering
    \includegraphics[width=8.6cm,height=3.7cm]{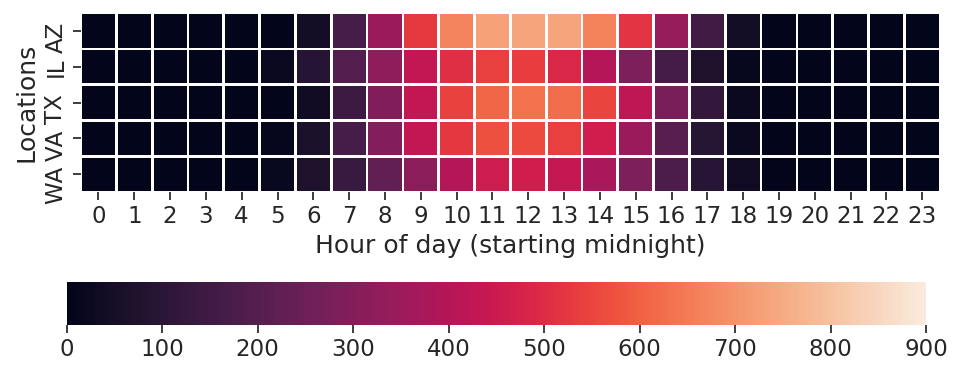}
    \caption{Irradiance profile}
    \end{subfigure}
    
    \begin{subfigure}{.49\textwidth}
    \centering
    \includegraphics[width=8.5cm,height=4.8cm]{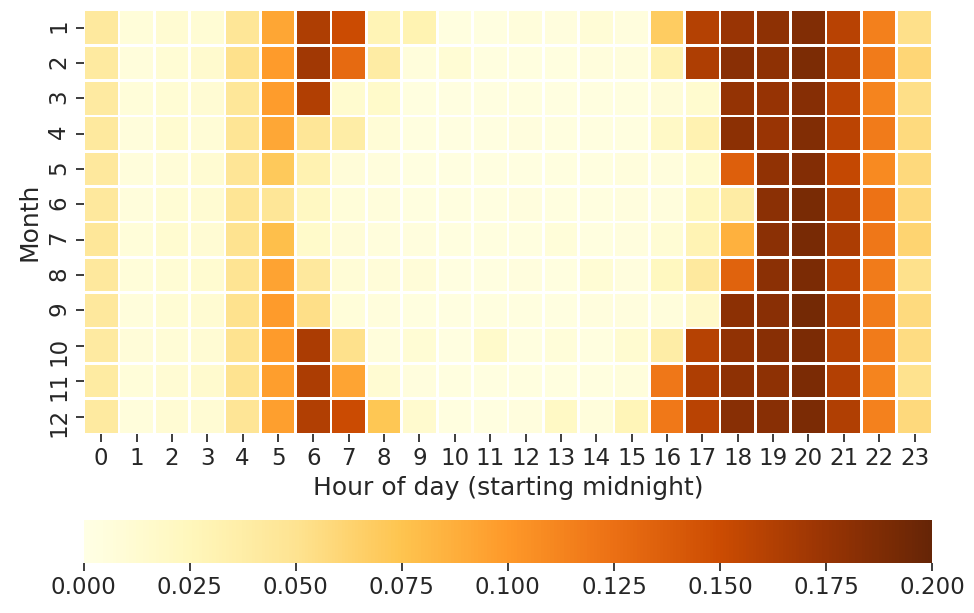}
    \caption{VA lighting profile}
    \end{subfigure}
    \begin{subfigure}{.49\textwidth}
    \centering
    \includegraphics[width=8.5cm,height=4.8cm]{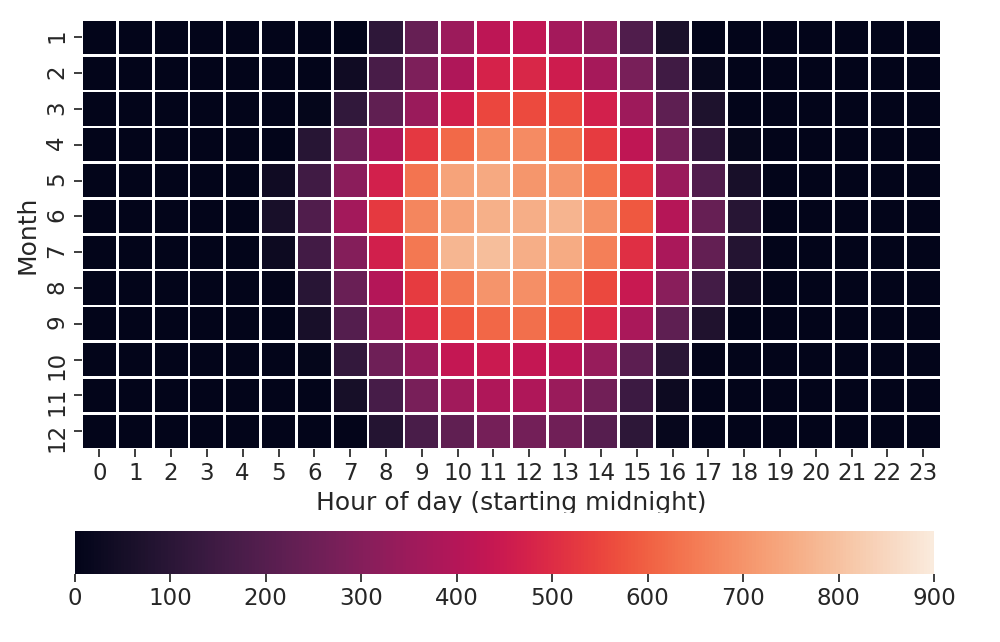}
    \caption{VA irradiance profile}
    \end{subfigure}
    
    \caption{\textbf{Heatmap depicting relation between hourly synthetic lighting usage and hourly irradiance.}  (a) shows average annual 24-hour lighting profiles of representative target locations. (b) shows average annual 24-hour irradiance profile of representative target locations. (c) and (d) present the variation in lighting usage and corresponding irradiance profiles at monthly level for Arlington, VA. (c) presents lighting consumption variation throughout the day in different months across the year. (d) shows variation in monthly irradiance profile. The units of measurements for energy usage is kWh and irradiance is Watts/m$^2$.  The lighting energy use is inversely proportional to the irradiance. The energy usage is higher in evening and night hours when the occupant is active in the dwelling. The average lighting and irradiance profiles show regional differences in irradiance availability and subsequent lighting energy usage. The VA profiles show that the day light is available for longer durations leading to lower lighting energy consumption as compared to winter.  }
    \label{fig:regional-lighting-heatmap}
\end{figure*}

Lighting energy-use varies by seasons in all regions as irradiance levels change with weather events and seasons.
Figure~\ref{fig:regional-lighting-heatmap}b shows average irradiance time series for the target locations. The corresponding lighting usage is shown in Figure~\ref{fig:regional-lighting-heatmap}a.
As an example, we look at monthly irrandiance profiles across 24 hours in Virginia for the year 2014~(Figure~\ref{fig:regional-lighting-heatmap}d). The corresponding monthly lighting energy use time series is shown in Figure~\ref{fig:regional-lighting-heatmap}c.
Example of lighting consumption w.r.t. household size is explored in Figure~\ref{fig:hvac-sqft-light-hhsize}b.

Figure~\ref{fig:regional-linechart-activities} shows the breakdown of appliance usage for different appliances and electronic devices.
Both figures show a line chart indicating average daily consumption for the month. 
The scatter plot in the background describes average daily consumption for an end-use for sampled days color coded by location, where 
the size of the markers denotes the standard deviation of the end-use consumption.
It is observed that appliance usage in activities such as cooking, dishwashing, performing laundry, watching TV, using computer, and cleaning are fairly similar in different regions.
The above comment is intuitively true since appliance use duration and their ratings may not vary across regions.
However, the occurrence timing throughout the day may vary from house to house depending upon occupant schedules irrespective of which geographic regions they belong to.

\begin{figure*}[!h]    
    \begin{subfigure}{.49\textwidth}
    \centering
    \includegraphics[width=8.5cm,height=3.8cm]{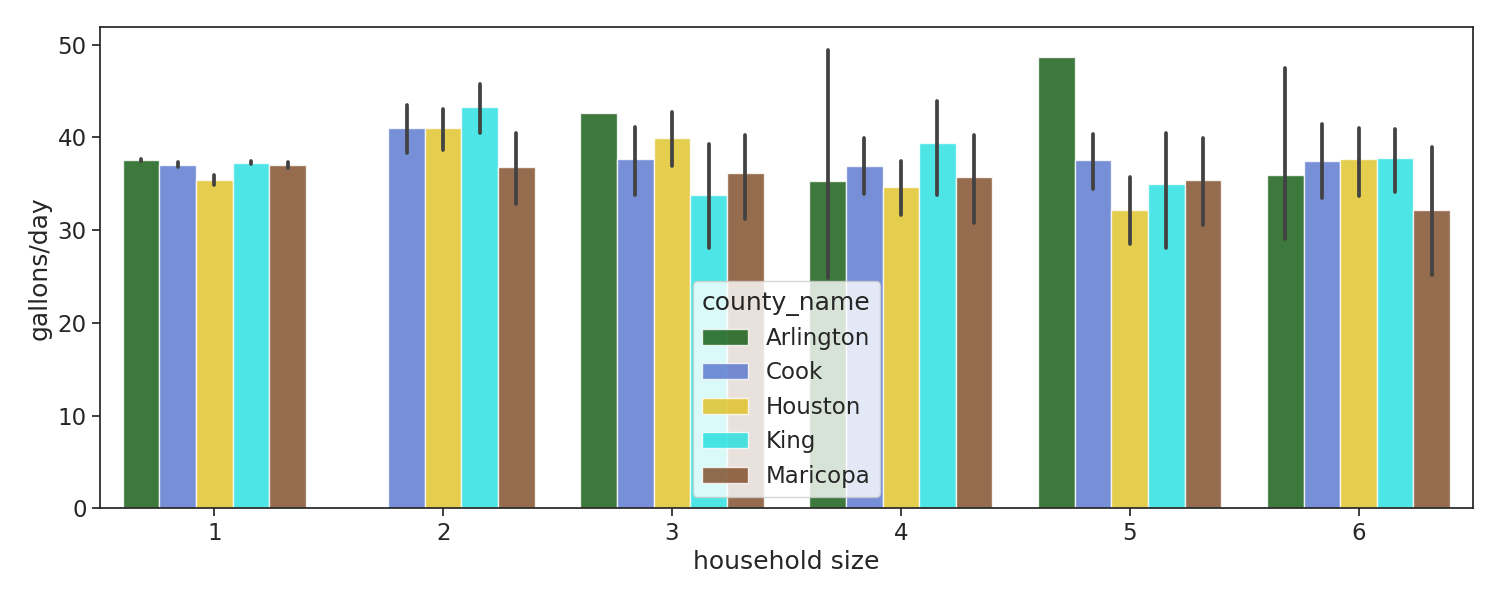}
    \caption{Hot water (gallons) vs. household size}
      \label{fig:}
    \end{subfigure}
    \begin{subfigure}{.49\textwidth}
    \centering
    \includegraphics[width=8.5cm,height=3.8cm]{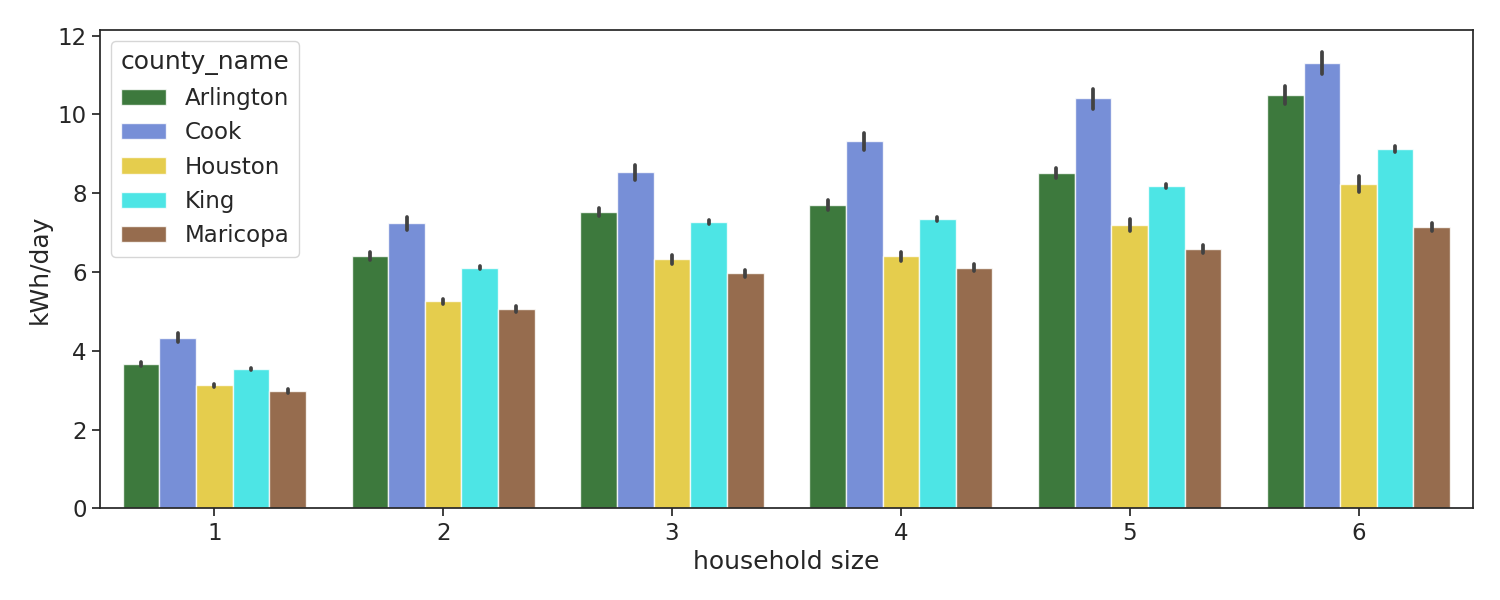}
    \caption{Hot water (energy) vs. household size}
      \label{fig:}
    \end{subfigure}
    \caption{\textbf{Synthetic hot water usage and energy vs. synthetic household size.} Household size indicates number of household members. The clustered bar charts show the amount of hot water consumed (in gallons in (a)) and corresponding energy usage in (b) according to household size in a winter day. The vertical black line on each bar shows the variation. Water usage and its variation increases with household size. The amount of energy for hot water end-use increases with household size and differs by region.} 
    \label{fig:regional-h2o-charts}
\end{figure*}

\section*{Usage Notes}

In order to analyze the dataset, researchers can use any programming languages such as Python, Java, Matlab, or R. As described in the `Data Records' section, the files are stored in csv format, so most of the file reading functions in the above languages can support reading/accessing the dataset. Next, we discuss the potential applications  of the released synthetic data.
We also highlight important challenges and limitations of this work.

\subsection*{Applicability and benefits of the dataset. }
We are releasing a  comprehensive household level dataset for energy use. 
In addition to the household level disaggregated energy use data, household composition is also included from census data.
This work was reviewed by the University of Virginia's Institutional Review Board (IRB) and was determined to be exempt from board IRB approval, as this research project did not involve human subject research.
The dataset can be effectively employed in various applications such as NILM (non-intrusive load monitoring), load profile analyses for observing similarities/differences between end use consumption of different regions and seasons, evaluating effects of retrofits in buildings, studying effects of temperature rise in different regions, and so on.
In addition, this data can also be used for energy model calibration, occupant behavior evaluation, implementing demand response strategies and policy interventions.
The dataset can be especially leveraged in training deep learning models where massive amount data is appreciated. Such models can be used for real-time residential demand forecasting. 
The dataset released are essentially time-series along with categorical and numerical attributes. Thus, any statistical tool or programming language can be used to analyze them.
Study III in the `Technical Validation' study illustrates examples of the possible uses of the dataset.

\subsection*{Challenges and limitations. }

The use of synthetic residential energy demand data has its pros and cons. National scale hourly synthetic data can be used to carry out national and even potentially international policy analysis. The spatio-temporal variability allows one to access important emerging questions
related to energy equity, fairness and accessibility at a fine scale. A systems level approach can be taken to vexing questions outlined in the 2030 Intergovernmental Panel on Climate Change (IPCC) goals.
On the other hand, synthetic data sets have their limitations as well. For instance, the fine-scale variability (minutes level as well as weekly variation) of usage  amongst households cannot be
captured easily in such synthetic data sets.
Additionally, the behavior exhibited by any single synthetic family might be biased by the data used for synthesis. Thus, any insight generated from high resolution analyses should be considered carefully.

An important challenge in developing the realistic synthetic residential load profiles at a national scale and at a high spatio-temporal resolution is to find appropriate datasets for representing different types of climates, demographics, appliances, and activity patterns. 
Accessibility and availability of all the above information from legitimate sources is crucial to maintain trustworthiness in the resulting models. 
A robust and extensible infrastructure is developed to synthesize diverse data sources into detailed information structure at various spatial resolutions (e.g. combining household level data with climate zone related data such as insulation values). 
The infrastructure consists of methods to compose multiple models and data sets.
The overall time to generate the synthetic data was reduced by using high performance computing capabilities.

Some of the limitations of our work are discussed. 
The current synthetic data does not include power consumption by electric vehicles and energy generation via renewable generation (e.g. solar panel, wind).
The ATUS data is available for a normative day for individuals.
Thus, activity and appliance related demands are generated for a normative day with minor variations coming from the activity model.
Hence, our synthetic data might not be able to capture daily activity variation appropriately (e.g. as observed in real-time smart metering).
This can be challenging to work with especially when studying demand response scenarios.
The building envelop considered for a synthetic household is simplified due to lack of information needed to represent a large population group, thus limiting our ability to employ state-of-the-art and sophisticated building modeling techniques. (e.g. we use a simple HVAC physics based model to generate heating and cooling related energy demand).

\subsection*{Concluding remarks.}
The paper describes a bottom up approach to generate 
large-scale digital twin data of dis-aggregated residential energy use hourly timeseries for the residential sector at household resolution across the contiguous United States for millions of households. 
The approach integrates diverse open-source surveys and datasets, where the end-use models are developed by either extending well-established methods or by building new models.  
Extensive validation of the synthetic datasets is conducted using real/recorded energy-use data across spatial and temporal resolutions.

\section*{Code availability}

Programming languages such as Python~3 and Java~8 are used for modeling, analyzing, and developing the framework.
The code is deposited in the repository~\cite{thorve_scidata2022_doi} alongwith the dataset.

\bibliography{main}

\begin{thebibliography}{100}
\urlstyle{rm}
\expandafter\ifx\csname url\endcsname\relax
  \def\url#1{\texttt{#1}}\fi
\expandafter\ifx\csname urlprefix\endcsname\relax\def\urlprefix{URL }\fi
\expandafter\ifx\csname doiprefix\endcsname\relax\def\doiprefix{DOI: }\fi
\providecommand{\bibinfo}[2]{#2}
\providecommand{\eprint}[2][]{\url{#2}}

\bibitem{Hart2008}
\bibinfo{author}{Hart, D.~G.}
\newblock \bibinfo{journal}{\bibinfo{title}{{Using AMI to realize the Smart
  Grid}}}.
\newblock {\emph{\JournalTitle{{2008 IEEE Power and Energy Society General
  Meeting - Conversion and Delivery of Electrical Energy in the 21st
  Century}}}} \bibinfo{pages}{1--2}, \url{10.1109/PES.2008.4596961}
  (\bibinfo{year}{2008}).

\bibitem{Mohassel2014}
\bibinfo{author}{Mohassel, R.~R.}, \bibinfo{author}{Fung, A.~S.},
  \bibinfo{author}{Mohammadi, F.} \& \bibinfo{author}{Raahemifar, K.}
\newblock \bibinfo{journal}{\bibinfo{title}{{A survey on advanced metering
  infrastructure and its application in Smart Grids}}}.
\newblock {\emph{\JournalTitle{2014 IEEE 27th Canadian Conference on Electrical
  and Computer Engineering (CCECE)}}} \bibinfo{pages}{1--8},
  \url{10.1109/CCECE.2014.6901102} (\bibinfo{year}{2014}).

\bibitem{hailegiorgis2018}
\bibinfo{author}{Hailegiorgis, A.}, \bibinfo{author}{Crooks, A.} \&
  \bibinfo{author}{Cioffi-Revilla, C.}
\newblock \bibinfo{journal}{\bibinfo{title}{An agent-based model of rural
  households\& adaptation to climate change}}.
\newblock {\emph{\JournalTitle{Journal of Artificial Societies and Social
  Simulation}}} \textbf{\bibinfo{volume}{21}}, \bibinfo{pages}{4},
  \url{10.18564/jasss.3812} (\bibinfo{year}{2018}).

\bibitem{Auffhammer2017}
\bibinfo{author}{Auffhammer, M.}, \bibinfo{author}{Baylis, P.} \&
  \bibinfo{author}{Hausman, C.~H.}
\newblock \bibinfo{journal}{\bibinfo{title}{Climate change is projected to have
  severe impacts on the frequency and intensity of peak electricity demand
  across the united states}}.
\newblock {\emph{\JournalTitle{Proceedings of the National Academy of
  Sciences}}} \textbf{\bibinfo{volume}{114}}, \bibinfo{pages}{1886--1891},
  \url{10.1073/pnas.1613193114} (\bibinfo{year}{2017}).
\newblock \eprint{https://www.pnas.org/doi/pdf/10.1073/pnas.1613193114}.

\bibitem{Rai2021}
\bibinfo{author}{Busby, J.~W.} \emph{et~al.}
\newblock \bibinfo{journal}{\bibinfo{title}{{Cascading risks: Understanding the
  2021 winter blackout in Texas}}}.
\newblock {\emph{\JournalTitle{Energy Research \& Social Science}}}
  \textbf{\bibinfo{volume}{77}}, \bibinfo{pages}{102106},
  \url{https://doi.org/10.1016/j.erss.2021.102106} (\bibinfo{year}{2021}).

\bibitem{Petri2015}
\bibinfo{author}{Petri, Y.} \& \bibinfo{author}{Caldeira, K.}
\newblock \bibinfo{journal}{\bibinfo{title}{Impacts of global warming on
  residential heating and cooling degree-days in the united states}}.
\newblock {\emph{\JournalTitle{Scientific Reports}}}
  \textbf{\bibinfo{volume}{5}}, \bibinfo{pages}{12427}, \url{10.1038/srep12427}
  (\bibinfo{year}{2015}).

\bibitem{Goldstein19122}
\bibinfo{author}{Goldstein, B.}, \bibinfo{author}{Gounaridis, D.} \&
  \bibinfo{author}{Newell, J.~P.}
\newblock \bibinfo{journal}{\bibinfo{title}{{The carbon footprint of household
  energy use in the United States}}}.
\newblock {\emph{\JournalTitle{Proceedings of the National Academy of
  Sciences}}} \textbf{\bibinfo{volume}{117}}, \bibinfo{pages}{19122--19130},
  \url{10.1073/pnas.1922205117} (\bibinfo{year}{2020}).
\newblock \eprint{https://www.pnas.org/content/117/32/19122.full.pdf}.

\bibitem{NAP2021}
\bibinfo{author}{{National Academies of Sciences, Engineering, and Medicine}}.
\newblock \emph{\bibinfo{title}{{Accelerating Decarbonization of the U.S.
  Energy System}}} (\bibinfo{publisher}{The National Academies Press},
  \bibinfo{address}{Washington, DC}, \bibinfo{year}{2021}).

\bibitem{Gillingham2021}
\bibinfo{author}{Gillingham, K.~T.}, \bibinfo{author}{Huang, P.},
  \bibinfo{author}{Buehler, C.}, \bibinfo{author}{Peccia, J.} \&
  \bibinfo{author}{Gentner, D.~R.}
\newblock \bibinfo{journal}{\bibinfo{title}{The climate and health benefits
  from intensive building energy efficiency improvements}}.
\newblock {\emph{\JournalTitle{Science Advances}}}
  \textbf{\bibinfo{volume}{7}}, \bibinfo{pages}{eabg0947},
  \url{10.1126/sciadv.abg0947} (\bibinfo{year}{2021}).
\newblock \eprint{https://www.science.org/doi/pdf/10.1126/sciadv.abg0947}.

\bibitem{Berrill2021_iop}
\bibinfo{author}{Berrill, P.}, \bibinfo{author}{Gillingham, K.~T.} \&
  \bibinfo{author}{Hertwich, E.~G.}
\newblock \bibinfo{journal}{\bibinfo{title}{Drivers of change in {US}
  residential energy consumption and greenhouse gas emissions,
  1990{\textendash}2015}}.
\newblock {\emph{\JournalTitle{Environmental Research Letters}}}
  \textbf{\bibinfo{volume}{16}}, \bibinfo{pages}{034045},
  \url{10.1088/1748-9326/abe325} (\bibinfo{year}{2021}).

\bibitem{Berrill2021_acs}
\bibinfo{author}{Berrill, P.}, \bibinfo{author}{Gillingham, K.~T.} \&
  \bibinfo{author}{Hertwich, E.~G.}
\newblock \bibinfo{journal}{\bibinfo{title}{Linking housing policy, housing
  typology, and residential energy demand in the united states}}.
\newblock {\emph{\JournalTitle{Environmental Science {\&} Technology}}}
  \textbf{\bibinfo{volume}{55}}, \bibinfo{pages}{2224--2233},
  \url{10.1021/acs.est.0c05696} (\bibinfo{year}{2021}).

\bibitem{MIT2011}
\bibinfo{author}{Kassakian, J.} \emph{et~al.}
\newblock \bibinfo{journal}{\bibinfo{title}{{The Future of the Electric Grid:
  An Interdisciplinary MIT Study}}}.
\newblock {\emph{\JournalTitle{Massachusetts Institute of Technology, MIT
  Energy Initiative}}}  (\bibinfo{year}{2011}).

\bibitem{DEB2021116990}
\bibinfo{author}{Deb, C.}, \bibinfo{author}{Dai, Z.} \&
  \bibinfo{author}{Schlueter, A.}
\newblock \bibinfo{journal}{\bibinfo{title}{{A machine learning-based framework
  for cost-optimal building retrofit}}}.
\newblock {\emph{\JournalTitle{Applied Energy}}}
  \textbf{\bibinfo{volume}{294}}, \bibinfo{pages}{116990},
  \url{https://doi.org/10.1016/j.apenergy.2021.116990} (\bibinfo{year}{2021}).

\bibitem{Nutkiewicz2021}
\bibinfo{author}{Nutkiewicz, A.}, \bibinfo{author}{Choi, B.} \&
  \bibinfo{author}{Jain, R.~K.}
\newblock \bibinfo{journal}{\bibinfo{title}{Exploring the influence of urban
  context on building energy retrofit performance: A hybrid simulation and
  data-driven approach}}.
\newblock {\emph{\JournalTitle{Advances in Applied Energy}}}
  \textbf{\bibinfo{volume}{3}}, \bibinfo{pages}{100038},
  \url{https://doi.org/10.1016/j.adapen.2021.100038} (\bibinfo{year}{2021}).

\bibitem{Muratori2018}
\bibinfo{author}{Muratori, M.}
\newblock \bibinfo{journal}{\bibinfo{title}{Impact of uncoordinated plug-in
  electric vehicle charging on residential power demand}}.
\newblock {\emph{\JournalTitle{Nature Energy}}} \textbf{\bibinfo{volume}{3}},
  \bibinfo{pages}{193--201}, \url{10.1038/s41560-017-0074-z}
  (\bibinfo{year}{2018}).

\bibitem{Mahdavi2021}
\bibinfo{author}{Mahdavi, A.} \emph{et~al.}
\newblock \bibinfo{journal}{\bibinfo{title}{{The Role of Occupants in
  Buildings’ Energy Performance Gap: Myth or Reality?}}}
\newblock {\emph{\JournalTitle{Sustainability}}} \textbf{\bibinfo{volume}{13}},
  \url{10.3390/su13063146} (\bibinfo{year}{2021}).

\bibitem{Tanaka2021}
\bibinfo{author}{Tanaka, K.}, \bibinfo{author}{Wilson, C.} \&
  \bibinfo{author}{Managi, S.}
\newblock \bibinfo{journal}{\bibinfo{title}{Impact of feed-in tariffs on
  electricity consumption}}.
\newblock {\emph{\JournalTitle{Environmental Economics and Policy Studies}}}
  \url{10.1007/s10018-021-00306-w} (\bibinfo{year}{2021}).

\bibitem{Tsaousoglou2019}
\bibinfo{author}{Tsaousoglou, G.}, \bibinfo{author}{Efthymiopoulos, N.},
  \bibinfo{author}{Makris, P.} \& \bibinfo{author}{Varvarigos, E.}
\newblock \bibinfo{journal}{\bibinfo{title}{{Personalized real time pricing for
  efficient and fair demand response in energy cooperatives and highly
  competitive flexibility markets}}}.
\newblock {\emph{\JournalTitle{Journal of Modern Power Systems and Clean
  Energy}}} \textbf{\bibinfo{volume}{7}}, \bibinfo{pages}{151--162},
  \url{10.1007/s40565-018-0426-0} (\bibinfo{year}{2019}).

\bibitem{nas2016}
\bibinfo{author}{{National Academies of Sciences, Engineering, and Medicine}}.
\newblock \emph{\bibinfo{title}{{Analytic Research Foundations for the
  Next-Generation Electric Grid.}}} (\bibinfo{publisher}{The National Academies
  Press. Washington, DC.}, \bibinfo{year}{2016}).

\bibitem{Klemenjak2020}
\bibinfo{author}{Klemenjak, C.}, \bibinfo{author}{Kovatsch, C.},
  \bibinfo{author}{Herold, M.} \& \bibinfo{author}{Elmenreich, W.}
\newblock \bibinfo{journal}{\bibinfo{title}{{A synthetic energy dataset for
  non-intrusive load monitoring in households}}}.
\newblock {\emph{\JournalTitle{Scientific Data}}} \textbf{\bibinfo{volume}{7}},
  \url{10.1038/s41597-020-0434-6} (\bibinfo{year}{2020}).

\bibitem{SynD_DOI}
\bibinfo{author}{Klemenjak, C.}, \bibinfo{author}{Kovatsch, C.},
  \bibinfo{author}{Herold, M.} \& \bibinfo{author}{Elmenreich, W.}
\newblock \bibinfo{journal}{\bibinfo{title}{{SynD: A Synthetic Energy Dataset
  for Non-Intrusive Load Monitoring in Households.}}}
\newblock {\emph{\JournalTitle{figshare}}}
  \url{https://doi.org/10.6084/m9.figshare.c.4716179} (\bibinfo{year}{2020}).

\bibitem{Kolter11redd}
\bibinfo{author}{Kolter, J.~Z.} \& \bibinfo{author}{Johnson, M.~J.}
\newblock \bibinfo{journal}{\bibinfo{title}{{REDD: A Public Data Set for Energy
  Disaggregation Research}}}.
\newblock {\emph{\JournalTitle{SustKDD workshop on Data Mining Applications in
  Sustainability}}}  (\bibinfo{year}{2011}).

\bibitem{REDD_DOI}
\bibinfo{author}{Kolter, J.~Z.} \& \bibinfo{author}{Johnson, M.~J.}
\newblock \bibinfo{journal}{\bibinfo{title}{{REDD: The Reference Energy
  Disaggregation Data Set}}}.
\newblock {\emph{\JournalTitle{MIT Initial REDD Release, Version 1.0}}}
  \url{http://redd.csail.mit.edu/} (\bibinfo{year}{2011}).

\bibitem{RAE-Makonin-17}
\bibinfo{author}{Makonin, S.}, \bibinfo{author}{Wang, Z.~J.} \&
  \bibinfo{author}{Tumpach, C.}
\newblock \bibinfo{journal}{\bibinfo{title}{{RAE:} the rainforest automation
  energy dataset for smart grid meter data analysis}}.
\newblock {\emph{\JournalTitle{CoRR}}}
  \textbf{\bibinfo{volume}{abs/1705.05767}},
  \url{http://arxiv.org/abs/1705.05767} (\bibinfo{year}{2017}).

\bibitem{Murray2017}
\bibinfo{author}{Murray, D.}, \bibinfo{author}{Stankovic, L.} \&
  \bibinfo{author}{Stankovic, V.}
\newblock \bibinfo{journal}{\bibinfo{title}{{An electrical load measurements
  dataset of United Kingdom households from a two-year longitudinal study}}}.
\newblock {\emph{\JournalTitle{Scientific Data}}} \textbf{\bibinfo{volume}{4}},
  \url{10.1038/sdata.2016.122} (\bibinfo{year}{2017}).

\bibitem{Murray2017_doi}
\bibinfo{author}{Murray, D.}, \bibinfo{author}{Stankovic, L.} \&
  \bibinfo{author}{Stankovic, V.}
\newblock \bibinfo{journal}{\bibinfo{title}{{REFIT: Electrical Load
  Measurements (Cleaned)}}}.
\newblock {\emph{\JournalTitle{University of Strathclyde, PURE}}}
  \url{http://dx.doi.org/10.15129/9ab14b0e-19ac-4279-938f-27f643078cec}
  (\bibinfo{year}{2015}).

\bibitem{pecanstreet}
\bibinfo{author}{Nagasawa, K.} \emph{et~al.}
\newblock \bibinfo{journal}{\bibinfo{title}{{Data Management for a Large-Scale
  Smart Grid Demonstration Project in Austin, Texas}}}.
\newblock {\emph{\JournalTitle{ASME 2012 6th International Conference on Energy
  Sustainability}}}  (\bibinfo{year}{2013}).

\bibitem{pecan_doi}
\bibinfo{author}{Webber, M.}
\newblock \bibinfo{journal}{\bibinfo{title}{{Pecan Street Dataport}}}.
\newblock {\emph{\JournalTitle{Pecan Street Inc.}}}
  \url{https://www.pecanstreet.org/dataport/} (\bibinfo{year}{2013}).

\bibitem{Rashid2019}
\bibinfo{author}{Rashid, H.}, \bibinfo{author}{Singh, P.} \&
  \bibinfo{author}{Singh, A.}
\newblock \bibinfo{journal}{\bibinfo{title}{{I-BLEND, a campus-scale commercial
  and residential buildings electrical energy dataset}}}.
\newblock {\emph{\JournalTitle{Scientific Data}}} \textbf{\bibinfo{volume}{6}},
  \url{10.1038/sdata.2019.15} (\bibinfo{year}{2019}).

\bibitem{iblend_doi}
\bibinfo{author}{Rashid, H.}, \bibinfo{author}{Singh, P.} \&
  \bibinfo{author}{Singh, A.}
\newblock \bibinfo{journal}{\bibinfo{title}{{I-BLEND, a campus-scale commercial
  and residential buildings electrical energy dataset}}}.
\newblock {\emph{\JournalTitle{figshare}}}
  \url{https://doi.org/10.6084/m9.figshare.c.3893581} (\bibinfo{year}{2019}).

\bibitem{Paige2019}
\bibinfo{author}{Paige, F.}, \bibinfo{author}{Agee, P.} \&
  \bibinfo{author}{Jazizadeh, F.}
\newblock \bibinfo{journal}{\bibinfo{title}{{flEECe, an energy use and occupant
  behavior dataset for net-zero energy affordable senior residential
  buildings}}}.
\newblock {\emph{\JournalTitle{Scientific Data}}} \textbf{\bibinfo{volume}{6}},
  \url{10.1038/s41597-019-0275-3} (\bibinfo{year}{2019}).

\bibitem{flEECe_doi}
\bibinfo{author}{Paige, F.} \& \bibinfo{author}{Agee, P.}
\newblock \bibinfo{journal}{\bibinfo{title}{{flEECe, an Energy Use and Occupant
  Behavior Dataset for Net Zero Energy Affordable Senior Residential
  Buildings.}}}
\newblock {\emph{\JournalTitle{Open Science Framework}}}
  \url{https://doi.org/10.17605/OSF.IO/2AX9D} (\bibinfo{year}{2019}).

\bibitem{Shin2019}
\bibinfo{author}{Shin, C.} \emph{et~al.}
\newblock \bibinfo{journal}{\bibinfo{title}{{The ENERTALK dataset, 15 Hz
  electricity consumption data from 22 houses in Korea}}}.
\newblock {\emph{\JournalTitle{Scientific Data}}} \textbf{\bibinfo{volume}{6}},
  \url{10.1038/s41597-019-0212-5} (\bibinfo{year}{2019}).

\bibitem{enertalk_doi}
\bibinfo{author}{Shin, C.} \emph{et~al.}
\newblock \bibinfo{journal}{\bibinfo{title}{{The ENERTALK dataset, 15 Hz
  electricity consumption data from 22 houses in Korea}}}.
\newblock {\emph{\JournalTitle{figshare}}}
  \url{https://doi.org/10.6084/m9.figshare.c.4502780} (\bibinfo{year}{2019}).

\bibitem{UK-DALE}
\bibinfo{author}{Kelly, J.} \& \bibinfo{author}{Knottenbelt, W.}
\newblock \bibinfo{journal}{\bibinfo{title}{{The {UK-DALE} dataset, domestic
  appliance-level electricity demand and whole-house demand from five {UK}
  homes}}}.
\newblock {\emph{\JournalTitle{Scientific Data}}} \textbf{\bibinfo{volume}{2}},
  \url{10.1038/sdata.2015.7} (\bibinfo{year}{2015}).

\bibitem{ukdale_doi}
\bibinfo{author}{Kelly, J.} \& \bibinfo{author}{Knottenbelt, W.}
\newblock \bibinfo{journal}{\bibinfo{title}{The {UK-DALE} dataset}}.
\newblock {\emph{\JournalTitle{{UKERC Energy Data Centre}}}}
  \url{https://doi.org/10.5286/UKERC.EDC.000002} (\bibinfo{year}{2015}).

\bibitem{Anderson2012BLUED}
\bibinfo{author}{Anderson, K.}, \bibinfo{author}{Ocneanu, A.},
  \bibinfo{author}{Carlson, D.~R.}, \bibinfo{author}{Rowe, A.~G.} \&
  \bibinfo{author}{Berg{\'e}s, M.}
\newblock \bibinfo{journal}{\bibinfo{title}{{BLUED : A Fully Labeled Public
  Dataset for Event-Based Non-Intrusive Load Monitoring Research}}}.
\newblock {\emph{\JournalTitle{Proceedings of the 2nd KDD Workshop on Data
  Mining Applications in Sustainability}}}  (\bibinfo{year}{2012}).

\bibitem{blued_doi}
\bibinfo{author}{Anderson, K.}
\newblock \bibinfo{journal}{\bibinfo{title}{{Dataset Name: Building-Level fUlly
  labeled Electricity Disaggregation dataset (BLUED)}}}.
\newblock {\emph{\JournalTitle{{github}}}}
  \url{https://tokhub.github.io/dbecd/links/Blued.html} (\bibinfo{year}{2011}).

\bibitem{Barker12anopen}
\bibinfo{author}{Barker, S.} \emph{et~al.}
\newblock \bibinfo{journal}{\bibinfo{title}{{An Open Data Set and Tools for
  Enabling Research in Sustainable Homes}}}.
\newblock {\emph{\JournalTitle{{Proceedings of the 1st KDD Workshop on Data
  Mining Applications in Sustainability (SustKDD)}}}}  (\bibinfo{year}{2012}).

\bibitem{smartstar_doi}
\bibinfo{author}{Barker, S.}
\newblock \bibinfo{journal}{\bibinfo{title}{{UMass Smart* Dataset - 2017
  release}}}.
\newblock {\emph{\JournalTitle{{UMassTraceRepository}}}}
  \url{https://traces.cs.umass.edu/index.php/smart/smart}
  (\bibinfo{year}{2017}).

\bibitem{Beckel2014_ECO}
\bibinfo{author}{Beckel, C.}, \bibinfo{author}{Kleiminger, W.},
  \bibinfo{author}{Cicchetti, R.}, \bibinfo{author}{Staake, T.} \&
  \bibinfo{author}{Santini, S.}
\newblock \bibinfo{journal}{\bibinfo{title}{{The ECO Data Set and the
  Performance of Non-Intrusive Load Monitoring Algorithms}}}.
\newblock {\emph{\JournalTitle{Proceedings of the 1st ACM Conference on
  Embedded Systems for Energy-Efficient Buildings}}} \bibinfo{pages}{80–89},
  \url{10.1145/2674061.2674064} (\bibinfo{year}{2014}).

\bibitem{Pereira2014SustDataAP}
\bibinfo{author}{Pereira, L.}, \bibinfo{author}{Quintal, F.},
  \bibinfo{author}{Gonçalves, R.} \& \bibinfo{author}{Nunes, N.}
\newblock \bibinfo{journal}{\bibinfo{title}{{SustData: A Public Dataset for
  ICT4S Electric Energy Research}}}.
\newblock {\emph{\JournalTitle{ICT4S}}}  (\bibinfo{year}{2014}).

\bibitem{SustData_doi}
\bibinfo{author}{Pereira, L.}
\newblock \bibinfo{journal}{\bibinfo{title}{{SustData: A Public Dataset for
  ICT4S Electric Energy Research}}}.
\newblock {\emph{\JournalTitle{Open Science Framework}}}
  \url{https://osf.io/2ac8q/} (\bibinfo{year}{2021}).

\bibitem{Pereira2022}
\bibinfo{author}{Pereira, L.}, \bibinfo{author}{Costa, D.} \&
  \bibinfo{author}{Ribeiro, M.}
\newblock \bibinfo{journal}{\bibinfo{title}{A residential labeled dataset for
  smart meter data analytics}}.
\newblock {\emph{\JournalTitle{Scientific Data}}} \textbf{\bibinfo{volume}{9}},
  \bibinfo{pages}{134}, \url{10.1038/s41597-022-01252-2}
  (\bibinfo{year}{2022}).

\bibitem{GRREND}
\bibinfo{author}{Monacchi, A.}, \bibinfo{author}{Egarter, D.},
  \bibinfo{author}{Elmenreich, W.}, \bibinfo{author}{D'Alessandro, S.} \&
  \bibinfo{author}{Tonello, A.~M.}
\newblock \bibinfo{journal}{\bibinfo{title}{{GREEND: An energy consumption
  dataset of households in Italy and Austria}}}.
\newblock {\emph{\JournalTitle{2014 IEEE International Conference on Smart Grid
  Communications (SmartGridComm)}}} \bibinfo{pages}{511--516},
  \url{10.1109/SmartGridComm.2014.7007698} (\bibinfo{year}{2014}).

\bibitem{greend_doi}
\bibinfo{author}{Monacchi, A.}, \bibinfo{author}{Egarter, D.},
  \bibinfo{author}{Elmenreich, W.}, \bibinfo{author}{D'Alessandro, S.} \&
  \bibinfo{author}{Tonello, A.~M.}
\newblock \bibinfo{journal}{\bibinfo{title}{{GREEND: An energy consumption
  dataset of households in Italy and Austria}}}.
\newblock {\emph{\JournalTitle{Duke Energy Initiative Lakeside Labs}}}
  \url{https://energy.duke.edu/content/greend-electrical-energy-dataset}
  (\bibinfo{year}{2021}).

\bibitem{Pullinger2021}
\bibinfo{author}{Pullinger, M.} \emph{et~al.}
\newblock \bibinfo{journal}{\bibinfo{title}{{The IDEAL household energy
  dataset, electricity, gas, contextual sensor data and survey data for 255 UK
  homes}}}.
\newblock {\emph{\JournalTitle{Scientific Data}}} \textbf{\bibinfo{volume}{8}},
  \bibinfo{pages}{146}, \url{10.1038/s41597-021-00921-y}
  (\bibinfo{year}{2021}).

\bibitem{ideal_doi}
\bibinfo{author}{Goddard, N.} \emph{et~al.}
\newblock \bibinfo{journal}{\bibinfo{title}{{The IDEAL Household Energy
  Dataset. }}}.
\newblock {\emph{\JournalTitle{Edinburgh DataShare}}}
  \url{https://doi.org/10.7488/ds/2836} (\bibinfo{year}{2021}).

\bibitem{Ruhnau2019}
\bibinfo{author}{Ruhnau, O.}, \bibinfo{author}{Hirth, L.} \&
  \bibinfo{author}{Praktiknjo, A.}
\newblock \bibinfo{journal}{\bibinfo{title}{{Time series of heat demand and
  heat pump efficiency for energy system modeling}}}.
\newblock {\emph{\JournalTitle{Scientific Data}}} \textbf{\bibinfo{volume}{6}},
  \url{10.1038/s41597-019-0199-y} (\bibinfo{year}{2019}).

\bibitem{Ruhnau_doi}
\bibinfo{author}{Ruhnau, O.}
\newblock \bibinfo{journal}{\bibinfo{title}{{When2Heat Heating Profiles}}}.
\newblock {\emph{\JournalTitle{Open Power System Data}}}
  \url{https://doi.org/10.25832/when2heat/2019-08-06} (\bibinfo{year}{2019}).

\bibitem{Kelly_2014_metadata}
\bibinfo{author}{Kelly, J.} \& \bibinfo{author}{Knottenbelt, W.}
\newblock \bibinfo{journal}{\bibinfo{title}{Metadata for energy
  disaggregation}}.
\newblock {\emph{\JournalTitle{2014 {IEEE} 38th International Computer Software
  and Applications Conference Workshops}}} \url{10.1109/compsacw.2014.97}
  (\bibinfo{year}{2014}).

\bibitem{Meyur2020}
\bibinfo{author}{Meyur, R.} \emph{et~al.}
\newblock \bibinfo{journal}{\bibinfo{title}{{Creating Realistic Power
  Distribution Networks using Interdependent Road Infrastructure}}}.
\newblock {\emph{\JournalTitle{2020 IEEE International Conference on Big Data
  (Big Data)}}} \bibinfo{pages}{1226--1235},
  \url{10.1109/BigData50022.2020.9377959} (\bibinfo{year}{2020}).

\bibitem{LiLBNL2021}
\bibinfo{author}{Li, H.}
\newblock \bibinfo{journal}{\bibinfo{title}{{AlphaBuilding Synthetic
  Dataset}}}.
\newblock {\emph{\JournalTitle{Lawrence Berkeley National Laboratory}}}
  (\bibinfo{year}{2021}).

\bibitem{Thorve2018}
\bibinfo{author}{Thorve, S.} \emph{et~al.}
\newblock \bibinfo{journal}{\bibinfo{title}{Simulating residential energy
  demand in urban and rural areas}}.
\newblock {\emph{\JournalTitle{Winter Simulation Conference}}}
  (\bibinfo{year}{2018}).

\bibitem{ROTH2020}
\bibinfo{author}{Roth, J.}, \bibinfo{author}{Martin, A.},
  \bibinfo{author}{Miller, C.} \& \bibinfo{author}{Jain, R.~K.}
\newblock \bibinfo{journal}{\bibinfo{title}{Syncity: Using open data to create
  a synthetic city of hourly building energy estimates by integrating
  data-driven and physics-based methods}}.
\newblock {\emph{\JournalTitle{Applied Energy}}}
  \textbf{\bibinfo{volume}{280}}, \bibinfo{pages}{115981},
  \url{https://doi.org/10.1016/j.apenergy.2020.115981} (\bibinfo{year}{2020}).

\bibitem{Tong2021}
\bibinfo{author}{Tong, K.}, \bibinfo{author}{Nagpure, A.} \&
  \bibinfo{author}{Ramaswami, A.}
\newblock \bibinfo{journal}{\bibinfo{title}{{All urban areas energy use data
  across 640 districts in India for the year 2011}}}.
\newblock {\emph{\JournalTitle{Scientific Data}}} \textbf{\bibinfo{volume}{8}},
  \url{https://doi.org/10.1038/s41597-021-00853-7} (\bibinfo{year}{2021}).

\bibitem{SPEW}
\bibinfo{author}{Bill, E.}, \bibinfo{author}{Shannon, G.},
  \bibinfo{author}{Lee, R.},  \& \bibinfo{author}{Sam, V.}
\newblock \bibinfo{title}{Synthetic populations and ecosystems of the world}.
\newblock \bibinfo{type}{Tech. Rep.}, \bibinfo{institution}{Department of
  Statistics, Carnegie Mellon University} (\bibinfo{year}{2017}).
\newblock \url{http://stat.cmu.edu/~spew/assets/spew_documentation.pdf}.

\bibitem{ATUS2015}
\bibinfo{author}{{ATUS Survey}}.
\newblock \bibinfo{title}{{U.S. Bureau of Labor Statistics: American Time Use
  Survey}}, \url{https://www.bls.gov/tus/datafiles_2015.htm}
  (\bibinfo{year}{2015}).
\newblock \bibinfo{note}{Accessed: Mar, 2018}.

\bibitem{Gallagher2018}
\bibinfo{author}{Gallagher, S.}, \bibinfo{author}{Richardson, L.~F.},
  \bibinfo{author}{Ventura, S.~L.} \& \bibinfo{author}{Eddy, W.~F.}
\newblock \bibinfo{journal}{\bibinfo{title}{{SPEW: Synthetic Populations and
  Ecosystems of the World}}}.
\newblock {\emph{\JournalTitle{{Journal of Computational and Graphical
  Statistics}}}} \textbf{\bibinfo{volume}{27}}, \bibinfo{pages}{773--784},
  \url{10.1080/10618600.2018.1442342} (\bibinfo{year}{2018}).

\bibitem{deming40ipf}
\bibinfo{author}{Deming, W.~E.} \& \bibinfo{author}{Stephan, F.~F.}
\newblock \bibinfo{journal}{\bibinfo{title}{{On a Least Squares Adjustment of a
  Sampled Frequency Table When the Expected Marginal Tables are Known}}}.
\newblock {\emph{\JournalTitle{Annals Math. Stats}}}
  \textbf{\bibinfo{volume}{11}}, \bibinfo{pages}{427--444}
  (\bibinfo{year}{1940}).

\bibitem{Fienberg}
\bibinfo{author}{Fienberg, S.~E.}
\newblock \bibinfo{journal}{\bibinfo{title}{{An Iterative Procedure for
  Estimation in Contingency Tables}}}.
\newblock {\emph{\JournalTitle{The Annals of Mathematical Statistics}}}
  \textbf{\bibinfo{volume}{41}}, \bibinfo{pages}{907--917}
  (\bibinfo{year}{1970}).

\bibitem{Pums2013}
\bibinfo{author}{{Public Use Microdata Sample (PUMS)}}.
\newblock \bibinfo{title}{{PUMS Documentation}},
  \url{https://www.census.gov/programs-surveys/acs/microdata/documentation.2013.html}
  (\bibinfo{year}{2013}).
\newblock \bibinfo{note}{Accessed: Nov, 2017}.

\bibitem{NLDAS}
\bibinfo{author}{{Land Data Assimilation System}}.
\newblock \bibinfo{title}{{North American Land Data Assimilation System (NLDAS)
  Climate Data}}, \url{https://ldas.gsfc.nasa.gov/nldas/}
  (\bibinfo{year}{2016}).
\newblock \bibinfo{note}{Accessed: Mar, 2018}.

\bibitem{RECS2015}
\bibinfo{author}{{United States Energy Information Administration}}.
\newblock \bibinfo{title}{{2015 RECS Survey Data}},
  \url{https://www.eia.gov/consumption/residential/data/2015/}
  (\bibinfo{year}{2015}).
\newblock \bibinfo{note}{Accessed: Nov, 2017}.

\bibitem{NSRDB2020}
\bibinfo{author}{{National Renewable Energy Laboratory (NREL)}}.
\newblock \bibinfo{title}{{National Solar Radiation Database (NSRDB)}},
  \url{https://nsrdb.nrel.gov/data-sets/us-data} (\bibinfo{year}{2014}).
\newblock \bibinfo{note}{Accessed: Nov, 2020}.

\bibitem{NIST_cook}
\bibinfo{author}{Nabinger, S.~J.}
\newblock \bibinfo{journal}{\bibinfo{title}{{Evaluation of Kitchen Cooking
  Appliance efficiency Test Procedures}}}.
\newblock {\emph{\JournalTitle{{National Institute of Standards and Technology,
  U.S. Department of Commerce}}}}  (\bibinfo{year}{1999}).

\bibitem{NIST_dw}
\bibinfo{author}{Castro, N.~S.}, \bibinfo{author}{Bowman, J.} \&
  \bibinfo{author}{Twigg, B.}
\newblock \bibinfo{journal}{\bibinfo{title}{{The New U.S. Department of Energy
  Dishwasher Test Procedure: Development and First Results}}}.
\newblock {\emph{\JournalTitle{{National Institute of Standards \&
  Technology}}}}  (\bibinfo{year}{2005}).

\bibitem{NRELhotwater129}
\bibinfo{author}{Jeff, M.}, \bibinfo{author}{Xia, F.} \& \bibinfo{author}{Eric,
  W.}
\newblock \bibinfo{journal}{\bibinfo{title}{{Comparison of Advanced Residential
  Water Heating Technologies in the United States}}}.
\newblock {\emph{\JournalTitle{{National Renewable Energy Laboratory Technical
  Reports}}}}  (\bibinfo{year}{2013}).

\bibitem{NRELhotwater61}
\bibinfo{author}{Wiehagen, J.} \& \bibinfo{author}{Sikora, J.}
\newblock \bibinfo{journal}{\bibinfo{title}{{Performance Comparison of
  Residential Hot Water Systems}}}.
\newblock {\emph{\JournalTitle{{National Renewable Energy Laboratory
  Reports}}}}  (\bibinfo{year}{2003}).

\bibitem{BECKMAN1996415}
\bibinfo{author}{Beckman, R.~J.}, \bibinfo{author}{Baggerly, K.~A.} \&
  \bibinfo{author}{McKay, M.~D.}
\newblock \bibinfo{journal}{\bibinfo{title}{Creating synthetic baseline
  populations}}.
\newblock {\emph{\JournalTitle{Transportation Research Part A:Policy and
  Practice}}} \textbf{\bibinfo{volume}{30(6)}}, \bibinfo{pages}{415--429},
  \url{https://doi.org/10.1016/0965-8564(96)00004-3} (\bibinfo{year}{1996}).

\bibitem{Lum2016}
\bibinfo{author}{Lum, K.}, \bibinfo{author}{Chungbaek, Y.},
  \bibinfo{author}{Eubank, S.} \& \bibinfo{author}{Marathe, M.}
\newblock \bibinfo{journal}{\bibinfo{title}{{A Two-stage, Fitted Values
  Approach to Activity Matching}}}.
\newblock {\emph{\JournalTitle{International Journal of Transportation}}}
  \textbf{\bibinfo{volume}{4}}, \bibinfo{pages}{41--56} (\bibinfo{year}{2016}).

\bibitem{ctreeR}
\bibinfo{author}{Torsten, H.}, \bibinfo{author}{Kurt, H.} \&
  \bibinfo{author}{Achim, Z.}
\newblock \emph{\bibinfo{title}{ctree: Conditional Inference Trees}}
  (\bibinfo{year}{2006}).
\newblock \bibinfo{note}{R package version 1.3-5}.

\bibitem{Hothorn2006ctree}
\bibinfo{author}{Hothorn, T.}, \bibinfo{author}{Hornik, K.} \&
  \bibinfo{author}{Zeileis, A.}
\newblock \bibinfo{journal}{\bibinfo{title}{Unbiased recursive partitioning: A
  conditional inference framework}}.
\newblock {\emph{\JournalTitle{Journal of Computational and Graphical
  Statistics}}} \textbf{\bibinfo{volume}{15}}, \bibinfo{pages}{651--674},
  \url{10.1198/106186006X133933} (\bibinfo{year}{2006}).

\bibitem{Barrett:2018:HPS:3195636.3158342}
\bibinfo{author}{Barrett, C.~L.}, \bibinfo{author}{Johnson, J.} \&
  \bibinfo{author}{Marathe, M.}
\newblock \bibinfo{journal}{\bibinfo{title}{{High Performance Synthetic
  Information Environments : An Integrating Architecture in the Age of
  Pervasive Data and Computing: Big Data (Ubiquity Symposium)}}}.
\newblock {\emph{\JournalTitle{Ubiquity}}} \textbf{\bibinfo{volume}{2018}},
  \bibinfo{pages}{1:1--1:11}, \url{10.1145/3158342} (\bibinfo{year}{2018}).

\bibitem{eia}
\bibinfo{author}{EIA}.
\newblock \bibinfo{title}{U.s. energy information administration}
  (\bibinfo{year}{2020}).

\bibitem{SWAN20091819}
\bibinfo{author}{Swan, L.~G.} \& \bibinfo{author}{Ugursal, V.~I.}
\newblock \bibinfo{journal}{\bibinfo{title}{{Modeling of end-use energy
  consumption in the residential sector: A review of modeling techniques}}}.
\newblock {\emph{\JournalTitle{Renewable and Sustainable Energy Reviews}}}
  \textbf{\bibinfo{volume}{13}}, \bibinfo{pages}{1819--1835},
  \url{https://doi.org/10.1016/j.rser.2008.09.033} (\bibinfo{year}{2009}).

\bibitem{MURATORI2013465}
\bibinfo{author}{Muratori, M.}, \bibinfo{author}{Roberts, M.~C.},
  \bibinfo{author}{Sioshansi, R.}, \bibinfo{author}{Marano, V.} \&
  \bibinfo{author}{Rizzoni, G.}
\newblock \bibinfo{journal}{\bibinfo{title}{{A highly resolved modeling
  technique to simulate residential power demand}}}.
\newblock {\emph{\JournalTitle{Applied Energy}}}
  \textbf{\bibinfo{volume}{107}}, \bibinfo{pages}{465--473},
  \url{https://doi.org/10.1016/j.apenergy.2013.02.057} (\bibinfo{year}{2013}).

\bibitem{SHIMODA20071617}
\bibinfo{author}{Shimoda, Y.}, \bibinfo{author}{Asahi, T.},
  \bibinfo{author}{Taniguchi, A.} \& \bibinfo{author}{Mizuno, M.}
\newblock \bibinfo{journal}{\bibinfo{title}{{Evaluation of city-scale impact of
  residential energy conservation measures using the detailed end-use
  simulation model}}}.
\newblock {\emph{\JournalTitle{Energy}}} \textbf{\bibinfo{volume}{32}},
  \bibinfo{pages}{1617--1633},
  \url{https://doi.org/10.1016/j.energy.2007.01.007} (\bibinfo{year}{2007}).

\bibitem{Tsuji2004}
\bibinfo{author}{Kiichiro, T.}, \bibinfo{author}{Fuminori, S.},
  \bibinfo{author}{Tsuyoshi, U.}, \bibinfo{author}{Osamu, S.} \&
  \bibinfo{author}{Takehiko, M.}
\newblock \bibinfo{journal}{\bibinfo{title}{{Bottom-Up Simulation Model for
  Estimating End-Use Energy Demand Profiles in Residential Houses}}}.
\newblock {\emph{\JournalTitle{{Proceedings from ACEEE Summer Studies on Energy
  Efficiency in Buildings}}}}  (\bibinfo{year}{2004}).

\bibitem{Subbiah2017}
\bibinfo{author}{Subbiah, R.}, \bibinfo{author}{Pal, A.},
  \bibinfo{author}{Nordberg, E.~K.}, \bibinfo{author}{Marathe, A.} \&
  \bibinfo{author}{Marathe, M.~V.}
\newblock \bibinfo{journal}{\bibinfo{title}{{Energy Demand Model for
  Residential Sector: A First Principles Approach}}}.
\newblock {\emph{\JournalTitle{IEEE Transactions on Sustainable Energy}}}
  \textbf{\bibinfo{volume}{8}}, \bibinfo{pages}{1215--1224},
  \url{10.1109/TSTE.2017.2669990} (\bibinfo{year}{2017}).

\bibitem{NREL2017-thermostat-study}
\bibinfo{author}{Chuck, B.} \emph{et~al.}
\newblock \bibinfo{journal}{\bibinfo{title}{{Residential Indoor Temperature
  Study}}}.
\newblock {\emph{\JournalTitle{National Renewable Energy Laboratory. Technical
  Report NREL/TP-5500-68019}}}  (\bibinfo{year}{2017}).

\bibitem{Vajen2001H2o}
\bibinfo{author}{Ulrike, J.} \& \bibinfo{author}{Klaus, V.}
\newblock \bibinfo{journal}{\bibinfo{title}{{Realistic Domestic Hot-Water
  Profiles in Different Time Scales}}}.
\newblock {\emph{\JournalTitle{{Universit\"{a}t Marburg}}}}
  \bibinfo{pages}{1--18} (\bibinfo{year}{2001}).

\bibitem{DESANTIAGO2017SwissH2O}
\bibinfo{author}{{de Santiago}, J.}, \bibinfo{author}{Rodriguez-Villalón, O.}
  \& \bibinfo{author}{Sicre, B.}
\newblock \bibinfo{journal}{\bibinfo{title}{The generation of domestic hot
  water load profiles in swiss residential buildings through statistical
  predictions}}.
\newblock {\emph{\JournalTitle{Energy and Buildings}}}
  \textbf{\bibinfo{volume}{141}}, \bibinfo{pages}{341--348},
  \url{https://doi.org/10.1016/j.enbuild.2017.02.045} (\bibinfo{year}{2017}).

\bibitem{Hendron2010H2o}
\bibinfo{author}{Bob, H.}, \bibinfo{author}{Jay, B.} \& \bibinfo{author}{Greg,
  B.}
\newblock \bibinfo{journal}{\bibinfo{title}{{Tool for Generating Realistic
  Residential Hot Water Event Schedules}}}.
\newblock {\emph{\JournalTitle{SimBuild Conference}}}  (\bibinfo{year}{2010}).

\bibitem{ROULEAU2019H2oCanada}
\bibinfo{author}{Rouleau, J.}, \bibinfo{author}{Ramallo-González, A.~P.},
  \bibinfo{author}{Gosselin, L.}, \bibinfo{author}{Blanchet, P.} \&
  \bibinfo{author}{Natarajan, S.}
\newblock \bibinfo{journal}{\bibinfo{title}{{A unified probabilistic model for
  predicting occupancy, domestic hot water use and electricity use in
  residential buildings}}}.
\newblock {\emph{\JournalTitle{{Energy and Buildings}}}}
  \textbf{\bibinfo{volume}{202}}, \bibinfo{pages}{109375},
  \url{https://doi.org/10.1016/j.enbuild.2019.109375} (\bibinfo{year}{2019}).

\bibitem{Hendron2008Building}
\bibinfo{author}{Hendron, R.}
\newblock \bibinfo{journal}{\bibinfo{title}{{Building America Research
  Benchmark Definition, Technical Report NREL/TP-550-44816}}}.
\newblock {\emph{\JournalTitle{National Renewable Energy Laboratory Reports}}}
  (\bibinfo{year}{2008}).

\bibitem{Capasso1994}
\bibinfo{author}{Capasso, A.}, \bibinfo{author}{Grattieri, W.},
  \bibinfo{author}{Lamedica, R.} \& \bibinfo{author}{Prudenzi, A.}
\newblock \bibinfo{journal}{\bibinfo{title}{A bottom-up approach to residential
  load modeling}}.
\newblock {\emph{\JournalTitle{IEEE Transactions on Power Systems}}}
  \textbf{\bibinfo{volume}{9}}, \bibinfo{pages}{957--964},
  \url{10.1109/59.317650} (\bibinfo{year}{1994}).

\bibitem{WIDEN2009Light}
\bibinfo{author}{Widén, J.}, \bibinfo{author}{Nilsson, A.~M.} \&
  \bibinfo{author}{Wäckelgård, E.}
\newblock \bibinfo{journal}{\bibinfo{title}{{A combined Markov-chain and
  bottom-up approach to modelling of domestic lighting demand}}}.
\newblock {\emph{\JournalTitle{Energy and Buildings}}}
  \textbf{\bibinfo{volume}{41}}, \bibinfo{pages}{1001--1012},
  \url{https://doi.org/10.1016/j.enbuild.2009.05.002} (\bibinfo{year}{2009}).

\bibitem{PALACIOSGARCIA2015Light}
\bibinfo{author}{Palacios-Garcia, E.} \emph{et~al.}
\newblock \bibinfo{journal}{\bibinfo{title}{{Stochastic model for lighting's
  electricity consumption in the residential sector. Impact of energy saving
  actions}}}.
\newblock {\emph{\JournalTitle{Energy and Buildings}}}
  \textbf{\bibinfo{volume}{89}}, \bibinfo{pages}{245--259},
  \url{https://doi.org/10.1016/j.enbuild.2014.12.028} (\bibinfo{year}{2015}).

\bibitem{STOKES2004103}
\bibinfo{author}{Stokes, M.}, \bibinfo{author}{Rylatt, M.} \&
  \bibinfo{author}{Lomas, K.}
\newblock \bibinfo{journal}{\bibinfo{title}{A simple model of domestic lighting
  demand}}.
\newblock {\emph{\JournalTitle{Energy and Buildings}}}
  \textbf{\bibinfo{volume}{36}}, \bibinfo{pages}{103--116},
  \url{https://doi.org/10.1016/j.enbuild.2003.10.007} (\bibinfo{year}{2004}).

\bibitem{RICHARDSON2009Light}
\bibinfo{author}{Richardson, I.}, \bibinfo{author}{Thomson, M.},
  \bibinfo{author}{Infield, D.} \& \bibinfo{author}{Delahunty, A.}
\newblock \bibinfo{journal}{\bibinfo{title}{{Domestic lighting: A
  high-resolution energy demand model}}}.
\newblock {\emph{\JournalTitle{Energy and Buildings}}}
  \textbf{\bibinfo{volume}{41}}, \bibinfo{pages}{781--789},
  \url{https://doi.org/10.1016/j.enbuild.2009.02.010} (\bibinfo{year}{2009}).

\bibitem{PaateroLund2006}
\bibinfo{author}{Paatero, J.~V.} \& \bibinfo{author}{Lund, P.~D.}
\newblock \bibinfo{journal}{\bibinfo{title}{A model for generating household
  electricity load profiles}}.
\newblock {\emph{\JournalTitle{International Journal of Energy Research}}}
  \textbf{\bibinfo{volume}{30}}, \bibinfo{pages}{273--290},
  \url{https://doi.org/10.1002/er.1136} (\bibinfo{year}{2006}).
\newblock \eprint{https://onlinelibrary.wiley.com/doi/pdf/10.1002/er.1136}.

\bibitem{BaselineLightingACEEE}
\bibinfo{author}{Tribwell, L.~S.} \& \bibinfo{author}{Lerman, D.~I.}
\newblock \bibinfo{journal}{\bibinfo{title}{{Baseline Residential Lighting
  Energy Use Study}}}.
\newblock {\emph{\JournalTitle{American Council for an Energy-Efficient Economy
  (ACEEE)}}}  (\bibinfo{year}{1996}).

\bibitem{DecadeReport2}
\bibinfo{author}{Boardman, B.} \emph{et~al.}
\newblock \bibinfo{journal}{\bibinfo{title}{{DECADE - Domestic Equipment and
  Carbon Dioxide Emissions}}}.
\newblock {\emph{\JournalTitle{{Energy and Environment Programme Environmental
  Change Unit University of Oxford}}}}  (\bibinfo{year}{1995}).

\bibitem{LBNLRefr2012}
\bibinfo{author}{Greenblatt, J.}, \bibinfo{author}{Hopkins, A.},
  \bibinfo{author}{Letschert, V.} \& \bibinfo{author}{Blasnik, M.}
\newblock \bibinfo{journal}{\bibinfo{title}{{Energy use of US residential
  refrigerators and freezers: function derivation based on household and
  climate characteristics}}}.
\newblock {\emph{\JournalTitle{Energy Analysis and Environmental Impacts
  Department Environmental Energy Technologies Division Lawrence Berkeley
  National Laboratory}}}  (\bibinfo{year}{2012}).

\bibitem{NIST_dw_cw_cd}
\bibinfo{author}{Christopher, I.}, \bibinfo{author}{Natascha~Milesi, F.} \&
  \bibinfo{author}{Michael~A., G.}
\newblock \bibinfo{journal}{\bibinfo{title}{{Consumer Use of Dishwashers,
  Clothes Washers, and Dryers: Data Needs and Availability}}}.
\newblock {\emph{\JournalTitle{{NIST Technical Note 1696, Mechanical Systems
  and Control Group Building Environment Division Engineering Laboratory,
  Department of Energy}}}}  (\bibinfo{year}{2011}).

\bibitem{energystar_computer_v5}
\bibinfo{author}{EnergyStar}.
\newblock \bibinfo{journal}{\bibinfo{title}{{ENERGY STAR Program Requirements
  for Computers}}}.
\newblock {\emph{\JournalTitle{{ENERGY STAR Report}}}}  (\bibinfo{year}{2010}).

\bibitem{energystar_TV_2021}
\bibinfo{author}{EnergyStar}.
\newblock \bibinfo{journal}{\bibinfo{title}{{Product Retrospective: TVs}}}.
\newblock {\emph{\JournalTitle{{ENERGY STAR Report}}}}  (\bibinfo{year}{2021}).

\bibitem{energystar_vacuum_2011}
\bibinfo{author}{Palmstedt, P.}
\newblock \bibinfo{journal}{\bibinfo{title}{{Vacuum Cleaners}}}.
\newblock {\emph{\JournalTitle{{ENERGY STAR Market \& Industry Scoping
  Report}}}}  (\bibinfo{year}{2011}).

\bibitem{electrolux_vacuum_2013}
\bibinfo{author}{Palmstedt, P.}
\newblock \bibinfo{journal}{\bibinfo{title}{{Electrolux Global Vacuuming Survey
  2013 Report}}}.
\newblock {\emph{\JournalTitle{Electrolux}}}  (\bibinfo{year}{2013}).

\bibitem{thorve_scidata2022_doi}
\bibinfo{author}{Thorve, S.}, \bibinfo{author}{Mortveit, H.} \&
  \bibinfo{author}{Marathe, M.}
\newblock \bibinfo{journal}{\bibinfo{title}{{Household-level disaggregated
  hourly synthetic residential energy use profiles for the United States}}}.
\newblock {\emph{\JournalTitle{{University of Virginia Dataverse}}}}
  \url{https://doi.org/10.18130/V3/VJUZSH} (\bibinfo{year}{2022}).

\bibitem{BAClimateZones}
\bibinfo{author}{Michael, B.~C.}, \bibinfo{author}{Theresa, G.~L.},
  \bibinfo{author}{C., P.~C.}, \bibinfo{author}{Marye, H.} \&
  \bibinfo{author}{Kathi, R.}
\newblock \bibinfo{journal}{\bibinfo{title}{{High-Performance Home
  Technologies: Guide to Determining Climate Regions by County}}}.
\newblock {\emph{\JournalTitle{Pacific Northwest National Laboratory}}}
  \textbf{\bibinfo{volume}{7.3}}, \bibinfo{pages}{1--50},
  \url{https://www.energy.gov/eere/buildings/downloads/building-america-best-practices-series-volume-73-guide-determining-climate}
  (\bibinfo{year}{2015}).

\bibitem{NEEAData}
\bibinfo{title}{Residential building stock assessment (rbsa) metering data,
  northwest energy efficiency alliance}.
\newblock
  \bibinfo{howpublished}{\url{https://neea.org/data/residential-building-stock-assessment}}.
\newblock \bibinfo{note}{Accessed: 2022-03-23}.

\bibitem{LADPU_dryad}
\bibinfo{author}{Souza, V.}, \bibinfo{author}{Estrada, T.},
  \bibinfo{author}{Bashir, A.} \& \bibinfo{author}{Mueen, A.}
\newblock \bibinfo{journal}{\bibinfo{title}{{LADPU Smart Meter Data}}}.
\newblock {\emph{\JournalTitle{Dryad}}}
  \url{https://doi.org/10.5061/dryad.m0cfxpp2c} (\bibinfo{year}{2020}).

\bibitem{JS1991}
\bibinfo{author}{{Lin}, J.}
\newblock \bibinfo{journal}{\bibinfo{title}{{Divergence measures based on the
  Shannon entropy}}}.
\newblock {\emph{\JournalTitle{IEEE Transactions on Information Theory}}}
  \textbf{\bibinfo{volume}{37}}, \bibinfo{pages}{145--151},
  \url{10.1109/18.61115} (\bibinfo{year}{1991}).

\bibitem{kwac2014}
\bibinfo{author}{Kwac, J.}, \bibinfo{author}{Flora, J.} \&
  \bibinfo{author}{Rajagopal, R.}
\newblock \bibinfo{journal}{\bibinfo{title}{Household energy consumption
  segmentation using hourly data}}.
\newblock {\emph{\JournalTitle{IEEE Transactions on Smart Grid}}}
  \textbf{\bibinfo{volume}{5}}, \bibinfo{pages}{420--430},
  \url{10.1109/TSG.2013.2278477} (\bibinfo{year}{2014}).

\end{thebibliography}

\section*{Acknowledgements} 
We thank the anonymous reviewers for their very helpful comments that helped us improve
the manuscript. We thank members of National Rural Electric Cooperative Association (NRECA)
for providing validation data for Rappahannock county, Virginia and Horry county, South Carolina. 
This work is partially supported by University of Virginia Strategic Investment Fund award number SIF160,
NSF EAGER CMMI-1745207,
NSF Grant OAC-1916805, 
and NSF BIGDATA IIS-1633028.

\section*{Author contributions statement}
S.T. collected data for the models, 
developed and implemented the modeling framework and all the individual energy use models,  prepared the manuscript;
Y.Y.B. implemented the ATUS model, read and edited the paper;
H.M and S.S. helped with model development,  data collection, 
read and edited the manuscript, provided guidance in writing Background and Summary;
M.M. worked on model development, validation, manuscript preparation, 
provided feedback for the Methodology and Validation section;
A.M. read and edited the manuscript and helped with validation;
A.V. edited the manuscript and helped with validation.
All authors participated in writing and reviewing the manuscript.

\section*{Competing interests}
The authors declare no competing interests.

\end{document}